\documentclass[twocolumn,english,aps,pra,superscriptaddress]{revtex4-1}
\usepackage[T1]{fontenc}
\usepackage[latin9]{inputenc}
\usepackage{color}
\usepackage{float}
\usepackage{amsmath}
\usepackage{amssymb}
\usepackage{graphicx}
\usepackage{esint}

\makeatletter

\newcommand{\lyxdot}{.}

\makeatother

\usepackage{babel}
\begin{document}

\title{Overcoming decoherence of cat-states formed in a cavity using squeezed-state
inputs}

\author{R. Y. Teh}

\affiliation{Centre for Quantum and Optical Science, Swinburne University of Technology,
Melbourne 3122, Australia}

\author{P. D. Drummond}

\affiliation{Centre for Quantum and Optical Science, Swinburne University of Technology,
Melbourne 3122, Australia}

\author{M. D. Reid}

\affiliation{Centre for Quantum and Optical Science, Swinburne University of Technology,
Melbourne 3122, Australia}
\begin{abstract}
A cat-state is a superposition of two coherent states with amplitudes
$\alpha_{0}$ and $-\alpha_{0}$.  Recent experiments create cat states
in a microwave cavity field using superconducting circuits. As with
degenerate parametric oscillation (DPO) in an adiabatic and highly
nonlinear limit, the states are formed in a signal cavity mode via
a two-photon dissipative process induced by the down conversion of
a pump field to generate pairs of signal photons. The damping of the
signal and the presence of thermal fluctuations rapidly decoheres
the state, and the effect on the dynamics is to either destroy the
possibility of a cat state, or else to sharply reduce the lifetime
and size of the cat-states that can be formed. In this paper, we
study the effect on both the DPO and microwave systems of a squeezed
reservoir coupled to the cavity. While the threshold nonlinearity
is not altered, we show that the use of squeezed states significantly
lengthens the lifetime of the cat states. This improves the feasibility
of generating cat states of large amplitude and with a greater degree
of quantum macroscopic coherence, which is necessary for many quantum
technology applications. Using current experimental parameters for
the microwave set-up, which requires a modified Hamiltonian, we further
demonstrate how squeezed states enhance the quality of the cat states
that could be formed in this regime. Squeezing also combats the significant
decoherence due to thermal noise, which is relevant for microwave
fields at finite temperature. By modeling a thermal squeezed reservoir,
we show that the thermal decoherence of the dynamical cat states can
be inhibited by a careful control of the squeezing of the reservoir.
To signify the quality of the cat state, we consider different signatures
including fringes and negativity, and the $C_{l_{1}}$ measure of
quantum coherence.
\end{abstract}
\maketitle

\section{Introduction}

Schr\"odinger raised the question of how to interpret a macroscopic
quantum superposition state in his essay of 1935 \citep{schrodinger_1935}.
In his analysis, a macroscopic object (a cat) becomes entangled with
a microscopic system, creating a paradoxical state that would seem
to defy some type of macroscopic reality. The paradoxical state is
a superposition of two macroscopically distinguishable states, like
a cat being dead or alive. The essay has motivated many papers, including
\citep{Brune_PRL1996,Monroe_Science1996,Friedman:2000aa,Ourjoumtsev_nature2007,Palacios-Laloy:2010aa,Vlastakis_Science2013,Wang_Science2016,PhysRevLett.57.13,knee2016strict,leggett1985quantum,marshall2003towards,vanner2011pulsed,vanner2011selective,asadian2014probing,Teh_PRA2018,budroni2015quantum,opanchuk2016quantifying,rosales2018leggett,thenabadu2019leggett}\textcolor{red}{}and
those that develop decoherence theories to eliminate the possibility
of such superposition states forming for massive objects \citep{Bassi_RMP2013}.

It remains a challenge to generate a macroscopic superposition where
the two states involved are truly macroscopically different. According
to quantum mechanics, the coupling between the system and its environment
tends to destroy the quantum coherence of the superposition state,
particularly as the two states involved in the superposition become
macroscopically distinct \citep{PhysRevLett.57.13,Brune_PRL1996,Caldeira_PRA1985,Walls_PRA1985}.
Quantum mechanics however predicts the existence of macroscopic superposition
states, of arbitrary size, in the absence of decoherence. A fundamental
question is whether macroscopic quantum superpositions states can
actually exist, or whether quantum mechanics needs modification. This
provides motivation to generate superposition states of a larger size,
and to test the predictions of quantum mechanics in the presence of
decoherence.

Cat states based on coherent states are promising for such studies,
as well as for applications in quantum information science \citep{Leghtas_PRA2013,Leghtas_Science2015,Mirrahimi_NJP2014}.
A cat-state is a quantum superposition of two single-mode coherent
states with amplitudes $\alpha_{0}$ and $-\alpha_{0}$, which become
widely separated in phase space as $\alpha_{0}\rightarrow\infty$.
Recent experiments are successful at creating such states in a microwave
cavity using superconducting circuits to enhance the nonlinearity.
Cat states with $\alpha_{0}\sim10$ and a high measure of quantum
coherence have been generated in these experiments \citep{Wang_Science2016,Vlastakis_Science2013,Leghtas_Science2015}.
The coupling of the system to the environment induces decoherence
however, which destroys the superposition nature of the cat state
as the separation in phase space of the two coherent states becomes
larger. Decoherence arises as a result of photon loss from the cavity
mode \citep{Wolinsky_PRL1988,Reid_PRA1992,Hach_PRA1994,Gilles_PRA1994,carmichael_QO4},
and from thermal noise which is relevant at microwave frequencies
even at low temperatures \citep{Kennedy_PRA1988,haroche2003quantum,el2003evolution,serafini2004decoherence,serafini2004minimum,serafini2005quantifying,paavola2011finite,rosales2015decoherence,Teh_PRA2018,tan2014enhanced,bennett2018rapid,xiong2019generation}.

In this paper, we demonstrate how squeezed states may enhance the
formation of cat states by modifying the decoherence. Squeezed states
are single mode states of a quantum field that have a variance in
one quadrature phase amplitude reduced below the standard quantum
limit \citep{yuen1976two,Caves_PRD1981}. Such states have been extensively
studied \citep{Caves_PRD1981,milburn1981production,Yurke_PRA1984,Yurke_Denker_PRA1984,Collett_PRA1984,reid1984quantum,reid1985squeezing,Fabre_1990detune,Movshovich_squeezing_PRL1990,squeezing_purdy_PRX2013,squeezing_Wollman_Science2015,castellanos2008amplification}
and were first created experimentally at optical frequencies \citep{Slusher_squeezing_PRL1985,Heidmann_PRL1987,squeezing_schnabel_PRL2008,squeezing_Schnabel_OptExp2011,mehmet2011squeezed,vahlbruch2016detection}.
Caves proposed squeezed states as a means to control the vacuum fluctuations
entering the input port of the LIGO interferometer, so that greater
sensitivities for detecting gravitational waves could be achieved
\citep{Caves_PRD1981}. This has recently been implemented, and there
are numerous other potential applications (including \citep{taylor2013biological,vahlbruch2016detection,Toscano_metrology_PRA2006,Tombesi_JosaB1987,lane1988absorption,Gardiner_PRL1986,tombesi1994physical,braunstein1998teleportation}).

Meccozi and Tombesi \citep{Mecozzi_PRL1987_squeeze_environ,Tombesi_JosaB1987},
and Kennedy and Walls \citep{Kennedy_PRA1988} first suggested using
squeezed states to engineer the environment coupled to a macroscopic
superposition state, and showed that squeezing in an optimally selected
quadrature can suppress the decoherence that otherwise destroys the
cat state. This was further explored by Munro and Reid \citep{Munro_PRA1995},
who studied the effect of a squeezed reservoir on the dynamical creation
of cat states generated in a highly nonlinear degenerate parametric
oscillator (DPO). That treatment however did not consider cat states
with amplitudes $\alpha_{0}>10$ and was limited to a model applicable
to optical rather than microwave systems.

We study the dynamical formation of cat states in a more general description
of a degenerate parametric oscillator (DPO) and show that the use
of squeezed states can significantly increase the lifetime of the
cat state in the presence of decoherence. By including an extra Kerr
term in the Hamiltonian for the DPO, we are able to model the current
microwave experiments, thus providing full solutions showing generation
of cat states with $\alpha_{0}\sim2$, in both the optical and microwave
regime. Our conclusions are that the Wigner negativity can be enhanced
by a factor of order $2$ for parameters corresponding to current
microwave experiments. While decoherence can be slowed, the squeezed
states have little effect on the threshold nonlinearity required for
the cat state to form.

The study of the dynamics of the formation of cat states is motivated
by potential application in quantum information \citep{Leghtas_PRA2013},
and also by the development of the Coherent Ising Machine (CIM), an
optimization device capable of solving NP-hard problems \citep{wang2013coherent,Marandi_CIM_Nature2014,shoji2017quantum,yamamura2017quantum,McMahon_CIM_science2019,Inagaki_CIM_science2019}.
The CIM is based on a network of DPOs which currently operate mainly
in an optical regime where the reduced nonlinearity makes formation
of cat states difficult. However, cat-state regimes may offer advantages
due to the enhancement of quantum coherence. Recent proposals \citep{Nigg_PRL2012,Goto_PRA2019}
and experiments \citep{Pfaff_Nature2017} exist for generating cat-states
in a DPO, which may be applied to a DPO network for adiabatic quantum
computation \citep{Goto_SciRep2016,Goto_PRA2016,Puri_Nature2017}
to solve these NP-hard problems. The application of squeezed states
to the CIM has been more recently studied and shown to potentially
enhance performance \citep{luo2020speed,maruo2016truncated}.\textcolor{red}{}

Cat states are predicted to be created in the signal cavity mode of
the DPO through a mechanism based on a two-photon dissipative process
\citep{Wolinsky_PRL1988,Gilles_PRA1994,Hach_PRA1994,Leghtas_Science2015}.
The two-photon process originates from the parametric interaction
in which pump photons are down-converted to correlated pairs of signal
photons \citep{Drummond_OpActa1981}, and can be realized in both
optical and microwave cavities. However, the presence of signal losses
significantly alters the dynamics, so that a cat state is not generated
unless there is a very strong parametric nonlinearity \citep{Wolinsky_PRL1988,Krippner_PRA1994,Munro_PRA1995,Sun_NJP2019,Sun_PRA2019,teh2020dynamics,carmichael_QO4}.
While not yet feasible for optical modes, this can be achieved for
microwave superconducting circuit experiments, where the cat-states
are generated as a transient state \citep{Leghtas_Science2015,teh2020dynamics}.
In this paper, we study the dynamics of the cat states and their decoherence
as the signal losses are increased. With squeezed input states into
the cavity, we then explain that although the threshold nonlinearity
is not reduced, decoherence can be compensated for to allow cat states
of longer lifetimes and with larger amplitudes $\alpha_{0}$ to be
formed,

Thermal noise also destroys the cat state, and we therefore investigate
the effect of squeezed thermal reservoirs on the feasibility of creating
the cat states. We model the squeezed inputs for temperatures corresponding
to the microwave experiment of Leghtas et al. \citep{Leghtas_Science2015}.
We allow for both a thermalized squeezed reservoir, where the cavity
evolves coupled to a finite temperature environment \citep{Fearn_JModOpt1988},
and also the input of a squeezed thermal state \citep{Kim_squeezed_PRA1989},
where the squeeze state is created using a finite temperature source
\citep{Movshovich_squeezing_PRL1990}. We show that an optimal amount
of squeezing will enhance the cat state that is formed. Our work complements
previous treatments by Kennedy and Walls \citep{Kennedy_PRA1988},
Serafini et al \citep{serafini2005quantifying,serafini2004minimum}
and Bennett and Bowen \citep{bennett2018rapid} who either analyze
the effect of squeezing applied to a cat state generated in an optomechanical
cavity mode, as in \citep{gilchrist1999contradiction}, or else examine
cat states coupled to thermal reservoirs where a squeezed thermal
state is injected.\textcolor{blue}{}\textcolor{red}{}

For any experimental situation, the cat states that are generated
are not pure states. This is particularly relevant for the cat states
generated in a cavity where signal loss is important. As such, it
is necessary to carefully signify the existence and quality of the
cat states \citep{Frowis_RMP2018}. In this paper, we certify cat
states using three different criteria: we use interference fringes
\citep{Ourjoumtsev_nature2007,PhysRevLett.57.13}, the negativity
of the Wigner function \citep{Kenfack_2004}, and the $C_{l_{1}}$
measure of quantum coherence \citep{Baumgratz_PRL2014}.

\section{Hamiltonian and Master equation with a squeezed reservoir}

We begin by reviewing the model of degenerate parametric oscillation
and the master equation formalism whereby the interaction between
the system and the external environment can be taken into account.
The degenerate parametric oscillator has two modes, the pump mode
and signal mode with a frequency of $2\omega$ and $\omega$ respectively.
These modes are resonant in a cavity. An external pumping field at
the frequency $2\omega$ creates photons in the pump mode, which in
turn interacts with a non-linear crystal that generates correlated
pairs of photons in the signal mode at half the frequency of the pump
mode.

There is single-photon damping in both the signal and pump modes with
rates $\gamma_{1}$ and $\gamma_{2}$ respectively. These damping
rates model the leakage of photons out of the cavity. We assume this
takes place through an end-mirror of the cavity. We consider a single-sided
cavity for the signal mode, where one mirror is totally reflecting,
thus not allowing signal photons to leak through. Typically, the pump
mode decay rate is much larger than the signal mode decay rate ($\gamma_{2}\gg\gamma_{1}$),
which allows the pump mode to be adiabatically eliminated. There is
also a two-photon loss process due to the conversion of two signal
mode photons back to a single pump mode photon.

The Hamiltonian that captures all these effects, after carrying out
the adiabatic elimination procedure, and also including a Kerr effect,
is given by \textcolor{black}{\citep{Drummond_OpActa1981,Kheruntsyan_OptCom1996,Sun_NJP2019,Sun_PRA2019}}\textcolor{red}{}
\begin{align}
H & =i\hbar\left(\frac{\bar{g}\epsilon}{\gamma_{2}}a_{1}^{\dagger2}-\frac{\bar{g}^{*}\epsilon^{*}}{\gamma_{2}}a_{1}^{2}\right)+\hbar\bar{\chi}a_{1}^{\dagger2}a_{1}^{2}\nonumber \\
 & +a_{1}^{\dagger}\Gamma_{1}+a_{1}\Gamma_{1}^{\dagger}+\frac{\left|\bar{g}\right|^{2}}{4\gamma_{2}}\left(a_{1}^{2}\Gamma_{2}^{\dagger}+a_{1}^{\dagger2}\Gamma_{2}\right)\,.\label{eq:simplified_hamiltonian}
\end{align}
Here, $\bar{g}$ is the parametric coupling strength between the pump
and signal, and $\bar{\chi}$ is an anharmonic Kerr effect nonlinearity
in the signal mode. The terms $\Gamma_{1}$ and $\Gamma_{2}$ are
operators for the external reservoirs that give rise to single-photon
and two-photon loss processes respectively, which also allows finite
temperature effects to be included.

Next, we discuss the nature of the environment that interacts with
the signal mode. This can be controlled via the input field to the
single end-mirror of the cavity, which allows leakage of the signal
cavity photons into the external environment, and can be formalized
using input-output cavity theory \citep{Yurke_PRA1984,Yurke_Denker_PRA1984,Collett_PRA1984,PhysRevA.31.3761}.
As shown by Gardiner and Collett \citep{PhysRevA.31.3761}, a bath
with quantum fluctuations corresponding to a squeezed state, at a
frequency $\omega$ has the following statistical moments:
\begin{align}
\langle\Gamma_{1}^{\dagger}\left(t\right)\Gamma_{1}\left(t'\right)\rangle & =2\gamma_{1}N\delta\left(t-t'\right)\nonumber \\
\langle\Gamma_{1}\left(t\right)\Gamma_{1}^{\dagger}\left(t'\right)\rangle & =2\gamma_{1}\left(N+1\right)\delta\left(t-t'\right)\nonumber \\
\langle\Gamma_{1}\left(t\right)\Gamma_{1}\left(t'\right)\rangle & =2\gamma_{1}Me^{-2i\omega t}\delta\left(t-t'\right)\nonumber \\
\langle\Gamma_{1}^{\dagger}\left(t\right)\Gamma_{1}^{\dagger}\left(t'\right)\rangle & =2\gamma_{1}M^{*}e^{2i\omega t}\delta\left(t-t'\right)\,,\label{eq:noise_statistics}
\end{align}
where $N$ is a positive, real number characterizing the mean photon
number of the state of the environment, and $M=|M|e^{i\Phi}$ is \textcolor{blue}{}\textcolor{red}{{}
} a complex number that characterizes the nature of the squeezing.
\textcolor{blue}{} When $M=0$, the bath fluctuations are not squeezed\textcolor{black}{.
When $M\neq0$} at zero temperature,\textcolor{black}{{} the reservoir
is in a squeezed state centered around the frequency $\omega$ and
$N$ is the mean photon number of the corresponding squeezed state.
The link between $N$ and $M$ depends on the precise model for the
squeezed reservoir that we adopt.}

Thermal noise is a secondary source of decoherence known to destroy
the quantum coherence of the cat states \citep{Kennedy_PRA1988}.
In this work, we mainly focus on the simple case where a squeezed
vacuum is applied to the system at zero temperature, in which case
$N$ will be zero if there is no squeezing the reservoir fluctuations.
The zero temperature limit is justified for optical frequencies where
thermal noise is low even at room temperature, and for microwave systems
is achievable by cooling.

However, to gain some insight into the sensitivity of the quantum
coherences on temperature, our equations can allow for thermal noise,
which may be modeled in two ways. We may consider the reservoir to
be a thermalized squeezed state or else a squeezed thermal state \citep{Fearn_JModOpt1988,Kim_squeezed_PRA1989}.
The statistics arising in each case, and the resulting effects on
the decoherence of an ideal cat state is summarized in the Appendix.
In both cases, the reservoir has a mean photon number $N=N_{th}+N_{s}$
that is a sum of a number $N_{th}$ due to finite temperature, and
the photon number due to the squeezed state $N_{s}$. The corresponding
statistical moments can be expressed in the form given by Eq. (\ref{eq:noise_statistics})\textcolor{red}{}.
However, for a squeezed thermal state (defined by a squeezing interaction
acting on a system prepared initially in a thermal state), the mean
photon number contains a cross-term between the squeezing and thermal
noise contribution \citep{Fearn_JModOpt1988} which leads to a stronger
reduction in quantum noise\textcolor{blue}{}. We note that while
thermal noise is significant at microwave frequencies, squeezing at
this frequency has been achieved experimentally \citep{Movshovich_squeezing_PRL1990,squeezing_Castellanos-Beltran_nature2008,squeezing_purdy_PRX2013,squeezing_Wollman_Science2015,Ockeloen-Korppi_squeezing_PRL2017}.
The states generated at a finite temperature by some squeezing mechanism
would be modeled by a squeezed thermal state.

We may also justify the model of the reservoir for the cavity field
where $N_{th}$ is due to the temperature $T$ of the environment
of the cavity as the cat states evolve. Here we suppose the squeezed
fields to be generated at zero temperature, so that the thermal noise
$N_{th}$ term is independent of the characterization of the squeezing
that enters the cavity. This model corresponds to a thermalized squeezed
reservoir. 

\textcolor{black}{We need to know how $N$ and $M$ relate to the
amount of squeezing of the reservoir input. For a squeezed state produced
at zero temperature from an optical parametric oscillator, we have
$N_{th}=0$, and the values of $N_{s}$ and $M$ are linked, the degree
of squeezing being determined by $N$. In particular, it is known
that the optimal squeezing produced from an optical parametric oscillator
at zero temperature is achieved when $\left|M\right|=\sqrt{N_{s}\left(N_{s}+1\right)}$
\citep{Collett_PRA1984,PhysRevA.31.3761,Gardiner_PRL1986,Kennedy_PRA1988,Munro_PRA1995,gardiner2004quantum}.}
To this end, we relate the amount of squeezing in the quadrature to
the parameters $N_{s}$ and $M$. Recall that the bath which interacts
with the signal mode is modeled by a harmonic oscillator satisfying
a set of noise correlations as given in Eq. (\ref{eq:noise_statistics}).
Defining the general rotated $X_{\theta}$ and $P_{\theta}$ quadratures
for this bath mode as
\begin{align}
X_{\theta} & \equiv\frac{1}{\sqrt{2}}\left(\Gamma\text{e}^{-i\theta}+\Gamma^{\dagger}\text{e}^{i\theta}\right)\nonumber \\
P_{\theta} & \equiv\frac{1}{i\sqrt{2}}\left(\Gamma\text{e}^{-i\theta}-\Gamma^{\dagger}\text{e}^{i\theta}\right)\,,\label{eq:noise_X_P-1}
\end{align}
we calculate the variances of $X_{\theta}$ and $P_{\theta}$ with
squeezing, with respect to their corresponding variances in $X_{\theta}$
and $P_{\theta}$ of a vacuum state. They are, using the statistical
moments in Eq. (\ref{eq:noise_statistics}), given by
\begin{align}
\Delta^{2}X_{\theta} & =\langle\Gamma^{\dagger}\Gamma\rangle+1+\frac{1}{2}\langle\Gamma^{2}\rangle e^{-2i\theta}+\frac{1}{2}\langle\Gamma^{\dagger2}\rangle e^{2i\theta}\nonumber \\
 & =2N_{s}+1+\left|M\right|\left[e^{i\left(\Phi-2\theta\right)}+e^{-i\left(\Phi-2\theta\right)}\right]\nonumber \\
\Delta^{2}P_{\theta} & =\langle\Gamma^{\dagger}\Gamma\rangle+1-\frac{1}{2}\langle\Gamma^{2}\rangle e^{-2i\theta}-\frac{1}{2}\langle\Gamma^{\dagger2}\rangle e^{2i\theta}\nonumber \\
 & =2N_{s}+1-\left|M\right|\left[e^{i\left(\Phi-2\theta\right)}+e^{-i\left(\Phi-2\theta\right)}\right]\,,\label{eq:vars-1}
\end{align}
where $M=\left|M\right|e^{i\Phi}$ as previously defined. Choosing
$M$ so that $\Phi=2\theta$  and $\left|M\right|=\sqrt{N_{s}\left(N_{s}+1\right)}$,
in the limit of large $N_{s}$, the variances of $X_{\theta=0}=X$
and $P_{\theta=0}=P$ are 
\begin{align}
\Delta^{2}X & \sim4\left(N_{s}+\frac{1}{2}\right)\nonumber \\
\Delta^{2}P & \sim\frac{1}{4N_{s}}\,.\label{eq:vars_largeN_limit}
\end{align}
The fluctuations in the $P$-quadrature can be negligibly small in
this limit. From Eq. (\ref{eq:vars-1}), squeezing in the quadrature
$P_{\theta}$ is obtained when $\Phi$ is chosen such that $\Phi=2\theta$.
We will consider here $\theta=0$, so that $M$ is real and positive.

The full dynamics including all the effects discussed in previous
paragraphs is dictated by the master equation, in the Markovian approximation,
as given below:\begin{widetext}
\begin{align}
\frac{\partial}{\partial t}\rho & =\frac{\left|\bar{g}\epsilon\right|}{2\gamma_{2}}\left[a^{\dagger2}-a^{2},\rho\right]+\frac{1}{2}\left(\frac{\bar{g}^{2}}{2\gamma_{2}}\right)\left(2a^{2}\rho a^{\dagger2}-a^{\dagger2}a^{2}\rho-\rho a^{\dagger2}a^{2}\right)-i\frac{\bar{\chi}}{2}\left[a^{\dagger2}a^{2},\rho\right]\nonumber \\
 & +\left(N_{th}+N_{s}+1\right)\gamma_{1}\left[2a\rho a^{\dagger}-a^{\dagger}a\rho-\rho a^{\dagger}a\right]+\left(N_{th}+N_{s}\right)\gamma_{1}\left[2a^{\dagger}\rho a-aa^{\dagger}\rho-\rho aa^{\dagger}\right]\nonumber \\
 & -M\gamma_{1}\left[2a\rho a-aa\rho-\rho aa\right]-M^{*}\gamma_{1}\left[2a^{\dagger}\rho a^{\dagger}-a^{\dagger}a^{\dagger}\rho-\rho a^{\dagger}a^{\dagger}\right]\,.\label{eq:master_eqn}
\end{align}
\end{widetext}With no loss of generality, we can choose the phase
of $\bar{g}$ such that $\bar{g}\epsilon=\bar{g}^{*}\epsilon^{*}$
\citep{carmichael_QO4,Kinsler_PRA1991}. Here, $\rho$ is the density
operator of the signal mode. It is convenient to scale all parameters
with respect to the signal mode single-photon decay rate $\gamma_{1}$.
This defines a dimensionless time $\tau=\gamma_{1}t$; a dimensionless
pump strength $\lambda=\left|\bar{g}\epsilon\right|/\left(\gamma_{1}\gamma_{2}\right)$,
dimensionless effective parametric interaction $g=\sqrt{\bar{g}^{2}/\left(2\gamma_{1}\gamma_{2}\right)}$
and a dimensionless Kerr strength $\chi'=\bar{\chi}/\gamma_{1}$.
Note that in the absence of single-photon damping ($\gamma_{1}=0$),
the steady state solution for the master equation (\ref{eq:master_eqn})
for an initial vacuum state corresponds to a cat state 
\begin{equation}
|\psi_{c}\rangle=\mathcal{N}(|\alpha_{0}\rangle+|-\alpha_{0}\rangle)\label{eq:cat-state}
\end{equation}
where $\alpha_{0}=\sqrt{\lambda/\left(g^{2}+i\chi'\right)}=\sqrt{\lambda/g^{2}\left(1+i\chi\right)}$,
$\chi\equiv\chi'/g^{2}$\textcolor{red}{{} }and $\mathcal{N=}\left[2\left(1+e^{-2\left|\alpha_{0}\right|^{2}}\right)\right]^{-1/2}$
is the normalization constant. This cat state has an even photon
number as required for a two-photon process arising from the vacuum
\citep{Hach_PRA1994,Gilles_PRA1994}.

In the presence of damping, the master equation (\ref{eq:master_eqn})
is best solved numerically. In this paper, the density operator is
expanded in the number state basis $\{|n\rangle\}$ with a suitable
photon number cutoff, which allows cat state signatures to be computed.
With the dimensionless parameters, the master equation (\ref{eq:master_eqn})
in the number state basis is reduced to a set of partial differential
equations of the form: 
\begin{align}
\frac{\partial}{\partial\tau}\rho_{n,m} & =\sum_{i}\sum_{j}\mathcal{L}_{nm}^{ij}\rho_{i,j}\,,\label{eq:number_state_expansion_2-1-1}
\end{align}
where \begin{widetext}
\begin{align}
\mathcal{L}_{nm}^{ij} & =\frac{\lambda}{2}\sqrt{n\left(n-1\right)}\delta_{n-2}^{i}\delta_{m}^{j}+\frac{\lambda}{2}\sqrt{m\left(m-1\right)}\delta_{n}^{i}\delta_{m-2}^{j}-\frac{\lambda}{2}\sqrt{\left(n+1\right)\left(n+2\right)}\delta_{n+2}^{i}\delta_{m}^{j}-\frac{\lambda}{2}\sqrt{\left(m+1\right)\left(m+2\right)}\delta_{n}^{i}\delta_{m+2}^{j}\nonumber \\
 & -i\frac{\chi'}{2}\left[n\left(n-1\right)-m\left(m-1\right)\right]\delta_{n}^{i}\delta_{m}^{j}+g^{2}\sqrt{\left(n+1\right)\left(n+2\right)\left(m+1\right)\left(m+2\right)}\delta_{n+2}^{i}\delta_{m+2}^{j}-\frac{g^{2}}{2}\left[n\left(n-1\right)+m\left(m-1\right)\right]\delta_{n}^{i}\delta_{m}^{j}\nonumber \\
 & +\left(N_{th}+N_{s}+1\right)\left[2\sqrt{\left(n+1\right)\left(m+1\right)}\delta_{n+1}^{i}\delta_{m+1}^{j}-n\delta_{n}^{i}\delta_{m}^{j}-m\delta_{n}^{i}\delta_{m}^{j}\right]\nonumber \\
 & +\left(N_{th}+N_{s}\right)\left[2\sqrt{nm}\delta_{n-1}^{i}\delta_{m-1}^{j}-\left(n+1\right)\delta_{n}^{i}\delta_{m}^{j}-\left(m+1\right)\delta_{n}^{i}\delta_{m}^{j}\right]\nonumber \\
 & -M\left[2\sqrt{\left(n+1\right)m}\delta_{n+1}^{i}\delta_{m-1}^{j}+\sqrt{\left(n+1\right)\left(n+2\right)}\delta_{n+2}^{i}\delta_{m}^{j}+\sqrt{m\left(m-1\right)}\delta_{n}^{i}\delta_{m-2}^{j}\right]\nonumber \\
 & -M^{*}\left[2\sqrt{n\left(m+1\right)}\delta_{n-1}^{i}\delta_{m+1}^{j}+\sqrt{n\left(n-1\right)}\delta_{n-2}^{i}\delta_{m}^{j}+\sqrt{\left(m+1\right)\left(m+2\right)}\delta_{n}^{i}\delta_{m+2}^{j}\right]\,.\label{eq:superoperator_squeezed_bath-1}
\end{align}
\end{widetext}Here, $\delta_{n}^{i}$ is a Kronecker delta function
with $\delta_{n}^{i}=1$ if $i=n$ and $\delta_{n}^{i}=0$ otherwise.

\section{Cat-state signatures}

In order to verify the presence of a cat state, we will compute several
different cat-state signatures. The objective is to distinguish the
cat state $|\psi_{cat}\rangle=N(|\alpha\rangle+|-\alpha\rangle)$
of Eq (\ref{eq:cat-state}) (where $\alpha$ is real and $N$ is a
normalization constant) from the classical mixture $\rho_{mix}$ of
the two coherent states, given as 
\begin{equation}
\rho_{mix}=\frac{1}{2}(|\alpha\rangle\langle\alpha|+|-\alpha\rangle\langle-\alpha|).\label{eq:mix}
\end{equation}
Alternative approaches to detecting macroscopic coherence are available
but are not employed in this paper \citep{Terhal_PRA2000,Brennen_QInfo_2003,Cavalcanti_PRL2006,Cavalcanti_PRA2008,Frowis_NJP2012,Sekatski_PRA2014,Frowis_PRL2016,Yadin_PRA2016,reid2019criteria,Frowis_RMP2018}.
These include signatures and measures for macroscopic quantum coherence
based on variances and quantum Fisher information.

\subsection{Interference fringes}

A commonly used signature is the presence of interference fringes.
A rotated quadrature operator is defined as $x_{\theta}=\left(e^{-i\theta}a+e^{i\theta}a^{\dagger}\right)/\sqrt{2}$
and $p_{\theta}=\left(e^{-i\theta}a-e^{i\theta}a^{\dagger}\right)/\sqrt{2}i$.
In particular, $\theta=0$ and $\theta=\pi/2$ correspond to the standard
$x$ and $p$-quadrature operators, respectively. The $x_{\theta}$-quadrature
probability distribution $P\left(x_{\theta}\right)$ for a density
operator $\rho=\sum_{n,m}\rho_{nm}|n\rangle\langle m|$ in the number
state basis is given by
\begin{align}
\langle x_{\theta}|\rho|x_{\theta}\rangle & =\sum_{n,m}\rho_{nm}\langle x_{\theta}|n\rangle\langle m|x_{\theta}\rangle\,\label{eq:prob_dist-1-1}
\end{align}
where 
\begin{align}
\langle x_{\theta}|n\rangle & =\frac{e^{-i\theta n}}{\sqrt{2^{n}n!\sqrt{\pi}}}e^{-\frac{x_{\theta}^{2}}{2}}H_{n}\left(x_{\theta}\right)\,.\label{eq:x_theta_n}
\end{align}
Observation of two well-separated Gaussian peaks in $P(x_{\theta})$
along with interference fringes in $P(p_{\theta})$ gives evidence
of the system being in a cat-state as opposed to the mixture of the
two coherent states.

\subsection{Wigner function negativity}

We also compute the Wigner function and its negativity. It has been
shown that the Wigner function in the number state basis has the following
expression \citep{Cahill_PhysRev1969,PATHAK2014117}
\begin{align}
W\left(\alpha,\alpha^{*}\right) & =\sum_{n}^{N_{c}}\rho_{nn}X_{nn}+2Re\left(e^{-2\left|\alpha\right|^{2}}\sum_{l=1}^{N_{c}}c_{l}\left(2\alpha\right)^{l}\right)\,,\label{eq:Wigner_new_expression}
\end{align}
where
\begin{align}
c_{l} & =\sum_{n=0}^{N_{c}-l}\rho_{n,l+n}\frac{2\left(-1\right)^{n}}{\pi}\sqrt{\frac{n!}{\left(l+n\right)!}}L_{n}^{l}\left(4\left|\alpha\right|^{2}\right)\,.\label{eq:c_l}
\end{align}
The corresponding negativity is defined as \citep{Kenfack_2004}

\begin{align}
\delta & =\frac{1}{2}\intop\left[\left|W\left(\alpha,\alpha^{*}\right)\right|-W\left(\alpha,\alpha^{*}\right)\right]d^{2}\alpha\,.\label{eq:wigner_negativity}
\end{align}
The Wigner negativity quantifies the non-classicality of a physical
state. If the Wigner function is positive, then the Wigner distribution
can be interpreted as a probability distribution for variable $x_{\theta}$
and $p_{\theta}$, thus giving an explanation for the observed quadrature
distributions $P(x_{\theta})$ and $P(p_{\theta})$. A nonzero negativity
excludes this interpretation \citep{Bell_Physics1964,Reid_PRA1992,teh2020dynamics}.
The negativity Eq. (\ref{eq:wigner_negativity}) of a pure, even (odd)
cat state $W_{+}$ ($W_{-}$) is calculated from the Wigner function
\citep{Teh_PRA2018,teh2020dynamics}

\begin{eqnarray}
W_{\pm}\left(\alpha,\alpha^{*}\right) & = & \frac{2}{\pi}\mathcal{N}_{\pm}^{2}\left\{ exp\left[-2\left(\alpha^{*}-\alpha_{0}^{*}\right)\left(\alpha-\alpha_{0}\right)\right]\right.\nonumber \\
 &  & +exp\left[-2\left(\alpha^{*}+\alpha_{0}^{*}\right)\left(\alpha+\alpha_{0}\right)\right]\nonumber \\
 &  & \pm\langle\alpha_{0}|-\alpha_{0}\rangle exp\left[-2\left(\alpha^{*}-\alpha_{0}^{*}\right)\left(\alpha+\alpha_{0}\right)\right]\nonumber \\
 &  & \left.\pm\langle-\alpha_{0}|\alpha_{0}\rangle exp\left[-2\left(\alpha^{*}+\alpha_{0}^{*}\right)\left(\alpha-\alpha_{0}\right)\right]\right\} \,.\nonumber \\
\label{eq:Wigner_cat_state}
\end{eqnarray}

However, we note that it is possible to have a positive Wigner function
even when significant quantum coherence exists, as shown in a previous
work \citep{Teh_PRA2018,Cavalcanti_PRA2008}. Other cat-state signatures
have to be inferred in order to conclusively verify the existence
of a cat-like state. We emphasize this by further pointing out that
an impure cat state or a broader class of mixtures can possess a non-zero
Wigner negativity.

\subsection{Quantum coherence}

With this motivation, we may also use quantum coherence as a signature
of the cat state. The quantum coherence is quantified by a non-negative
number, assigned as a measure of the degree of coherence. Two particular
measures have been developed \citep{Baumgratz_PRL2014}, the relative
entropy of coherence and $l_{1}$ norm of coherence. Here, we consider
 the $l_{1}$ norm defined as
\begin{align}
C_{l_{1}}(\rho) & =\sum_{i\neq j}\sum_{j}\left|\rho_{ij}\right|\,\label{eq:l1_norm-2}
\end{align}
where $\rho$ is the density operator and $|i\rangle$ a basis set.
One may evaluate the measure of total quantum coherence, using the
$l_{1}$ norm quantum coherence measure defined for continuous variables

\begin{align}
C_{l_{1}}\left(\rho\right) & \equiv\intop_{-\infty}^{\infty}\intop_{-\infty}^{\infty}\left|\rho_{xx'}\right|\,dxdx'-\intop_{-\infty}^{\infty}\left|\rho_{xx}\right|\,dx\nonumber \\
 & =\intop_{-\infty}^{\infty}\intop_{-\infty}^{\infty}\left|\langle x|\rho|x'\rangle\right|\,dxdx'-\intop_{-\infty}^{\infty}\left|\langle x|\rho|x\rangle\right|\,dx
\end{align}
Taking $\alpha_{0}$ to be real, we see that for the mixture, $C_{l_{1}}\left(\rho_{mix}\right)=2\sqrt{\pi}-1$,
while for the cat state (\ref{eq:cat-state}), $C_{l_{1}}\left(\rho_{cat}\right)=8\sqrt{\pi}\mathcal{N}-1$
($\mathcal{N=}1/\left[2\left(1+e^{-2\left|\alpha_{0}\right|^{2}}\right)\right]$
is the normalization factor previously defined). The quantum coherence
for the mixture arises from the quantum coherence of the coherent
states $|\pm\alpha\rangle$ involved in the mixture. The quantum coherence
for a cat state contains an extra contribution due to the state being
a macroscopic superposition.

Quantum coherence does not indicate negativity of the Wigner function
and is thus a measure of a different type of non-classicality. It
has been shown that squeezed states possess a high degree of quantum
coherence, yet also possess a positive Wigner function, thus admitting
local hidden variable theories for experiments involving quadrature
phase amplitude measurements.

\subsection{Purity and number distribution}

Finally, states created in the presence of signal damping, $\gamma_{1}\neq0$,
will not be pure. A full discussion of this is given in Refs. \citep{Gilles_PRA1994,Hach_PRA1994,Wolinsky_PRL1988,carmichael_QO4,Sun_PRA2019,teh2020dynamics}.
The degree of purity of the cat-like states that are created can be
extracted from the measure of purity $\mathcal{P}$
\begin{align}
\mathcal{P} & =\text{Tr}\left(\rho^{2}\right)\,.\label{eq:purity}
\end{align}
In the limit of long times where there are signal cavity losses, a
system initially prepared in the vacuum state will become equivalent
to a nearly equal mixture of two coherent states, and the long-time
purity of the system then approaches 50\%. \textbf{}An indication
of the purity of the cat state is also given by the photon number
distribution $P\left(n\right)$. This distribution $P(n)$ has zero
values for odd $n$ for the cat state (\ref{eq:cat-state}), which
is generated by the parametric process from a vacuum initial state.

\section{Degenerate parametric oscillation with a squeezed reservoir}

\subsection{Squeezed reservoir fields}

From the master equation, Eq. (\ref{eq:master_eqn}), the free parameters
are the pump strength $\lambda$, the effective parametric interaction
$g$, the Kerr nonlinearity $\chi$, the two parameters that characterize
the reservoir $N$ and $M$, and the dimensionless time $\tau$. In
particular, it has been shown, in the limit of zero single-photon
damping, that a cat-state formed is given as (\ref{eq:cat-state})
with amplitude is $|\alpha_{0}|=\sqrt{\lambda/g^{2}\sqrt{1+\chi^{2}}}$.

\begin{figure}
\begin{centering}
\includegraphics[width=0.7\columnwidth]{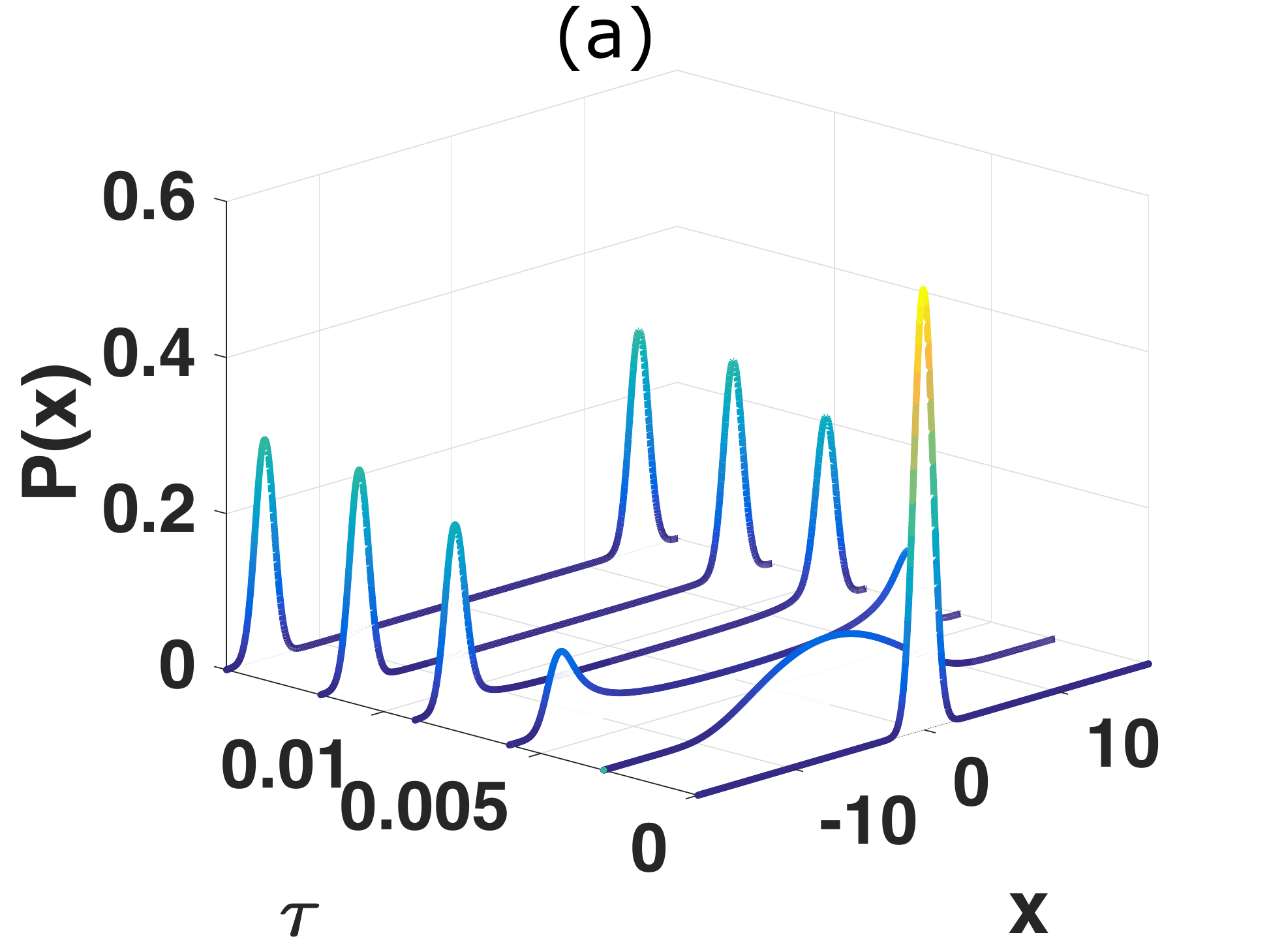}
\par\end{centering}
\begin{centering}
\includegraphics[width=0.7\columnwidth]{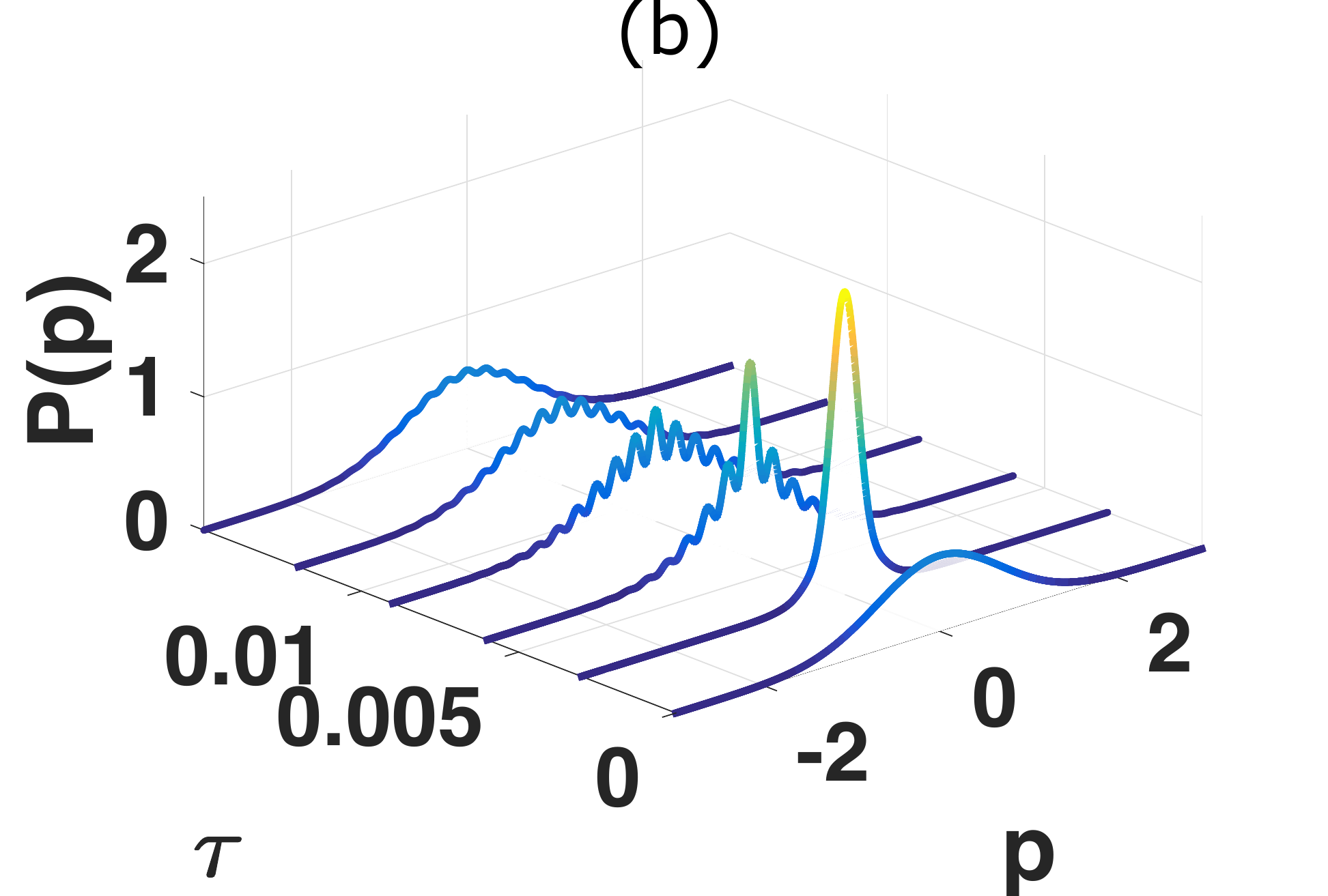}
\par\end{centering}
\begin{centering}
\includegraphics[width=0.7\columnwidth]{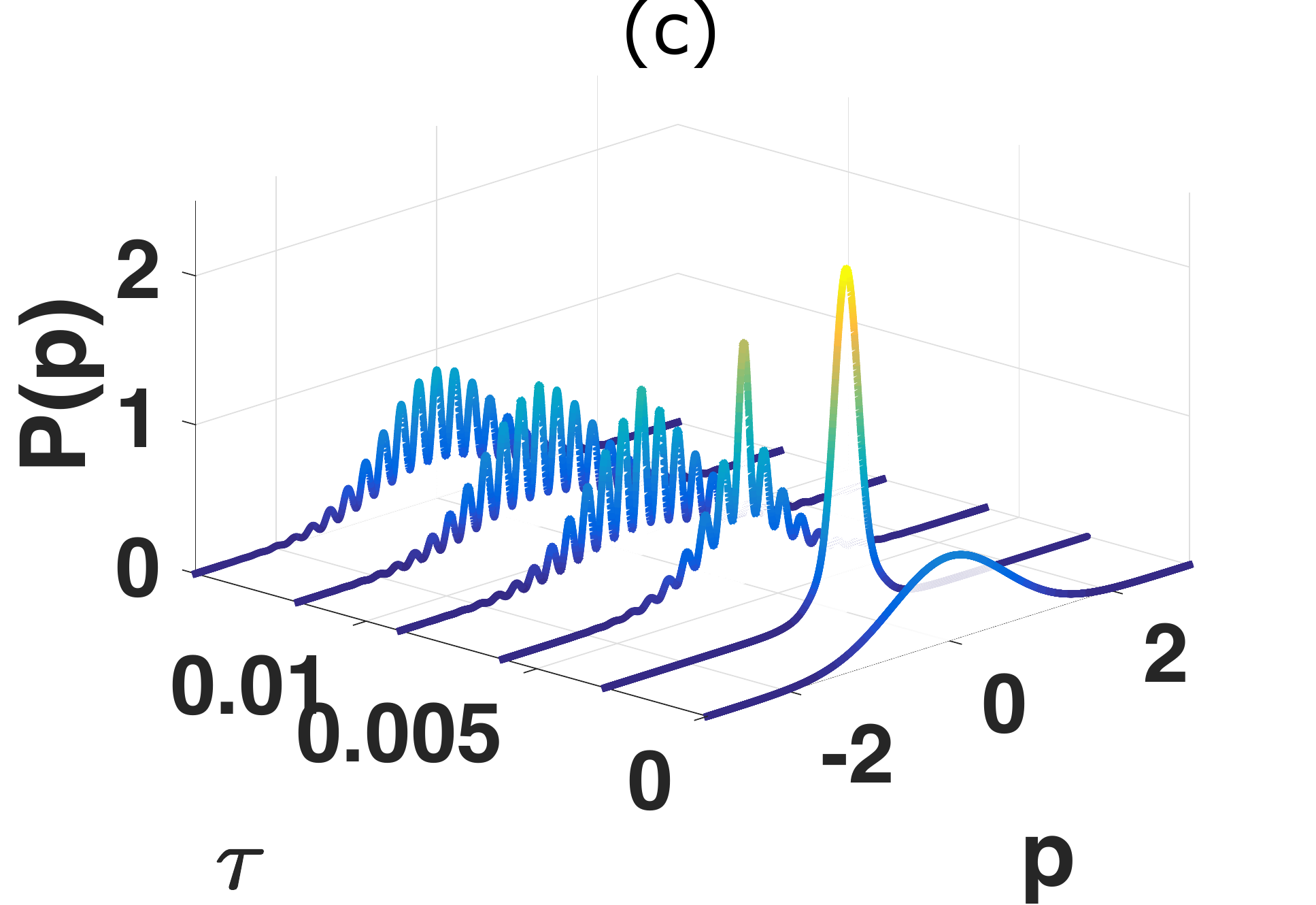}
\par\end{centering}
\caption{Effect of squeezed reservoirs on the cat state formation in a highly
nonlinear DPO without Kerr nonlinearity, at zero temperature. The
top plot (a) shows the $x$-quadrature probability distribution as
a function of time $\tau$. The case without squeezing ($N_{s}=0$)
and \textcolor{black}{with squeezing} ($N_{s}=1$, $M=\sqrt{2}$)
share similar time evolution of the $x$-quadrature probability distribution.
The lower plots {[}(b) and (b){]} show the $p$-quadrature probability
distribution (b) without squeezing $N_{s}=0$, and (c) with squeezing
$N_{s}=1$. The parameters are $g=2.5$, $\alpha_{0}=\lambda/g^{2}=100$,
$N_{th}=0$ and $M=\sqrt{N_{s}\left(N_{s}+1\right)}$.\label{fig:P_X_times-squeeze}}
\end{figure}

First, we take $\chi=0$, in which case the amplitude is real.\textcolor{red}{{}
}We assume zero temperature, $T=0$, which is a good approximation
to the DPO at room temperature for optical frequencies where thermal
noise is negligible. Here, we fix $|\alpha_{0}|=10$ and $g=2.5$,
and study the effect of different squeezing strengths at zero temperature
($N_{th}=0$). The initial state of the system is taken to be the
vacuum state.\textcolor{red}{{} }Fig. \ref{fig:P_X_times-squeeze}
shows the evolution of the two peaks in the $x$ quadrature, reflecting
the possible presence of the cat state. Here, given that the ideal
cat state has a real amplitude, as in \textcolor{black}{Meccozi and
Tombesi \citep{Mecozzi_PRL1987_squeeze_environ,Tombesi_JosaB1987},}
the optimal direction of squeezing in order to combat the decoherence
due to loss is directed along the $P$ axis. This corresponds to the
choice that $M$ is real and positive.\textcolor{blue}{}\textcolor{red}{}\textcolor{blue}{}

Fig. \ref{fig:P_X_times-squeeze} shows the time evolution of the
$p$-quadrature probability distribution at zero temperature with
and without a squeezed input. We see that interference fringes start
to appear after $\tau=0.005$. However, for a squeezed input, the
interference fringes are more refined. The visibility of the interference
fringes is significantly higher than the fringes without a squeezed
reservoir at the same dimensionless time $\tau$. In Fig. \ref{fig:P_X_times-squeeze},
we see that the $x$-quadrature probability distribution for the cases
with and without squeezing are similar. For real $\alpha_{0}$, the
existence of two peaks in the $x$-quadrature in Fig. \ref{fig:P_X_times-squeeze}
implies well-separated coherent state amplitudes, while the interference
fringes in the corresponding $p$-quadrature in Fig. \ref{fig:P_X_times-squeeze}
are an indication of the quantum nature of the underlying state. These
fringes rule out the possibility of the state being in a classical
mixture of two coherent states with amplitudes $\pm\alpha_{0}$. Clearly,
from the Figures, a squeezed input enhances the lifetime of the cat
state.
\begin{figure}

\begin{raggedright}
\includegraphics[width=0.51\columnwidth]{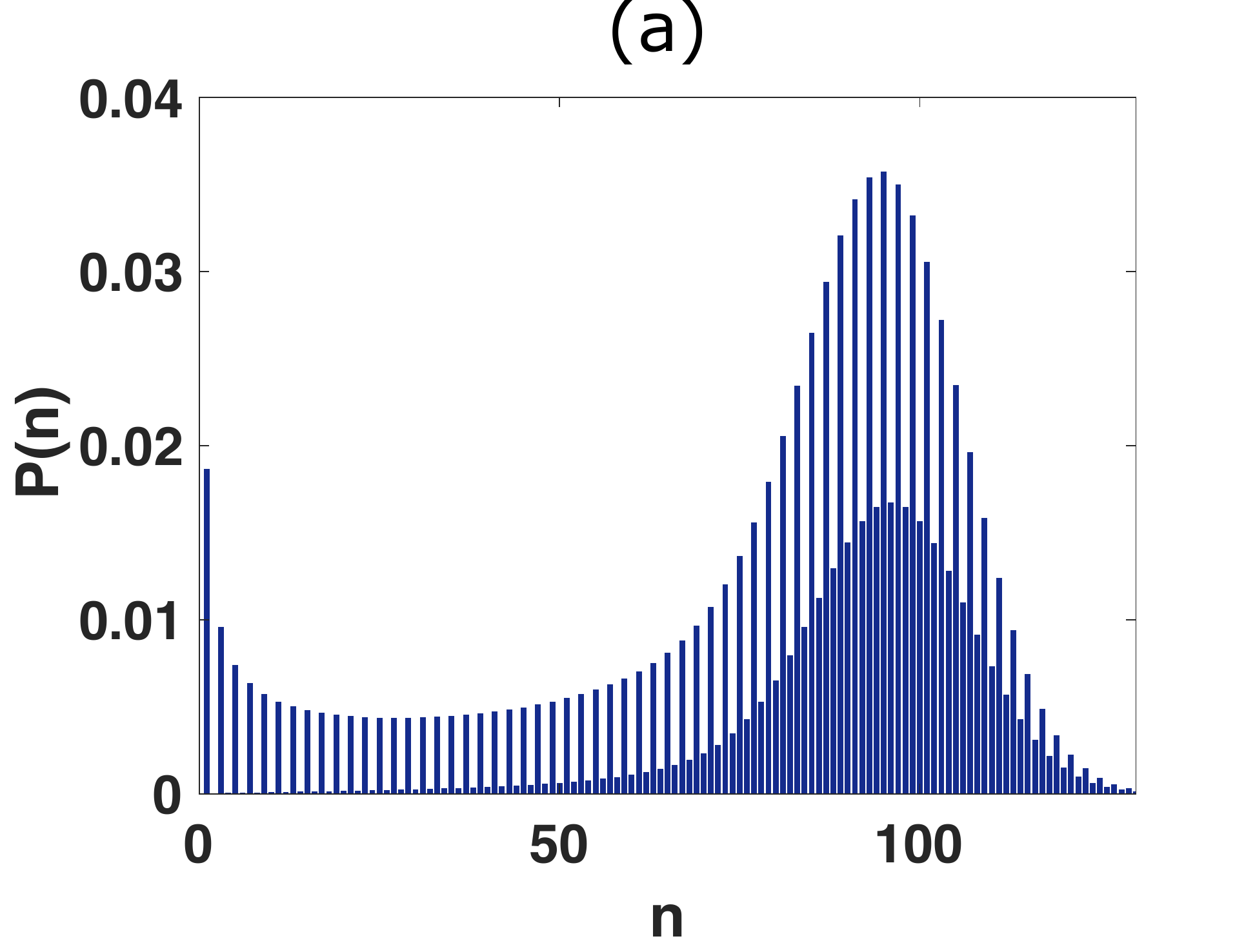}\includegraphics[width=0.51\columnwidth]{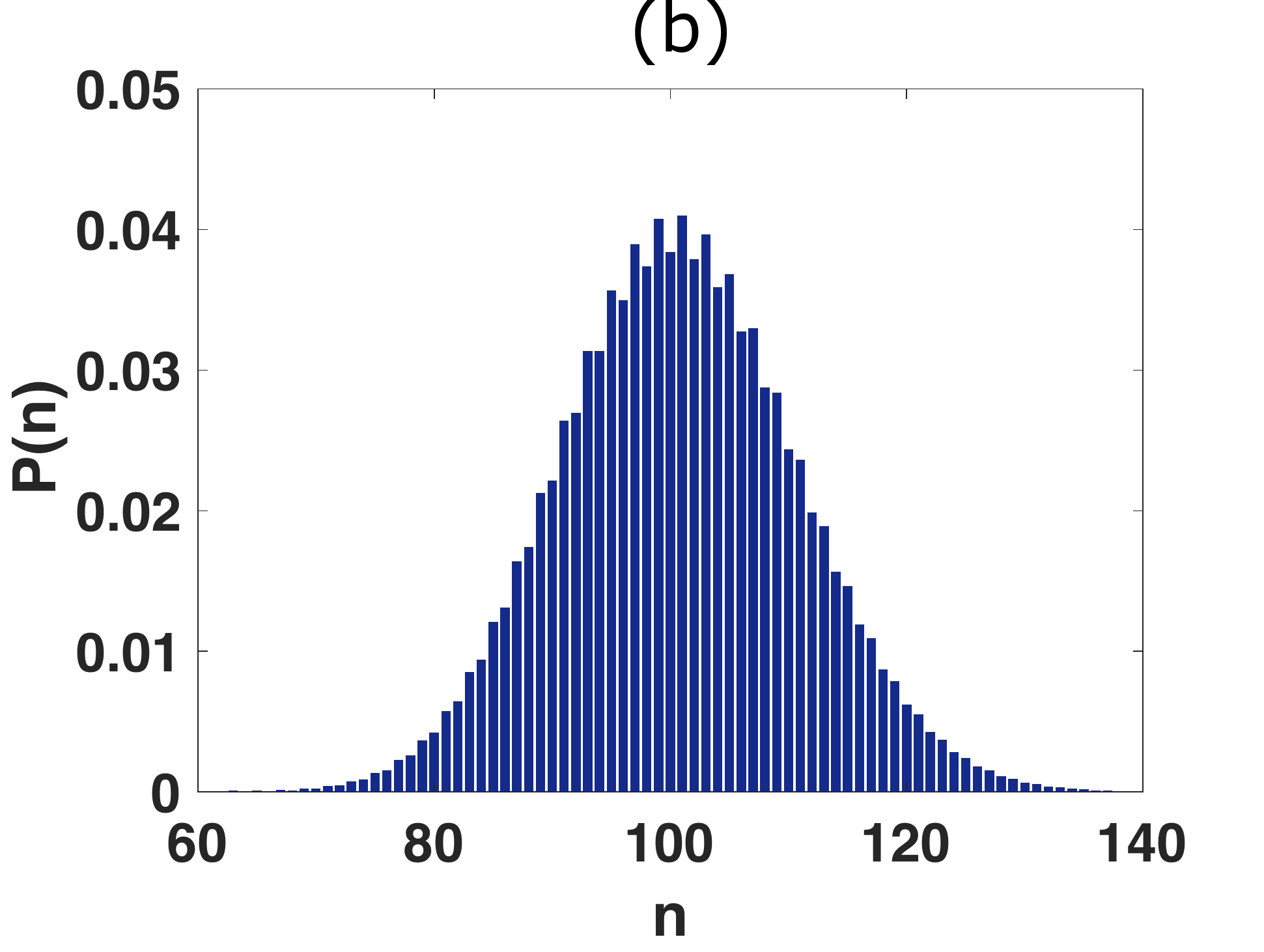}
\par\end{raggedright}
\bigskip{}

\begin{raggedright}
\includegraphics[width=0.51\columnwidth]{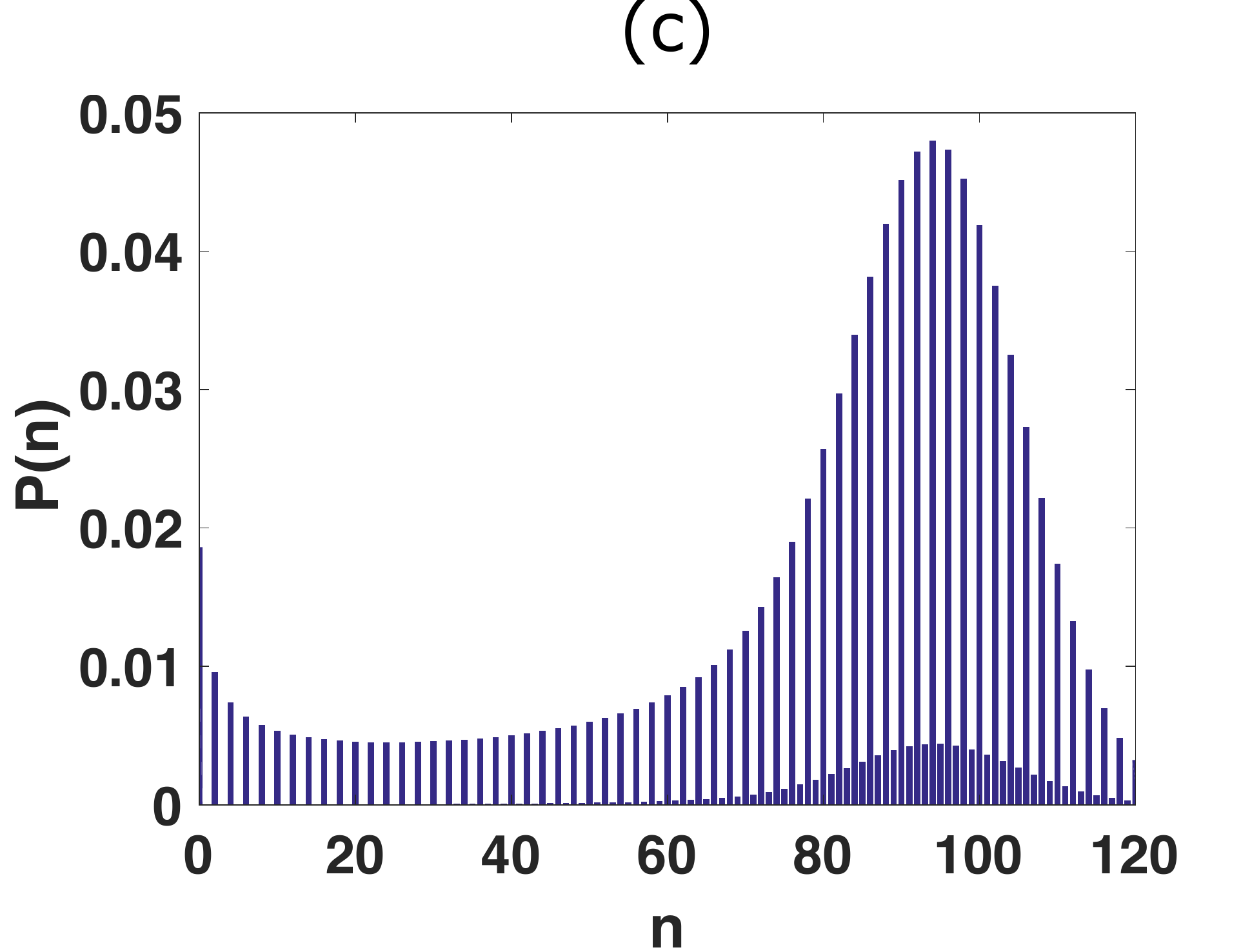}\includegraphics[width=0.51\columnwidth]{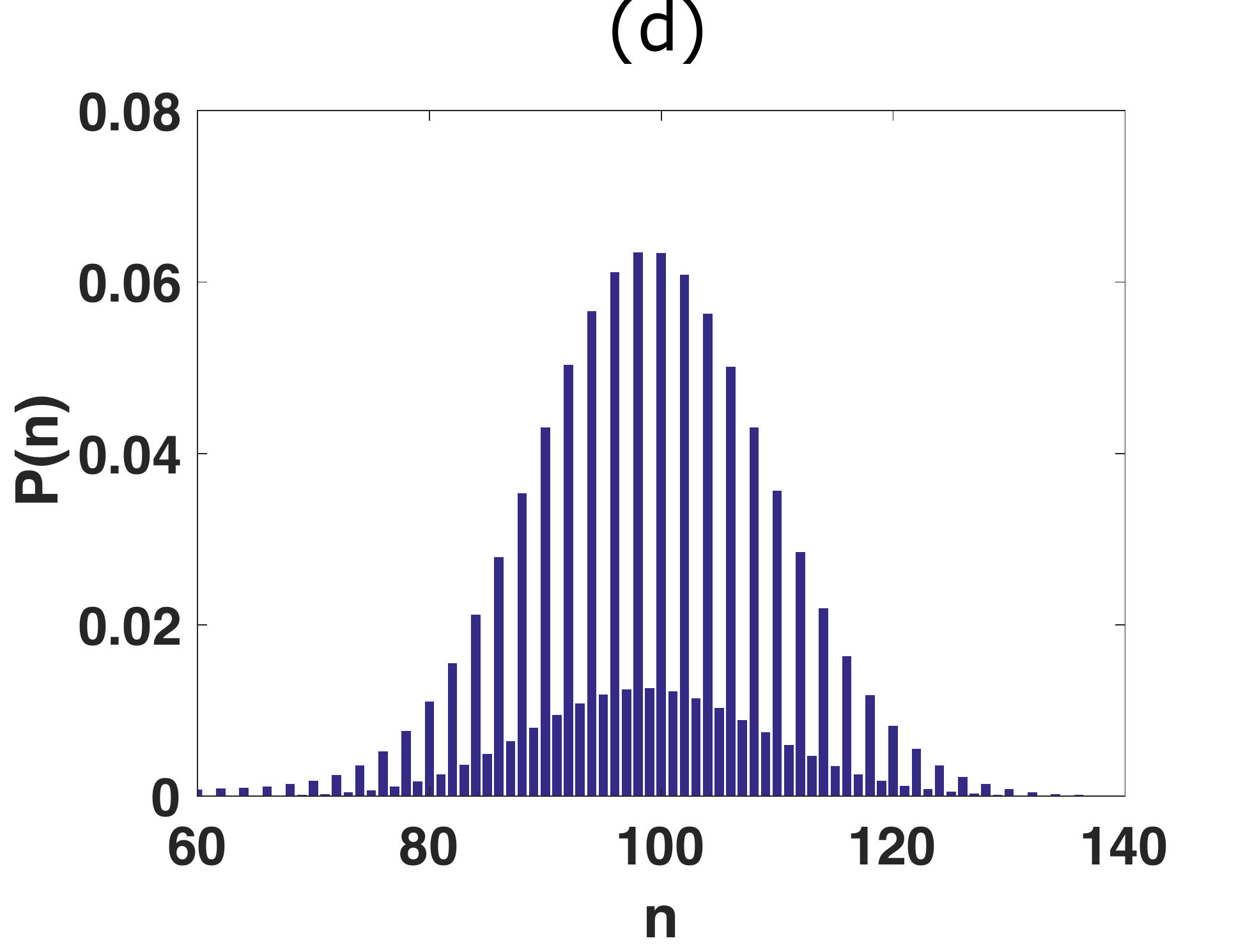}
\par\end{raggedright}
\caption{The effect of squeezed reservoirs on the cat-state formation as shown
by the photon number distribution. Plots show the photon number probability
distribution at times $\tau=0.0075$ {[}(a) and (c){]} and the later
time $\tau=0.015$ {[}(b) and (d){]}. The parameters are $g=2.5$,
$\lambda/g^{2}=100$, $N_{th}=0$ and $M=\sqrt{N_{s}\left(N_{s}+1\right)}$.
Figs. (a) and (b) (top) correspond to a reservoir with no squeezing.
Figs. (c) and (d) (lower) correspond to a squeezed reservoir ($N_{s}=1$,
$M=\sqrt{2}$), where we see that the probability for an odd photon
number is highly suppressed. \label{fig:P_n_times-squeeze}}
\end{figure}

\begin{figure}
\includegraphics[width=0.7\columnwidth]{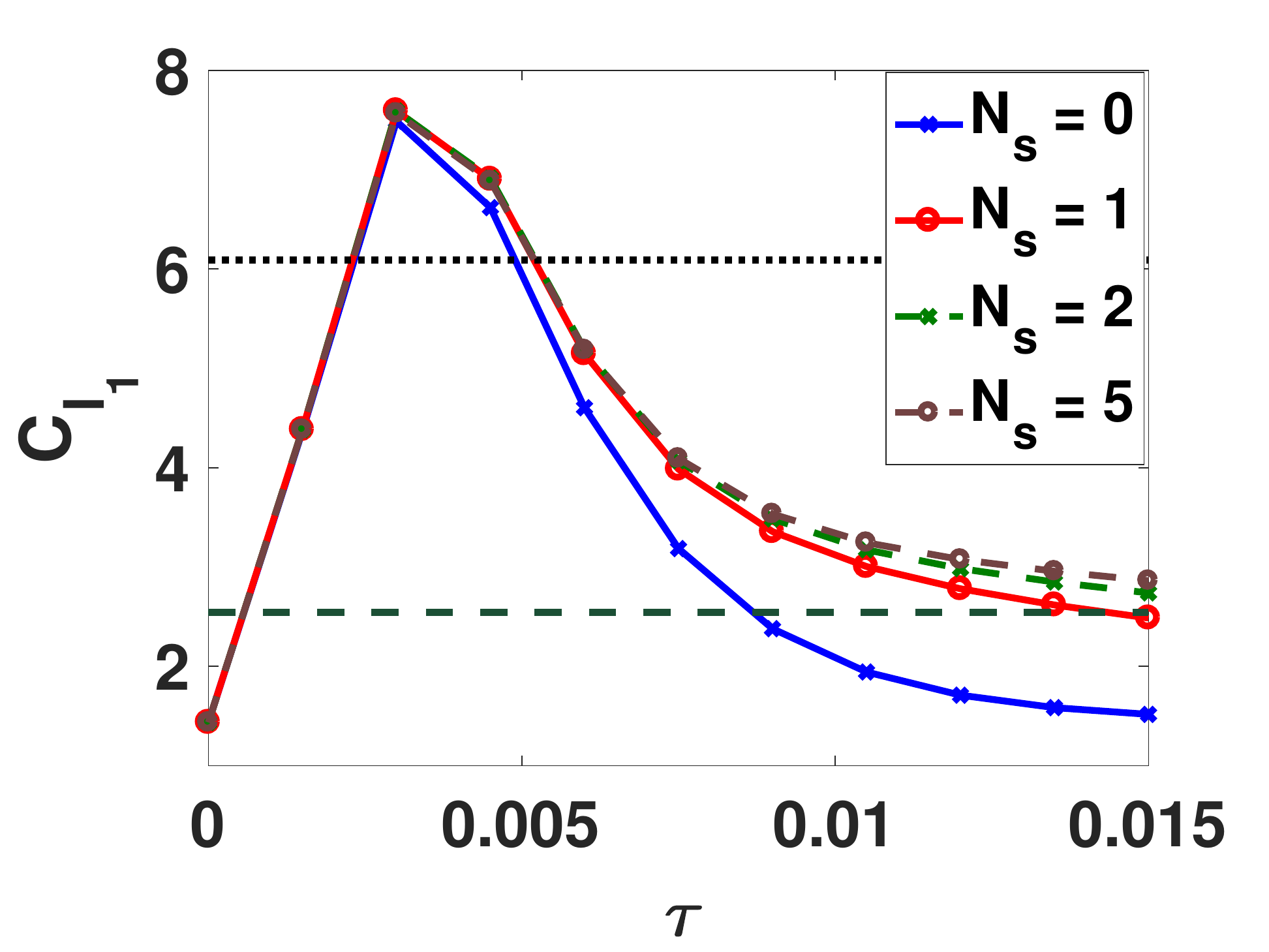}

\caption{The time evolution of $C_{l_{1}}$ at zero temperature ($N_{th}=0$)
for different squeezing strengths. Parameters are $g=2.5$, $\lambda/g^{2}=100$,
$\chi=0$, and $M=\sqrt{N_{s}\left(N_{s}+1\right)}$. The upper dashed
horizontal line gives the value for a pure cat state eq. (\ref{eq:cat-state}),
and the lower horizontal line gives the value for a mixture (\ref{eq:mix}).
\label{fig:cl1_nth0}}
\end{figure}
\begin{figure}
\begin{centering}
\includegraphics[width=0.7\columnwidth]{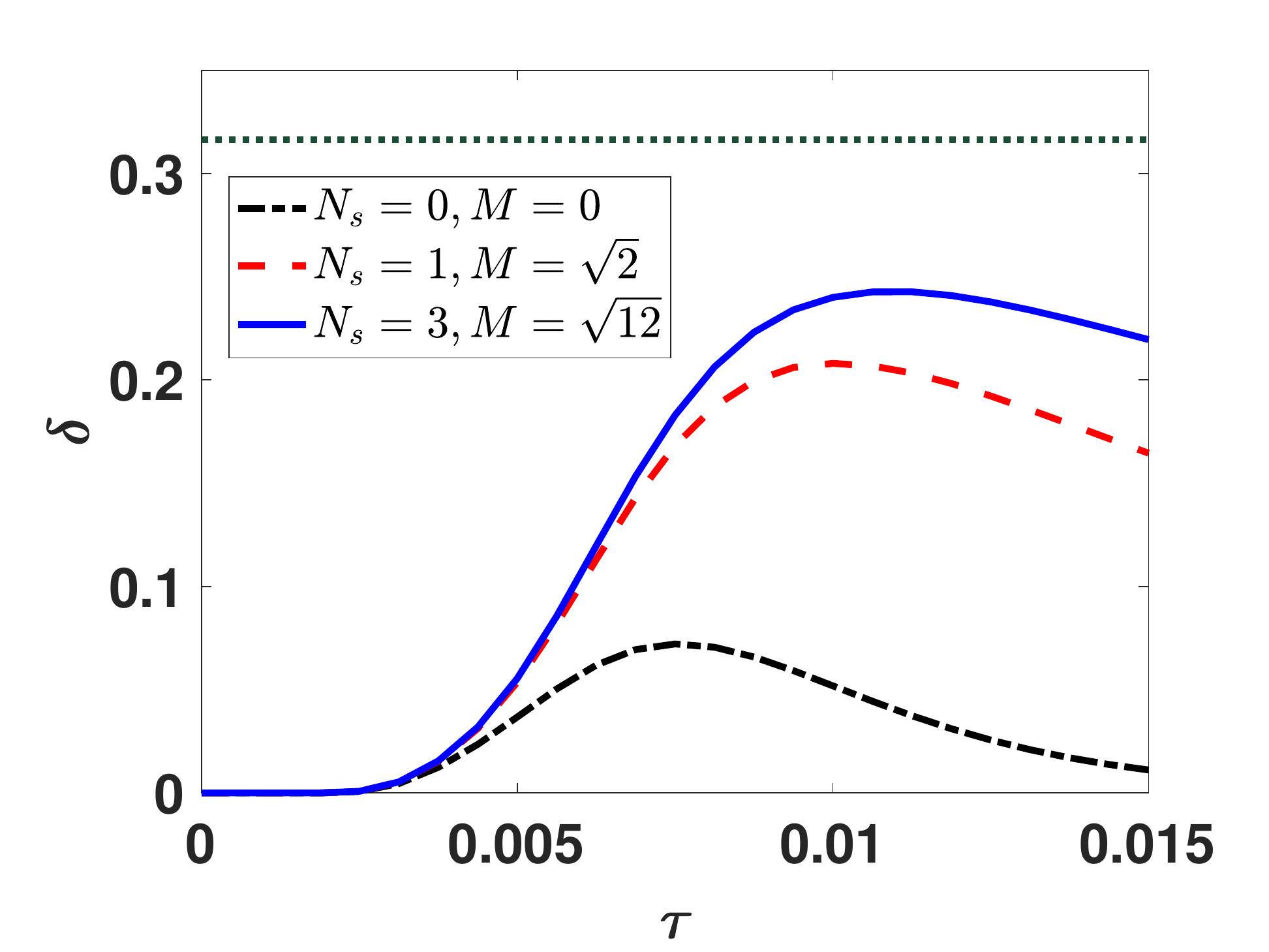}
\par\end{centering}
\begin{centering}
\includegraphics[width=0.7\columnwidth]{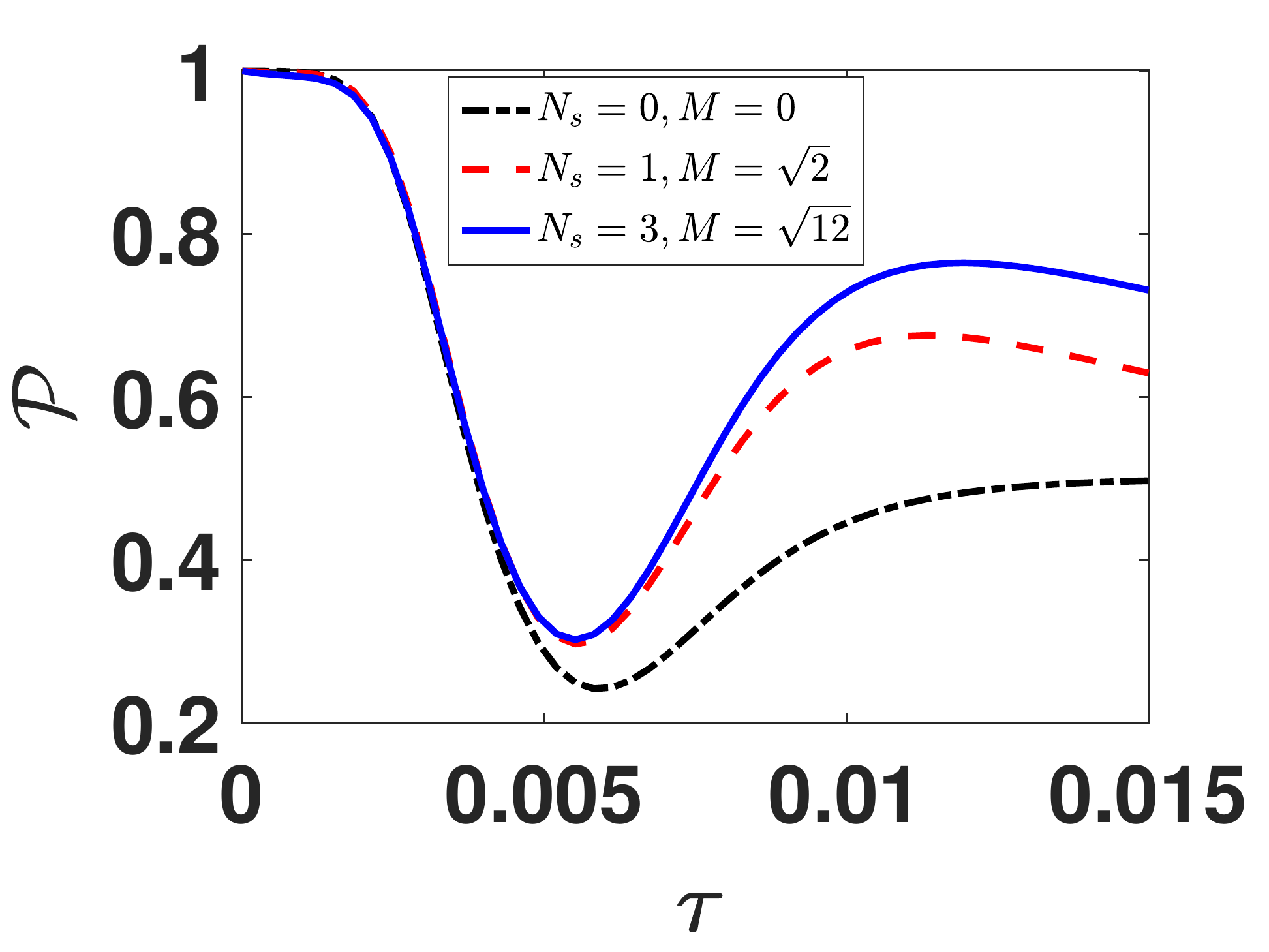}
\par\end{centering}
\caption{The effect of squeezed reservoirs on the cat-state formation as shown
by negativity and purity. The plots give the Wigner negativity $\delta$
and purity $\mathcal{P}$ as a function of the dimensionless time
$\tau$, for different degrees of squeezing, at zero temperature $N_{th}=0$.
$N_{s}$ is the mean photon number of the squeezed state and $M=\sqrt{N_{s}\left(N_{s}+1\right)}$
determines the strength of squeezing. Parameters are $g=2.5$ and
$\lambda/g^{2}=100$. The upper dashed horizontal line in the top
figure gives the negativity value for a pure cat state eq. (\ref{eq:cat-state}).
The mixture (\ref{eq:mix}) has zero negativity. \label{fig:purity}}
\end{figure}
According to Kennedy and Walls \citep{Kennedy_PRA1988}, the single
photon loss can be totally suppressed if the condition $\text{ln}N_{s}\gg2\tau_{int}$
is satisfied, where $\tau_{int}$ is the dimensionless time of interest
before measuring the state of the system. We note that this is only
true for the zero temperature case. In practice, the energy required
to achieve the amount of squeezing does not scale favorably with the
number, $N_{s}$. For instance, from Eq. (\ref{eq:vars-1}), we see
that the $p$-quadrature is squeezed and a squeezing of $82.84\%$
is needed for $N_{s}=1$ while a squeezing of $89.90\%$ is needed
for $N_{s}=2$.  Hence, the amount of squeezing shown in Figure \ref{fig:P_X_times-squeeze}
is $N_{s}=1$ corresponding to $\Delta^{2}P=0.17\Delta^{2}P_{\text{vac}}$,
where $\Delta^{2}P_{\text{vac}}=1$ is the variance of the $p$-quadrature
for a vacuum state.

The photon number distribution provides a clear perspective on why
the interference fringes have higher visibility for a squeezed input.
By comparing the photon number distribution at $\tau=0.0075$ and
$0.0150$ with squeezing in Fig. \ref{fig:P_n_times-squeeze}(b, d)
and without squeezing in Fig. \ref{fig:P_n_times-squeeze}(a, c),
we see that the probability for an odd number is suppressed in the
presence of a squeezed reservoir. In other words, the system retains
a high probability to be in an even cat-state for a longer time, as
compared to the case without the squeezed reservoir. The photon number
distribution results suggest that the effect of the squeezing is to
maintain the purity of the state for longer times. We quantify this
by evaluating the quantum coherence, negativity and purity, in Figures
\ref{fig:cl1_nth0} and \ref{fig:purity}. The quantum coherence and
Wigner negativity signatures are improved with the increase of squeezing
in the reservoir. 

The results for purity as in Eq. (\ref{eq:purity}) are plotted in
Fig. \ref{fig:purity}. Larger squeezing in the input reservoir leads
to higher purity, at a given time. We note the dip in the purity during
the time evolution before increasing again. This is also observed
in the case where the initial and final states are pure states, implying
that the state during evolution is not generally a pure one \citep{Hach_PRA1994}.
Eventually, for long times, the purity approaches $50\%$, as the
cat state becomes a nearly equal mixture of two coherent states.\textcolor{red}{}

\begin{figure}
\begin{centering}
\includegraphics[width=0.7\columnwidth]{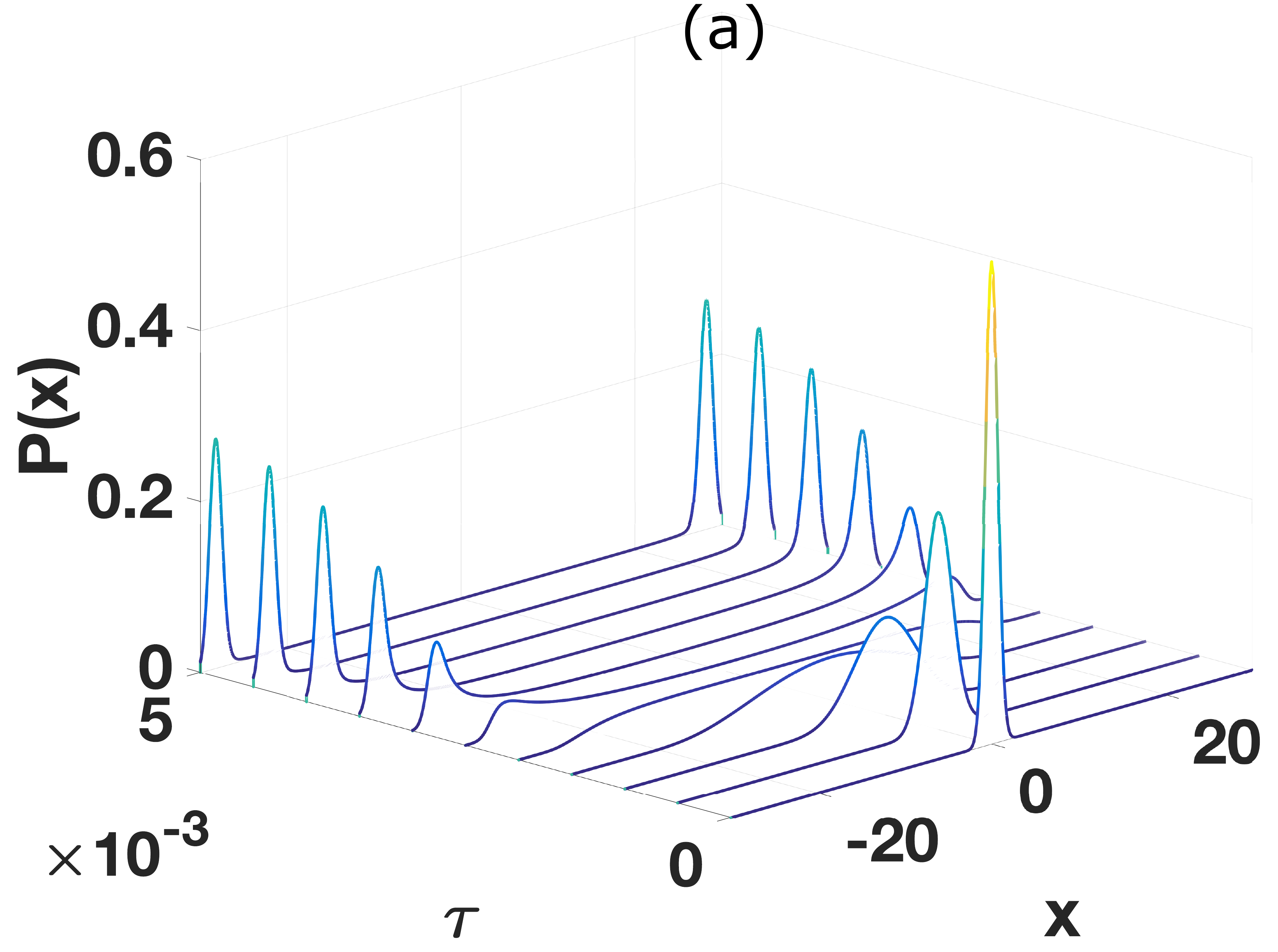}
\par\end{centering}
\begin{centering}
\includegraphics[width=0.7\columnwidth]{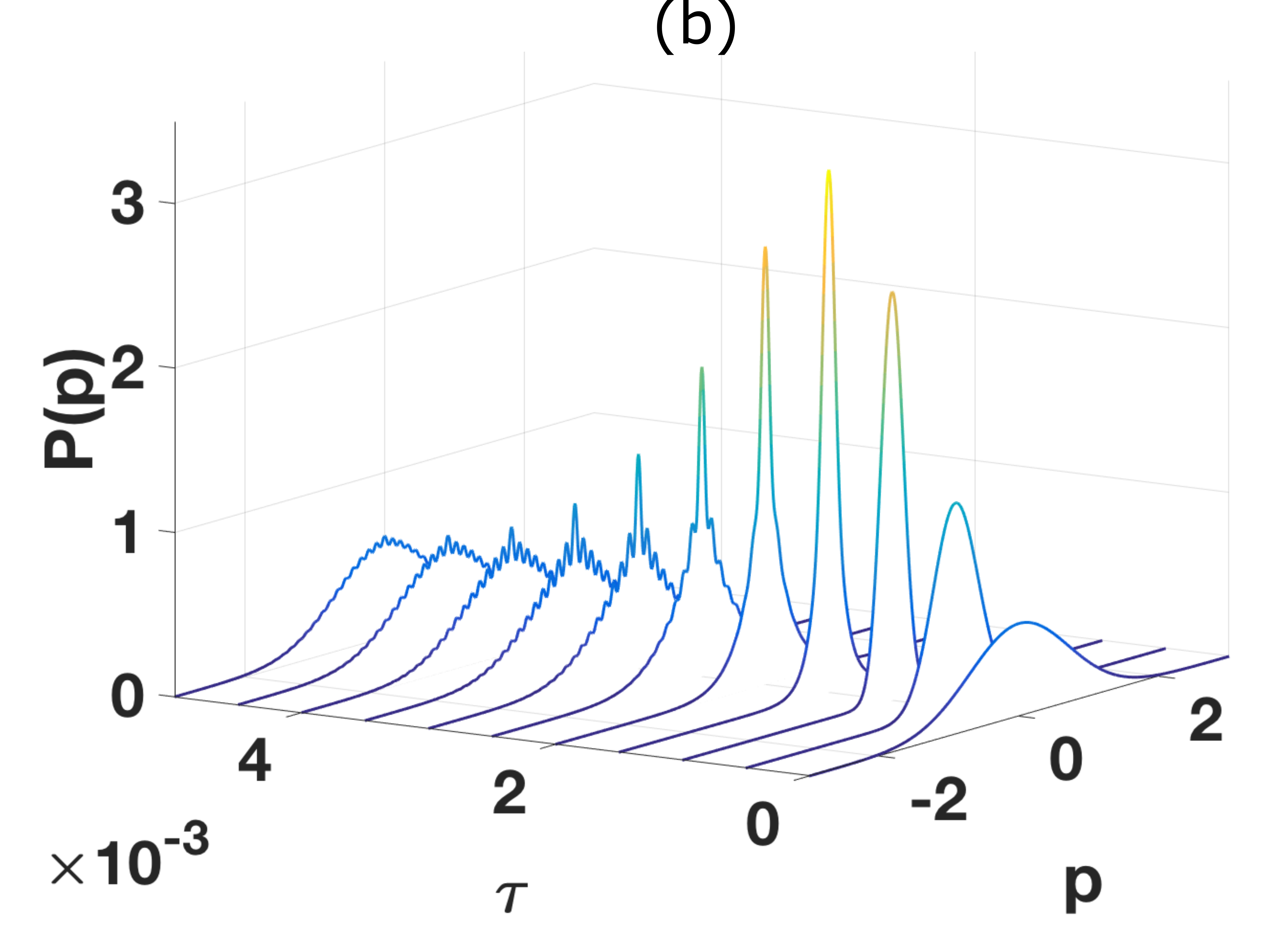}
\par\end{centering}
\bigskip{}

\begin{centering}
\par\end{centering}
\begin{centering}
\includegraphics[width=0.7\columnwidth]{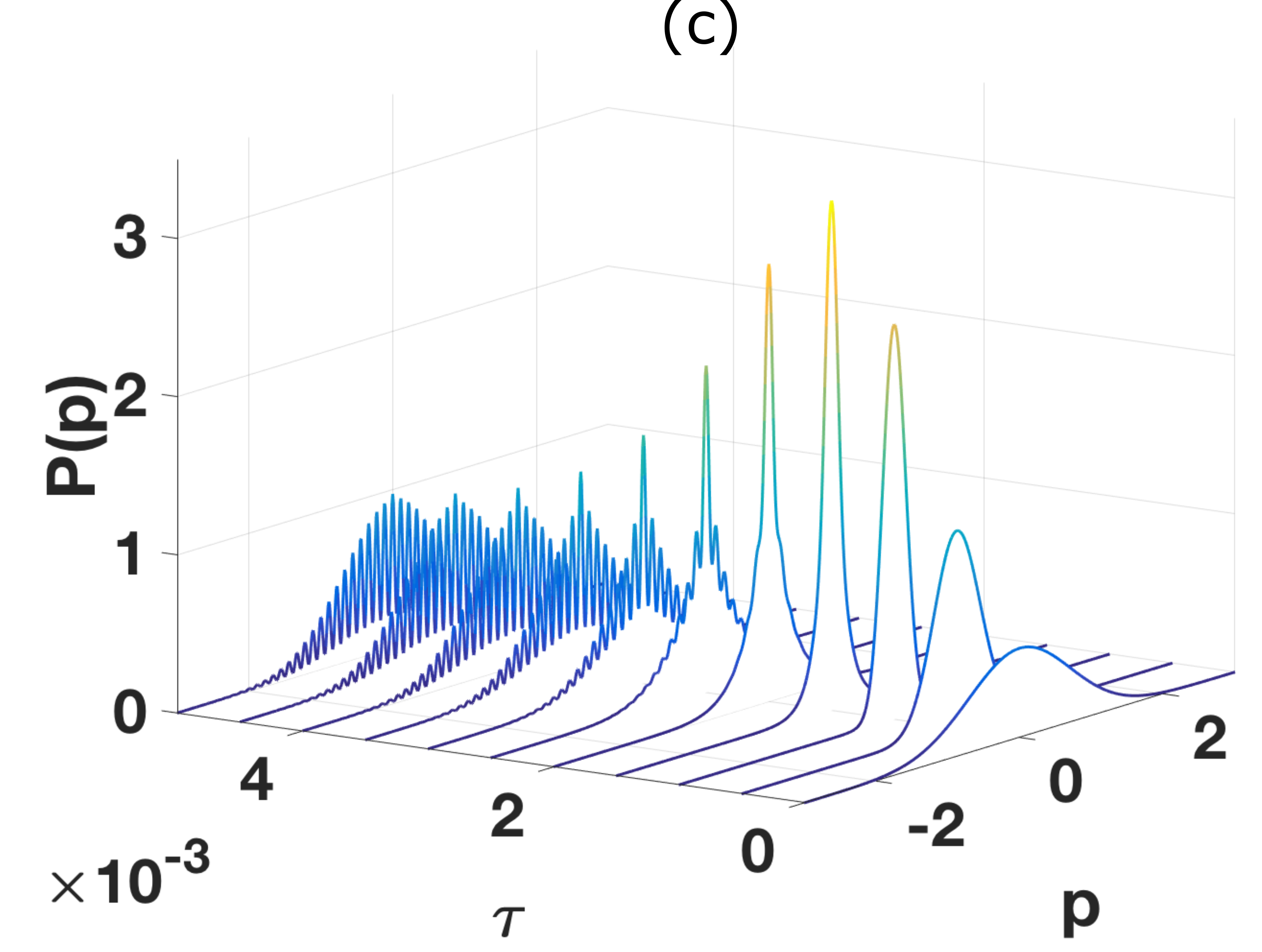}
\par\end{centering}
\caption{Formation of a large cat state in a DPO at zero temperature with the
aid of squeezing. The time evolution of the $x$ and $p$-quadratures
without a squeezed reservoir are shown in Plots (a) and (b). The time
evolution with a squeezed reservoir $N_{s}=2$, $M_{s}=\sqrt{6}$
is shown (c). The plot for the $x$-quadrature is indistinguishable
from plot (a), with no squeezing. The parameters are $g=2$ and $\lambda/g^{2}=400$
which corresponds to a coherent amplitude of $\alpha_{0}=20$.\label{fig:large_alpha_P_p_manytimes_zero_T}}
\end{figure}

\subsection{Large cat amplitudes}

The results presented so far suggest that a fragile mesoscopic/macroscopic
quantum state could well be preserved under the influence of a squeezed
reservoir. In the following, we consider a large cat-like state with
an amplitude $|\alpha_{0}|=20$, which corresponds to a photon number
of $400$, and investigate its quantum features. Again, we take the
vacuum state as an initial state. The results for the time evolution
of the $p$-quadrature probability distributions for a large amplitude
$\alpha_{0}=20$, for zero temperature and $\chi=0$, are shown in
Fig. \ref{fig:large_alpha_P_p_manytimes_zero_T}. We compare the results
without squeezing and with significant squeezing $(N=2,\,M=\sqrt{6},$
$\Delta^{2}p=0.1\Delta^{2}p_{\text{vac}}$). In accordance with the
earlier results for a smaller coherent amplitude in Fig. \ref{fig:P_X_times-squeeze},
we see an enhancement of the interference fringes in the $p$-quadrature
probability distribution when the reservoir is squeezed. Different
to the case of no squeezing, the interference fringes persist to the
end of our simulation at dimensionless time $\tau=0.0050$.

\section{cat-states for microwave fields with squeezed-state reservoirs}

We now include the thermal contribution and compare cases with and
without a squeezed reservoir. While thermal noise at room temperature
for optical fields is negligible, the effect of such noise is significant
if present. This helps us to understand the results for the microwave
regime, where thermal noise will be significant at room temperature.
For concreteness, we calculate the mean thermal photon number at room
temperature for the signal mode using the experimental parameters
of Leghtas et al. \citep{Leghtas_Science2015}. The signal mode frequency
is taken to be half the pump mode frequency, which is $\omega_{s}=2\pi\times4.01\text{GHz}$.
The mean thermal number is then given by the expression $N_{th}=1/\left[\text{exp}\left(\hbar\omega_{s}/kT\right)-1\right]\approx1522$.
At cryogenic temperatures of $T\sim100mK$, $N_{th}$ is of course
much lower, and therefore values of $N_{th}=.05-1$ are more typical
of these experiments, but thermal effects can still be very significant.

\begin{figure}
\begin{centering}
\includegraphics[width=0.7\columnwidth]{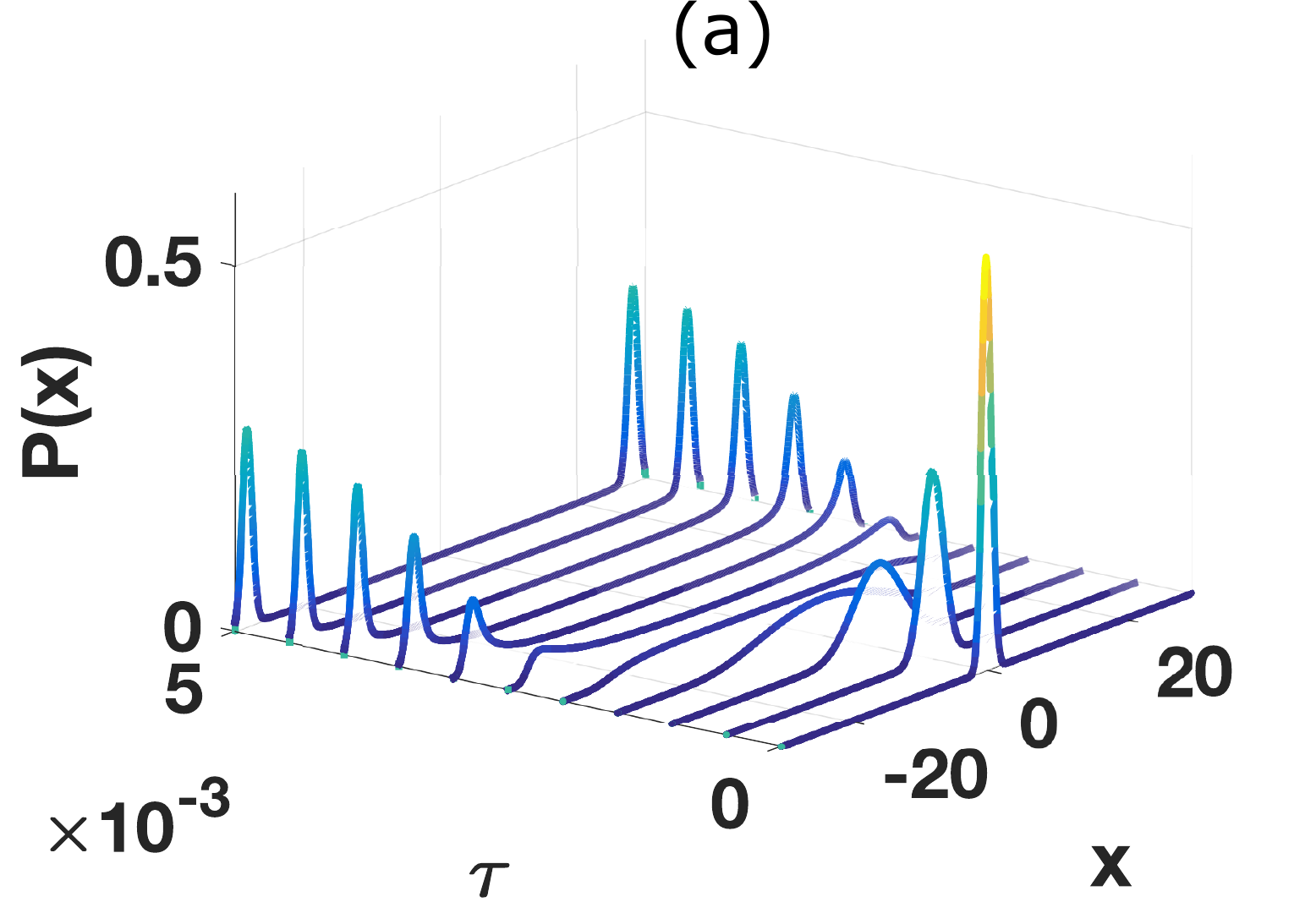}
\par\end{centering}
\begin{centering}
\includegraphics[width=0.7\columnwidth]{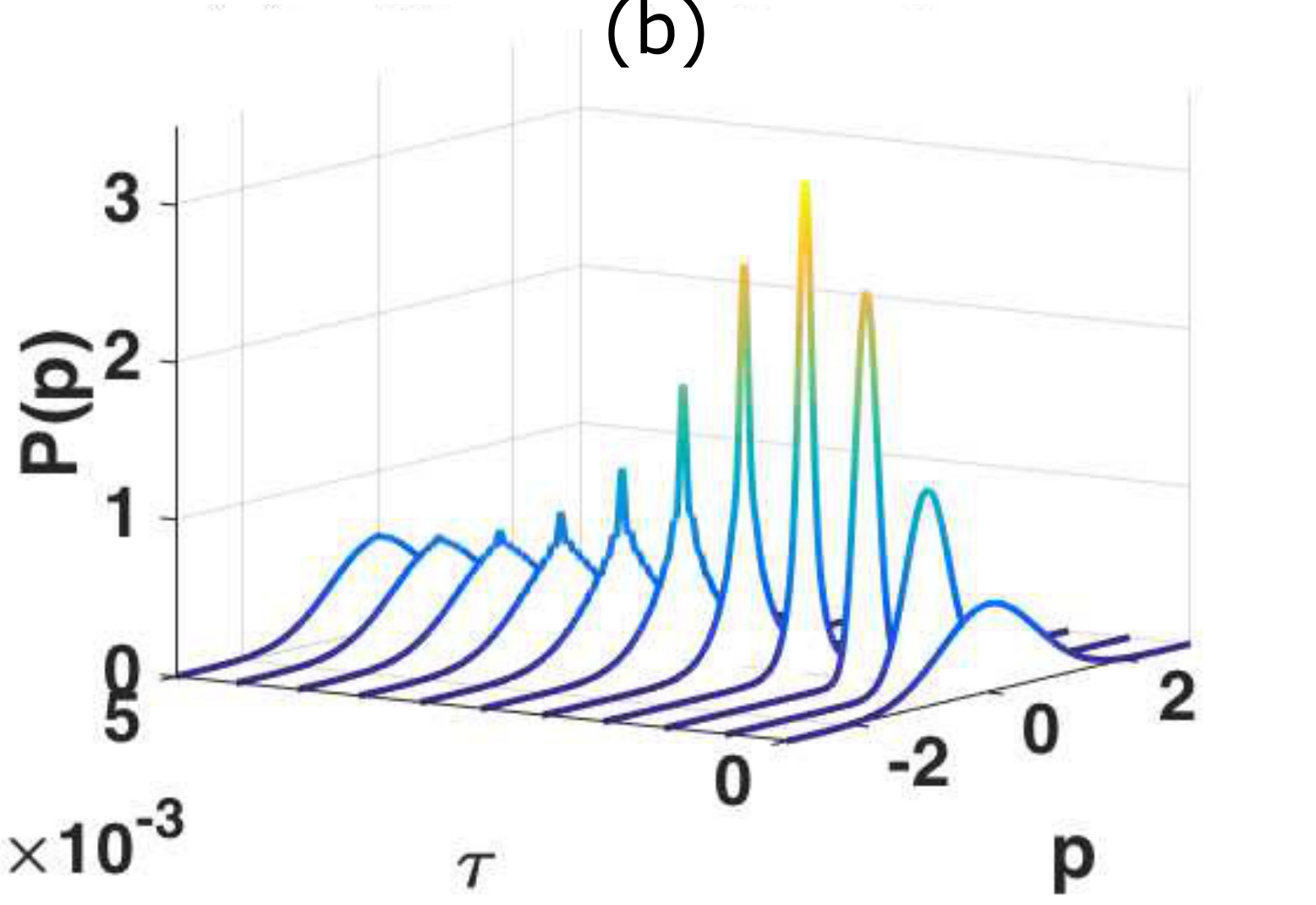}
\par\end{centering}
\begin{centering}
\par\end{centering}
\begin{centering}
\includegraphics[width=0.7\columnwidth]{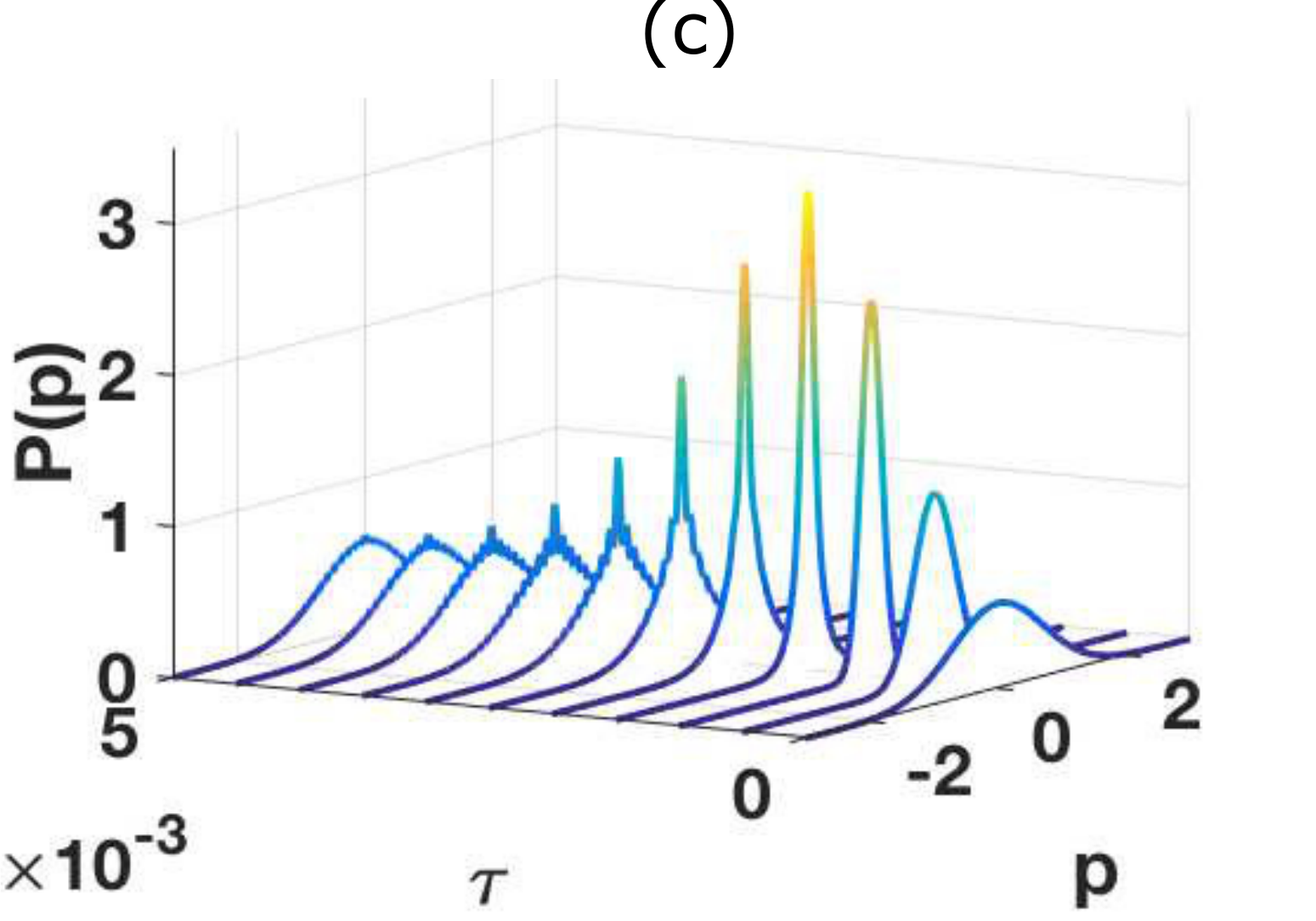}
\par\end{centering}
\caption{Effect of thermal noise. The time evolution the $x$-quadrature without
squeezing ($N_{s}=M=0$) is given by the top graph (a), and is unchanged
for squeezing given by $N_{s}=2$, $M=\sqrt{6}$. The middle graph
(b) shows the $p$-quadrature without squeezing, and the lower graph
(c) shows the $p$-quadrature with squeezing $N_{s}=2$, $M=\sqrt{6}$.
Here $N_{th}=0.5$. The parameters are as for Figure \ref{fig:large_alpha_P_p_manytimes_zero_T}.\label{fig:large_alpha_P_p_manytimes_finite_T}}
\end{figure}

In the Appendix, we summarize results based on the solutions of Kennedy
and Walls \citep{Kennedy_PRA1988}, analyzing the decoherence of a
system initially in a cat state, which is then coupled to a thermal
squeezed reservoir. We consider both types of thermal squeezing. Without
squeezing, thermal noise increases the decoherence rate, and the decoherence
rate increases with the size $\alpha$ of the cat state. With squeezing,
the decoherence is slowed. However, for a reservoir in a thermalized
squeezed state, one cannot overcome the decoherence for finite $N_{th}$
by simply increasing the amount of squeezing. A different result is
obtained for the squeezed thermal input, where enhanced squeezing
will eventually reduce the decoherence. This is expected, since in
the latter, the thermal noise is from the preparation of the squeezed
state, and does not arise from the immediate environment of the cavity.

\subsection{Thermal effects on cat-states}

\begin{figure}
\includegraphics[width=0.51\columnwidth]{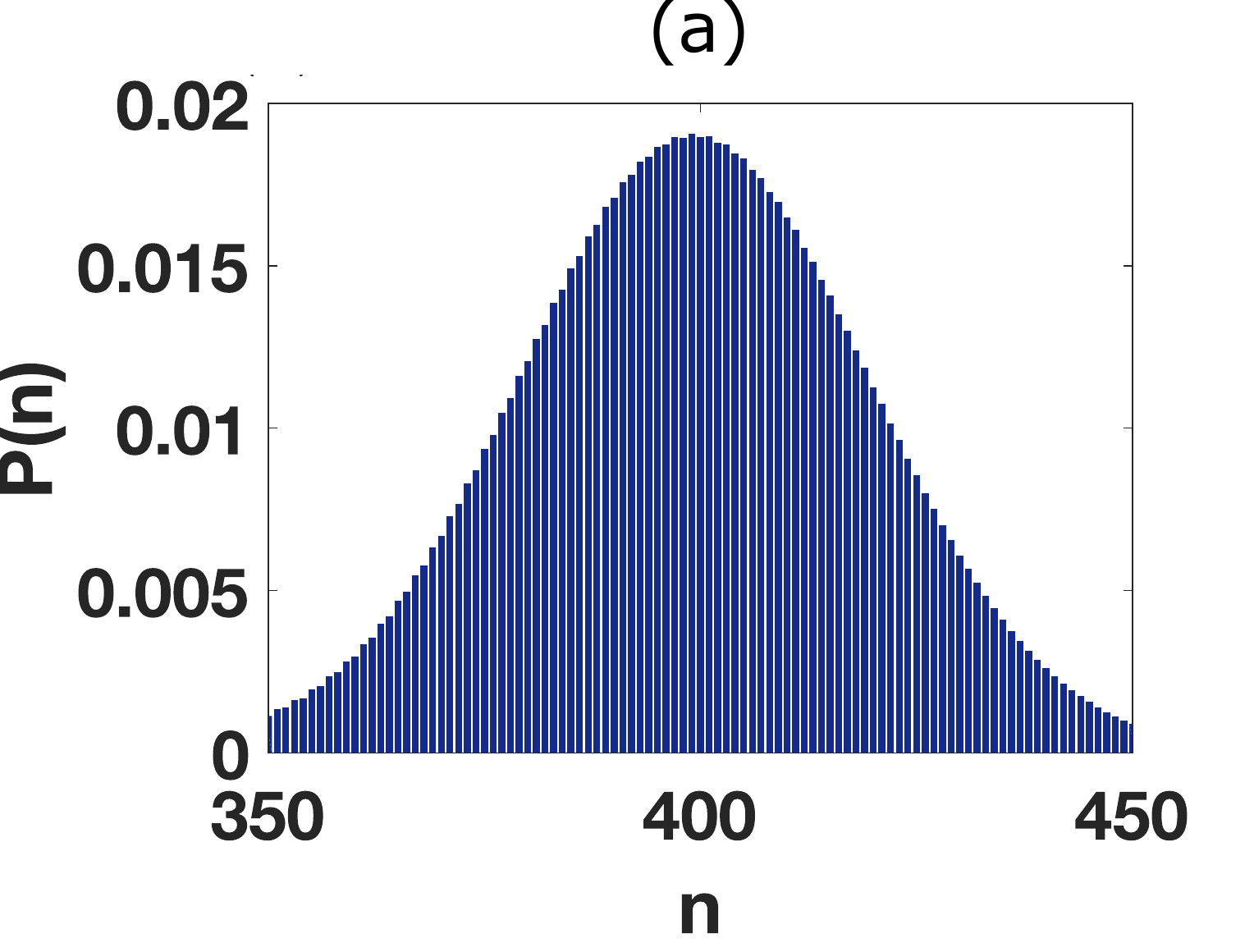}\includegraphics[width=0.51\columnwidth]{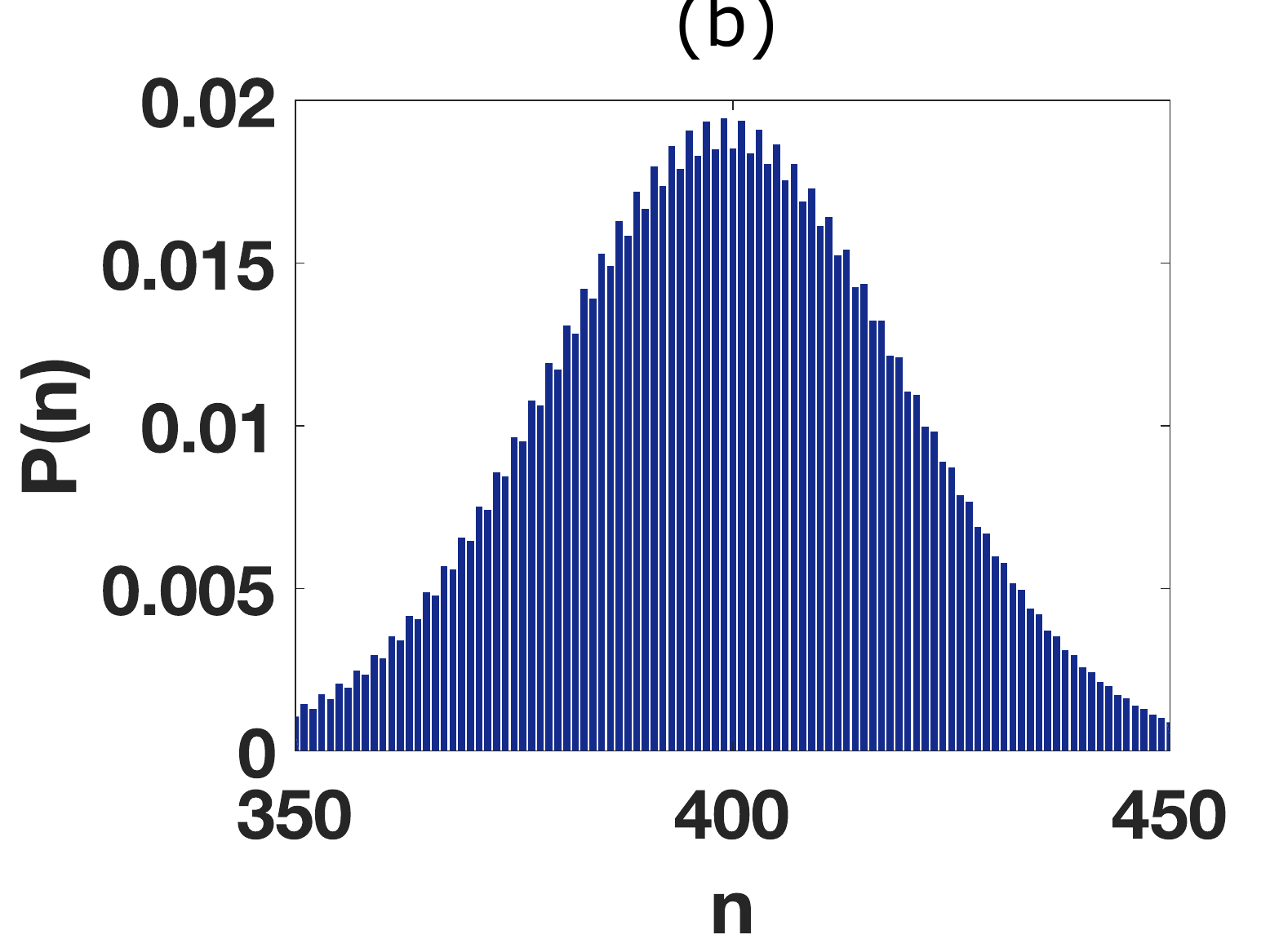}

\includegraphics[width=0.51\columnwidth]{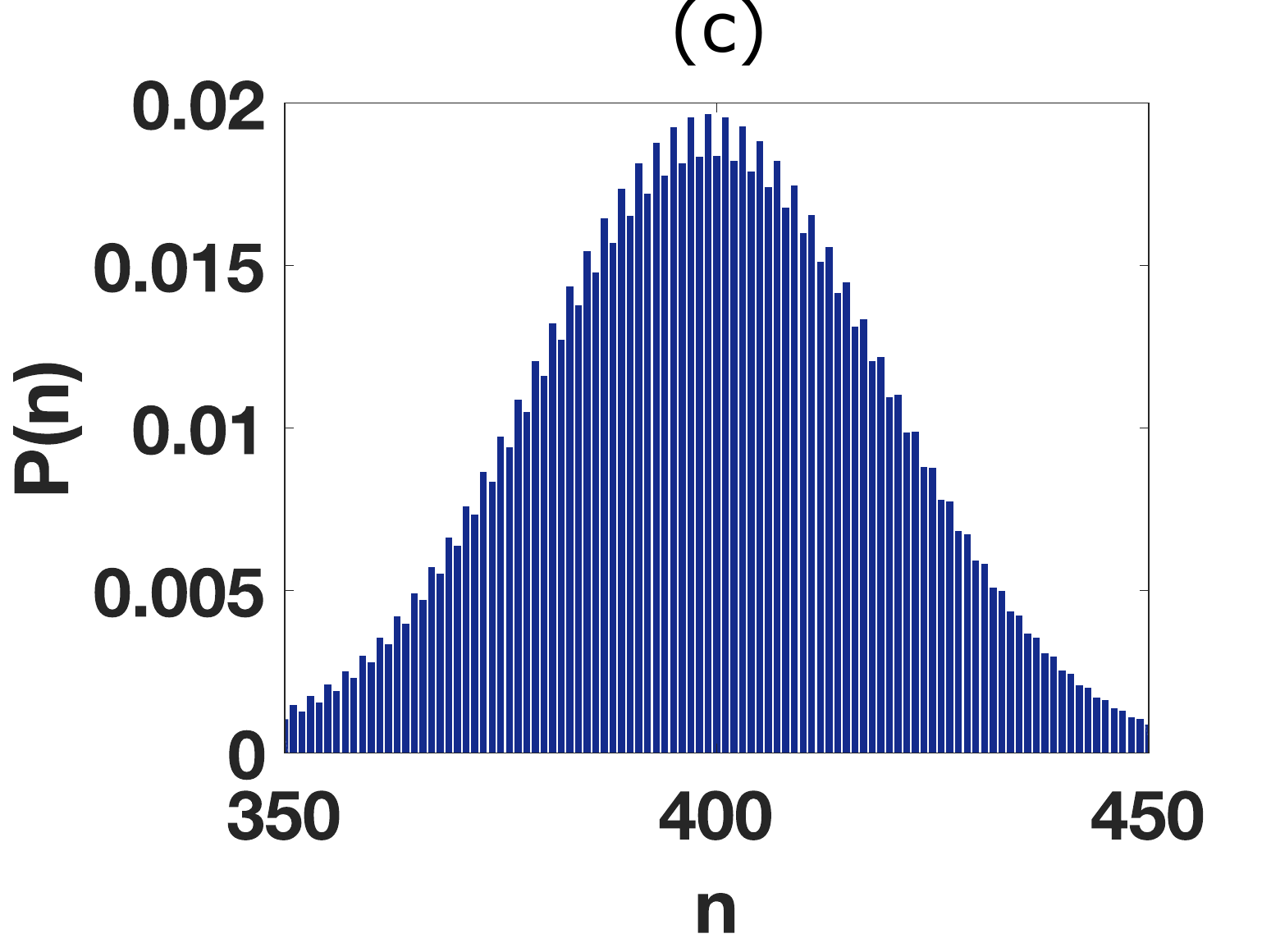}\includegraphics[width=0.51\columnwidth]{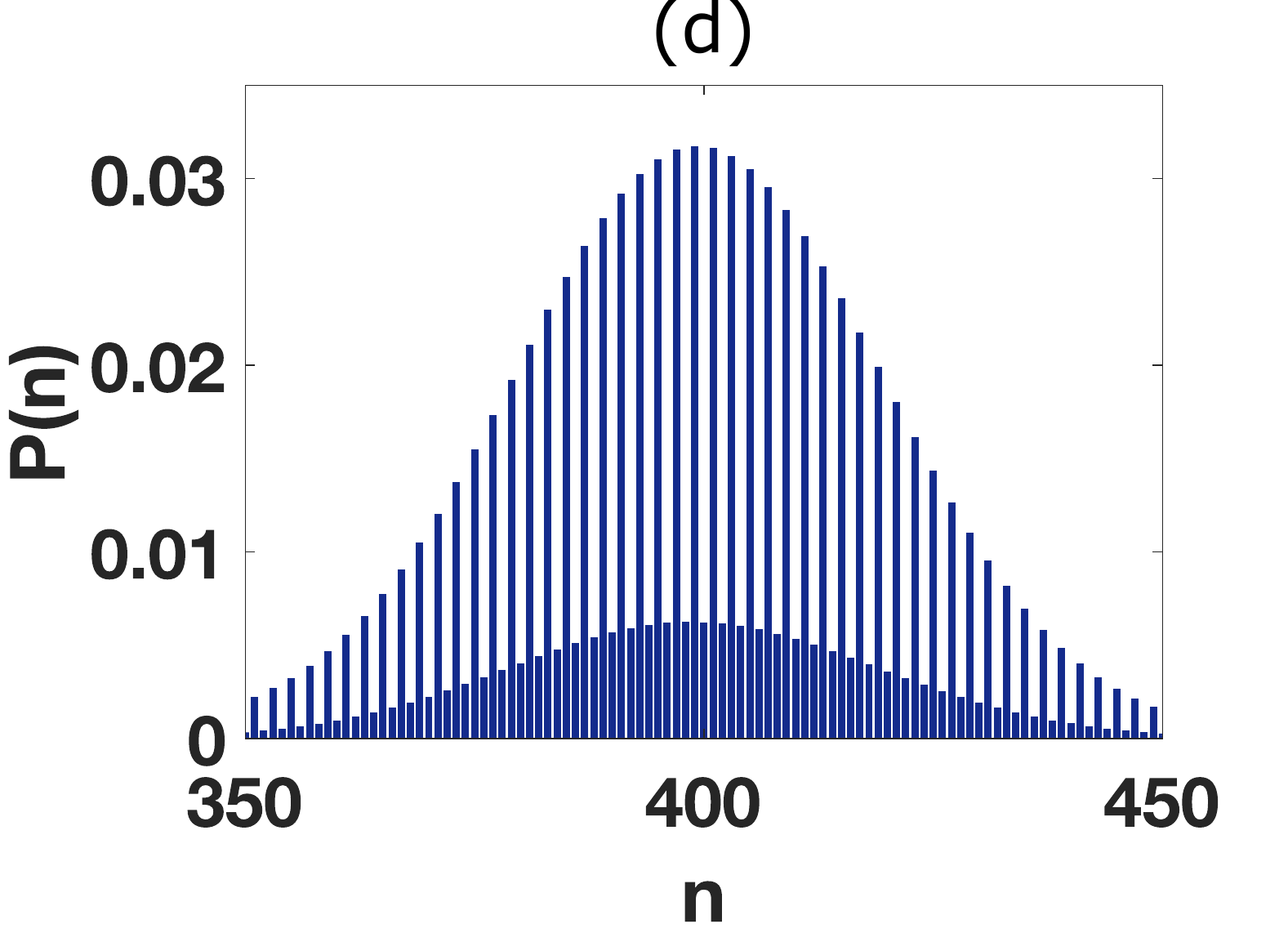}

\caption{Effect of thermal noise. The photon number probability distribution
at $\tau=0.0050$ for the case (a) without squeezing ($N=2,\,M=0$)
and (b) with squeezing ($N_{s}=2,\,M=\sqrt{6}$), at a finite temperature,
$N_{th}=0.5$. The non-Poissonian character is significantly reduced
compared to the case with zero temperature $N_{th}=0$, plotted directly
below, for the same squeezing parameters $N_{s}=0$ (c) and $N_{s}=2,$$M=\sqrt{6}$
(d).\label{large_alpha_Pn_finite_T}}
\end{figure}

\begin{figure}
\includegraphics[width=0.61\columnwidth]{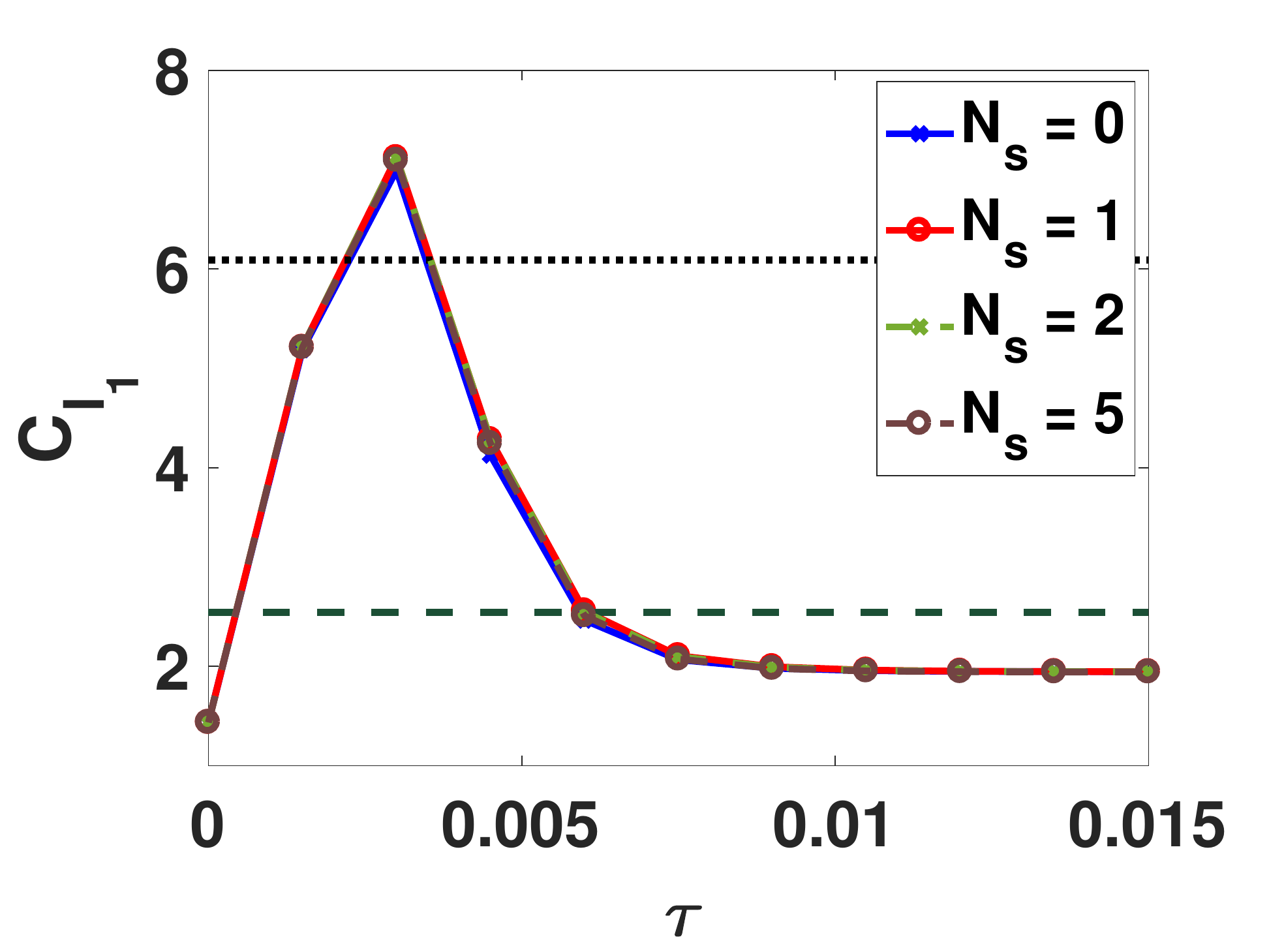}

\caption{The time evolution of $C_{l_{1}}$ at a finite temperature corresponding
to $N_{th}=1$, for different squeezing strengths. Parameters are
$g=2.5$, $\lambda/g^{2}=100$, $\chi=5$, and $M=\sqrt{N_{s}\left(N_{s}+1\right)}$,
as in Fig. \ref{fig:wigner_neg_purity_nth1}. The upper dashed horizontal
line gives the value for a pure cat state eq. (\ref{eq:cat-state}),
and the lower horizontal line gives the value for a mixture (\ref{eq:mix}).\label{fig:cl1_nth1}}
\end{figure}

First, we examine the case of a thermalized squeezed reservoir, which
models a system where the environment is at finite temperature as
the cat states evolve, but assuming the squeezed input is generated
at zero temperature. Results for the quadrature probability distributions
and the photon number probability distributions are shown in Figures
\ref{fig:large_alpha_P_p_manytimes_finite_T} and \ref{large_alpha_Pn_finite_T}
respectively. Even for $N_{th}\sim0.5-1$, thermal noise has a dramatic
effect in enhancing the ratio of odd to even photon numbers, thus
reducing the purity. In the presence of thermal noise, the squeezing
parameters that allow significant improvement of the fringe visibility
at zero temperature are no longer sufficient to suppress the loss
of quantum coherence. The plots in Figure \ref{fig:cl1_nth1} indicate
that the effect of thermal noise is not overcome by simply introducing
a large amount of squeezing, as consistent with the results given
in the Appendix. More optimistic results would be expected for the
squeezed thermal input.\textbf{\textcolor{blue}{}}

\textcolor{red}{}

\subsection{Effects of Kerr nonlinearities}

Next, we extend the study to a superconducting circuit experiment
where Kerr nonlinearity, represented by $\bar{\chi}$ in equation
(\ref{eq:simplified_hamiltonian}), is present. Typically, thermal
noise is not negligible. As shown in previous work, \citep{Sun_NJP2019,Sun_PRA2019,teh2020dynamics},
the Kerr nonlinearity rotates the steady state about the origin in
phase space. We will see that thermal noise induces decoherence and
shortens the cat-state lifetime. 

\begin{figure}[t]
\begin{centering}
\includegraphics[width=0.7\columnwidth]{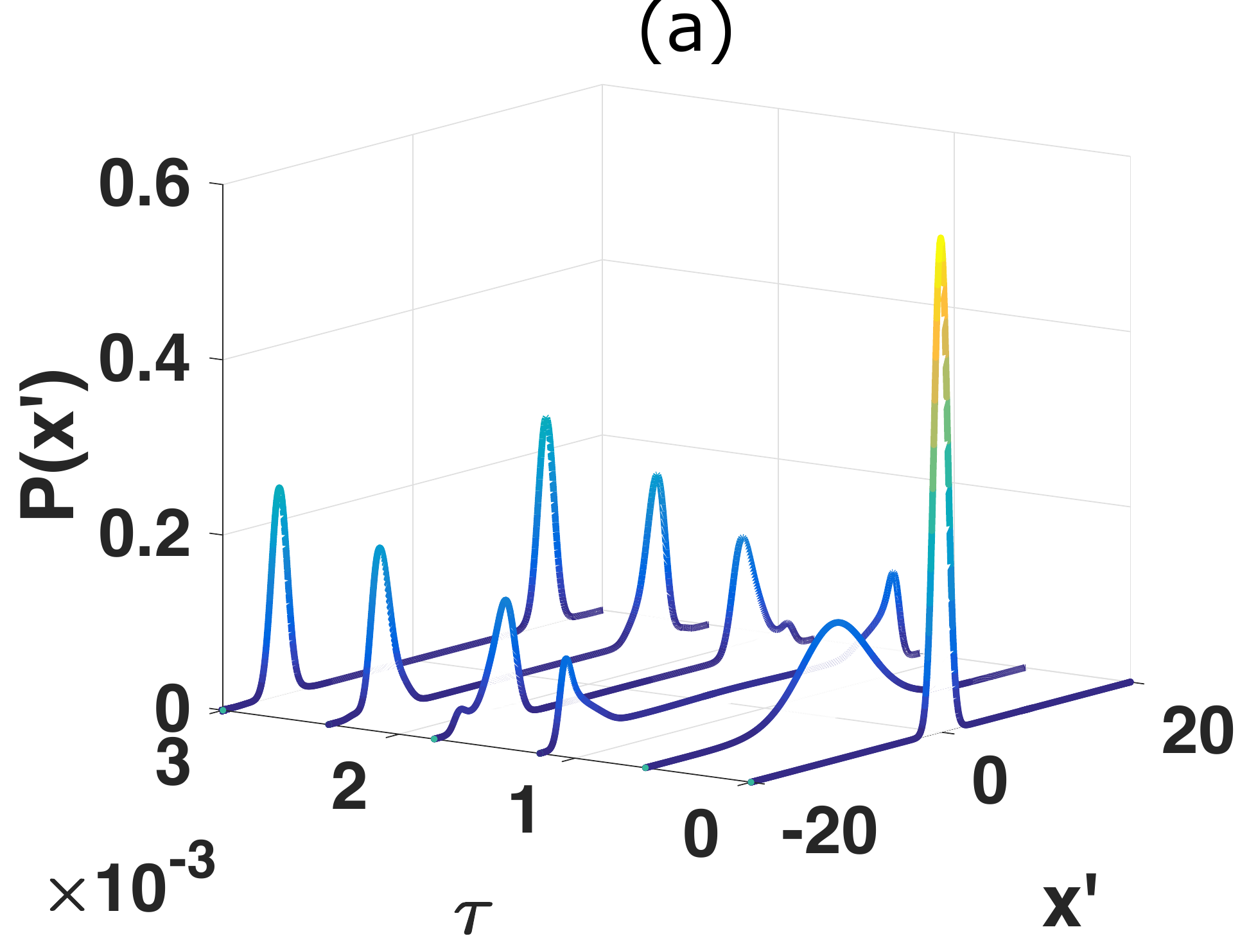}
\par\end{centering}
\includegraphics[width=0.7\columnwidth]{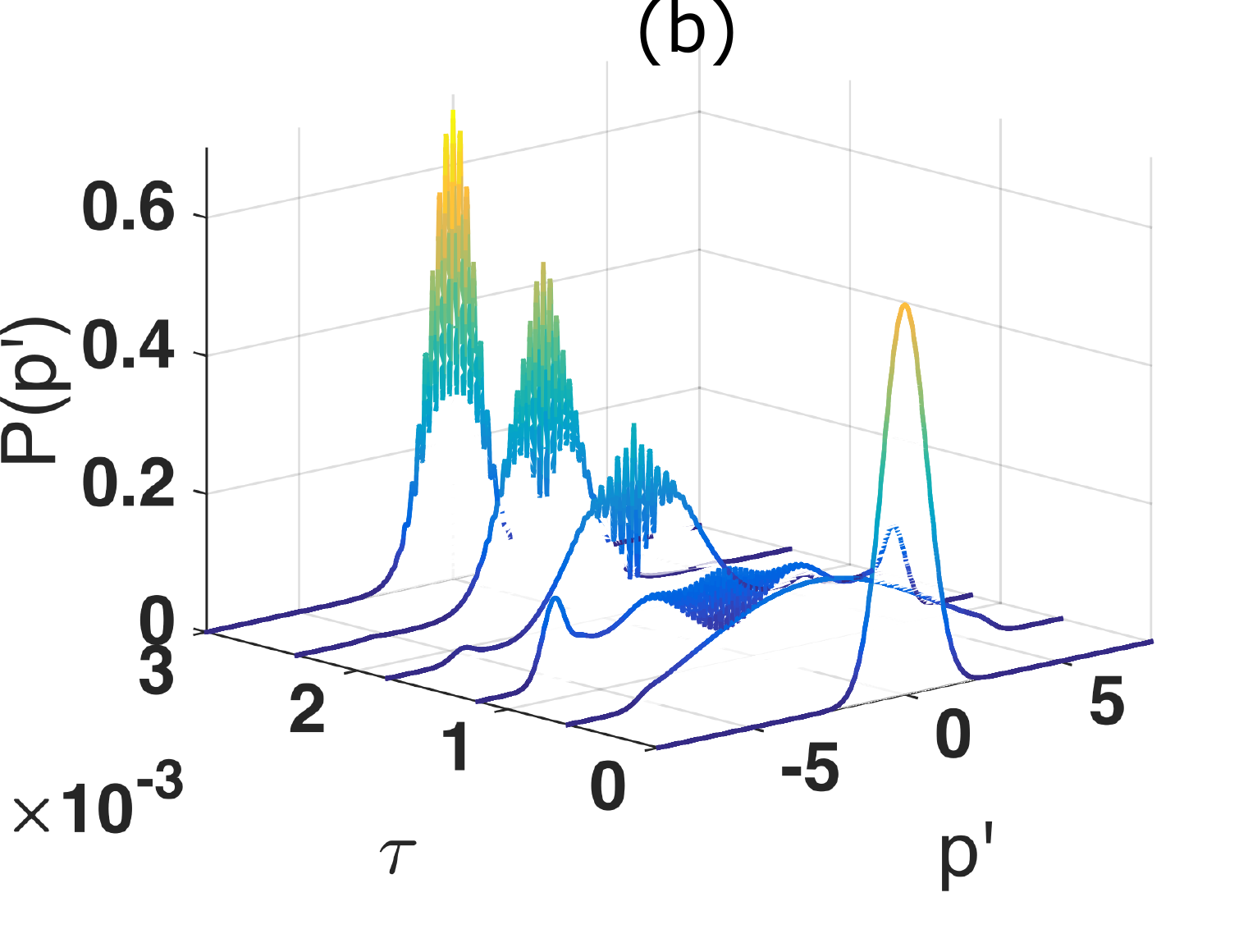}

\includegraphics[width=0.7\columnwidth]{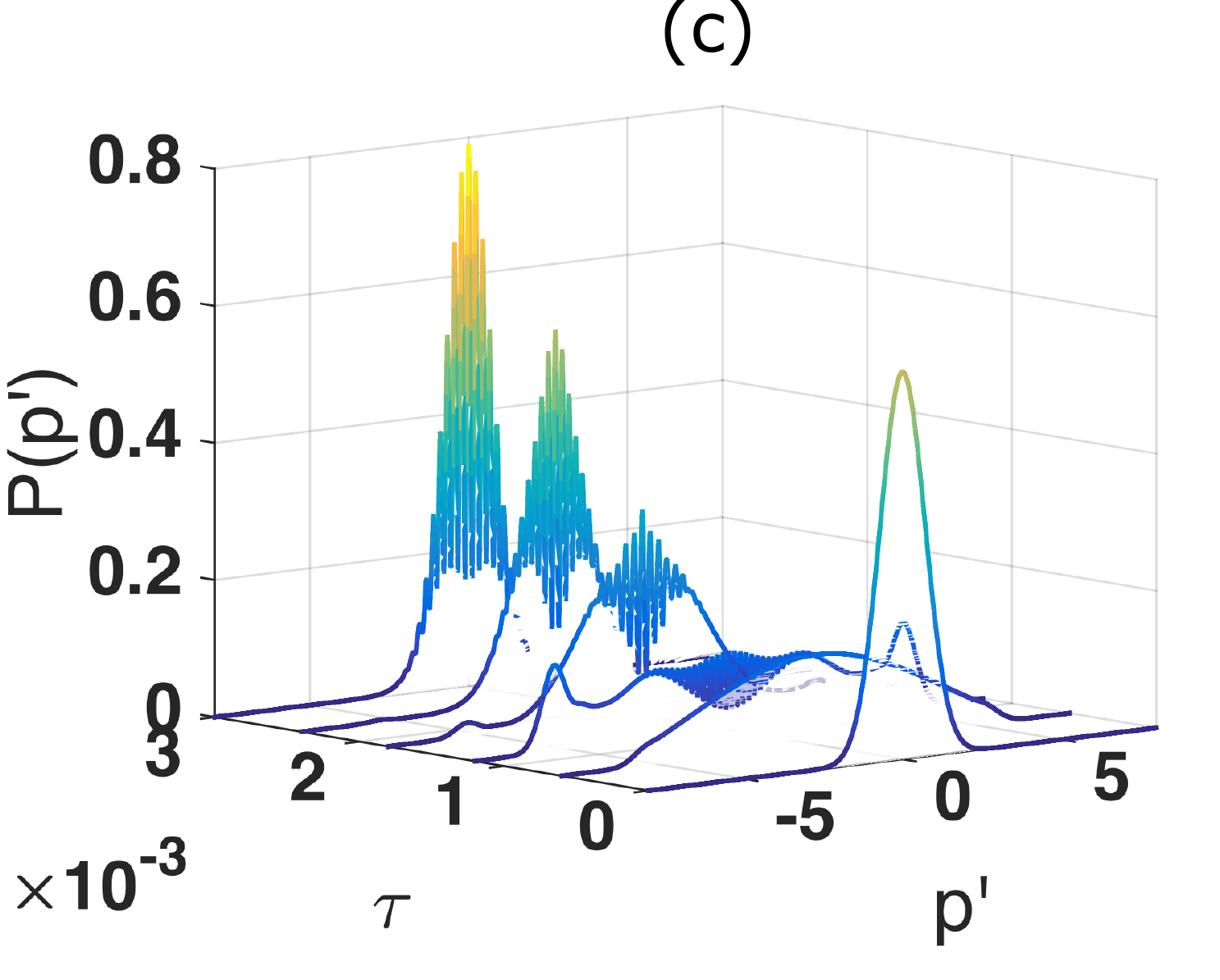}

\caption{Formation of cat states using squeezed reservoirs for the microwave
system. The $x'$-quadrature probability distribution and $p'$-quadrature
probability distribution are plotted as a function of dimensionless
time $\tau$, assuming zero temperature $N_{th}=0$ and with Kerr
nonlinearity $\chi=5$. The parameters are $g=2.5$, $\lambda/g^{2}=100$.
The amount of squeezing is determined by $M=\sqrt{N_{s}\left(N_{s}+1\right)}$.
Plot (b) is for $N_{s}=0$ (no squeezing). Plot (c) is for $N_{s}=2$,
$M=\sqrt{6}$ (squeezing is present). The $x'$-quadrature graphs
(a) are indistinguishable for each case.\label{fig:P_X_times-squeeze-Kerr}}
\end{figure}

In the presence of Kerr nonlinearity, the steady state of the system
in the limit of zero signal damping and for an initial vacuum state
corresponds to a cat state with complex amplitude $\alpha_{0}=\pm\sqrt{\lambda/\left(g^{2}+i\chi'\right)}$
\citep{Sun_PRA2019}. The interference fringes in this case are observed
along the $P_{\theta}$ quadrature, where $\theta$ is the phase angle
of the complex number $\alpha_{0}=|\alpha_{0}|\text{e}^{i\theta}$.
It is therefore desirable to have the reservoir in a squeezed state
along the direction of the $P_{\theta}$ quadrature. This can be achieved
by choosing $M=|M|\text{e}^{2i\theta}$, as can be seen in Eq. (\ref{eq:vars-1}),
with $|M|=\sqrt{N_{s}\left(N_{s}+1\right)}$ giving the optimal squeezing
strength.

\begin{figure}
\includegraphics[width=0.51\columnwidth]{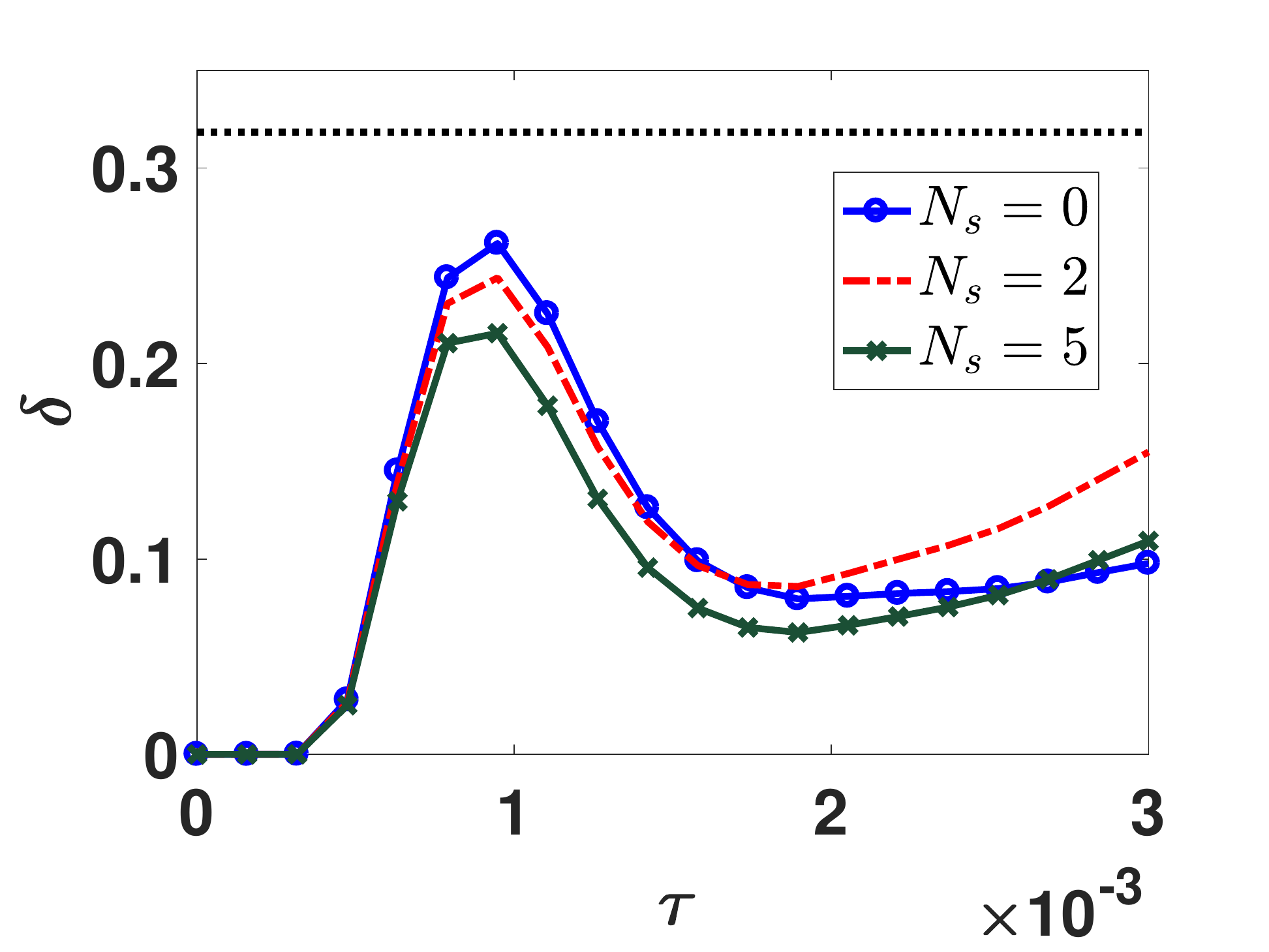}\includegraphics[width=0.51\columnwidth]{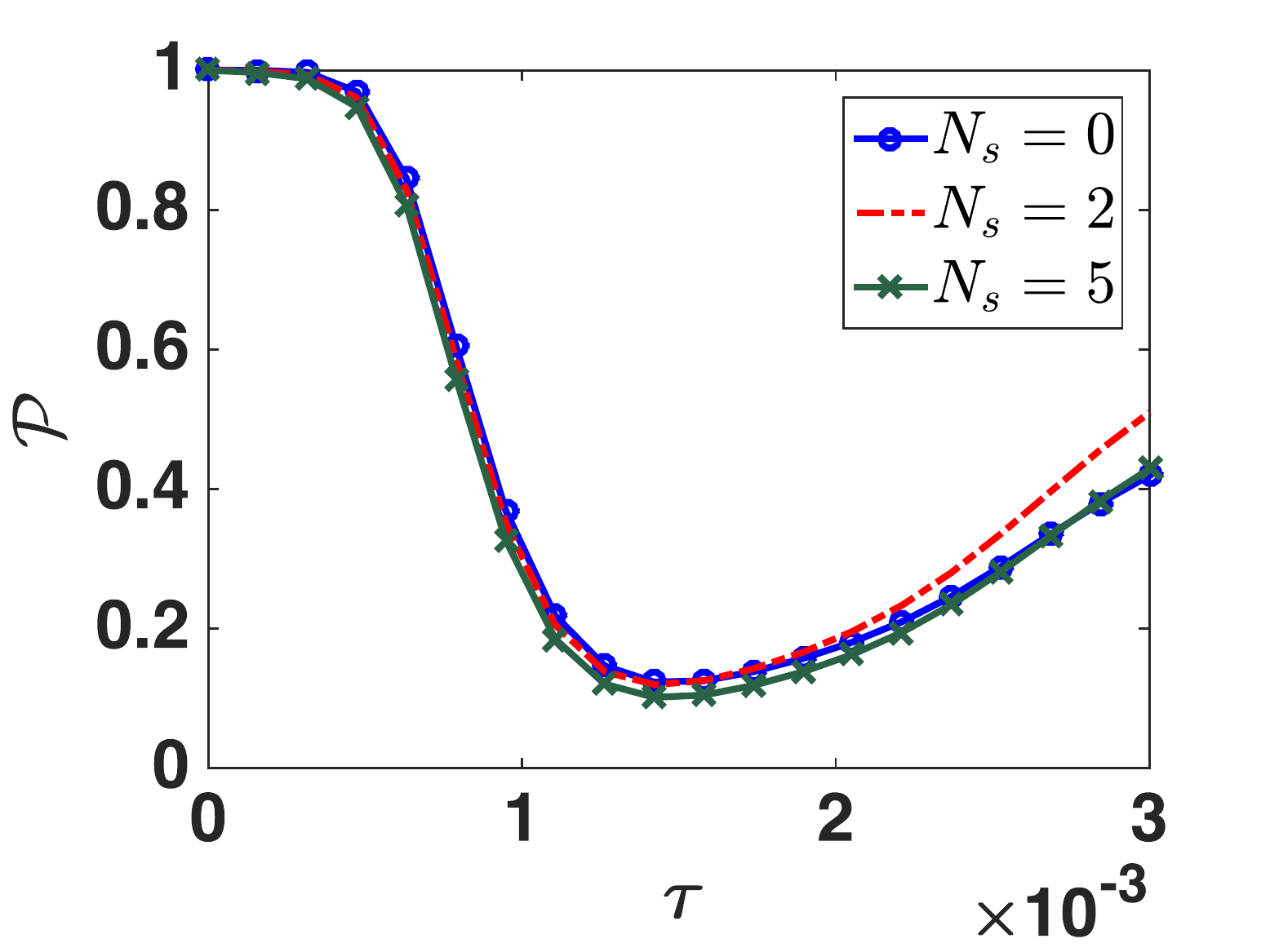}

\caption{The time evolution of the Wigner negativity (left) and purity (right)
at zero temperature ($N_{th}=0$) and for $\chi=5$, for different
squeezing strengths. Here, the parameters are $g=2.5$, $\lambda/g^{2}=100$
and $M=\sqrt{N_{s}\left(N_{s}+1\right)}$. The negativity value corresponding
to the pure cat state eq. (\ref{eq:cat-state}) is $\delta=0.318$
as shown in Figure \ref{fig:purity}. The mixture (\ref{eq:mix})
has zero negativity. \label{fig:wigner_neg_purity_nth0}}
\end{figure}

To study the effect of the squeezed reservoir, we first consider the
zero temperature case. In Fig. \ref{fig:P_X_times-squeeze-Kerr},
we observe two Gaussian peaks along the direction of the $x'$-quadrature,
indicating the formation of two coherent amplitudes with opposite
phases. As previously mentioned, the quantum nature of the state
is inferred from the $p'$-quadrature probability distribution. Fig.
\ref{fig:P_X_times-squeeze-Kerr} shows interference fringes. The
visibility of these fringes is higher with squeezing $N_{s}=2$. This
is confirmed by the time evolution of the Wigner negativity as presented
in Fig. \ref{fig:wigner_neg_purity_nth0}. We see that the Wigner
negativity is larger for $N_{s}=2$ as compared to $N_{s}=0$, after
the two peaks have formed along the $x'$-quadrature around $\tau=0.002$.
The corresponding purity is also higher with squeezing.

\begin{figure}
\includegraphics[width=0.51\columnwidth]{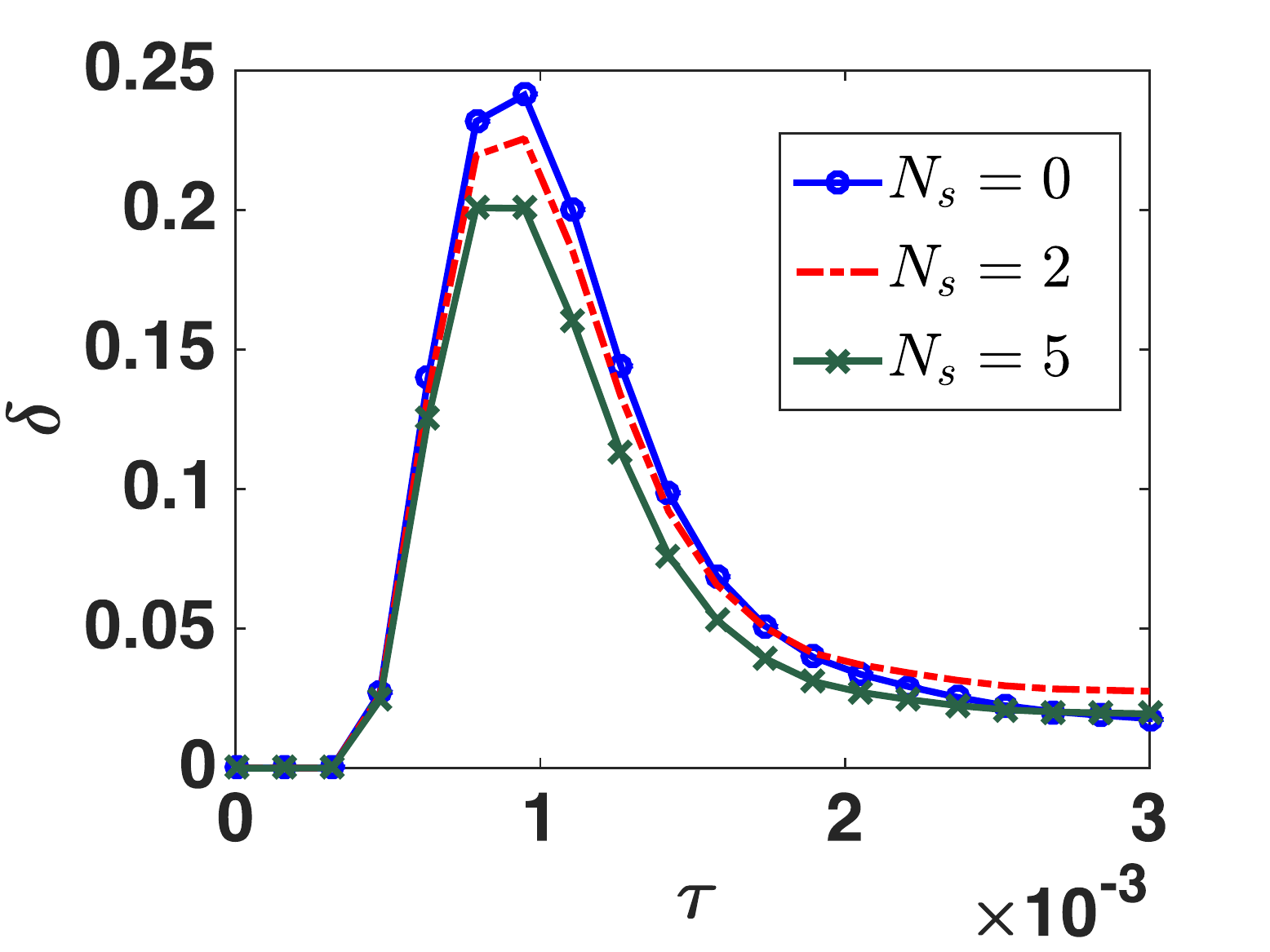}\includegraphics[width=0.51\columnwidth]{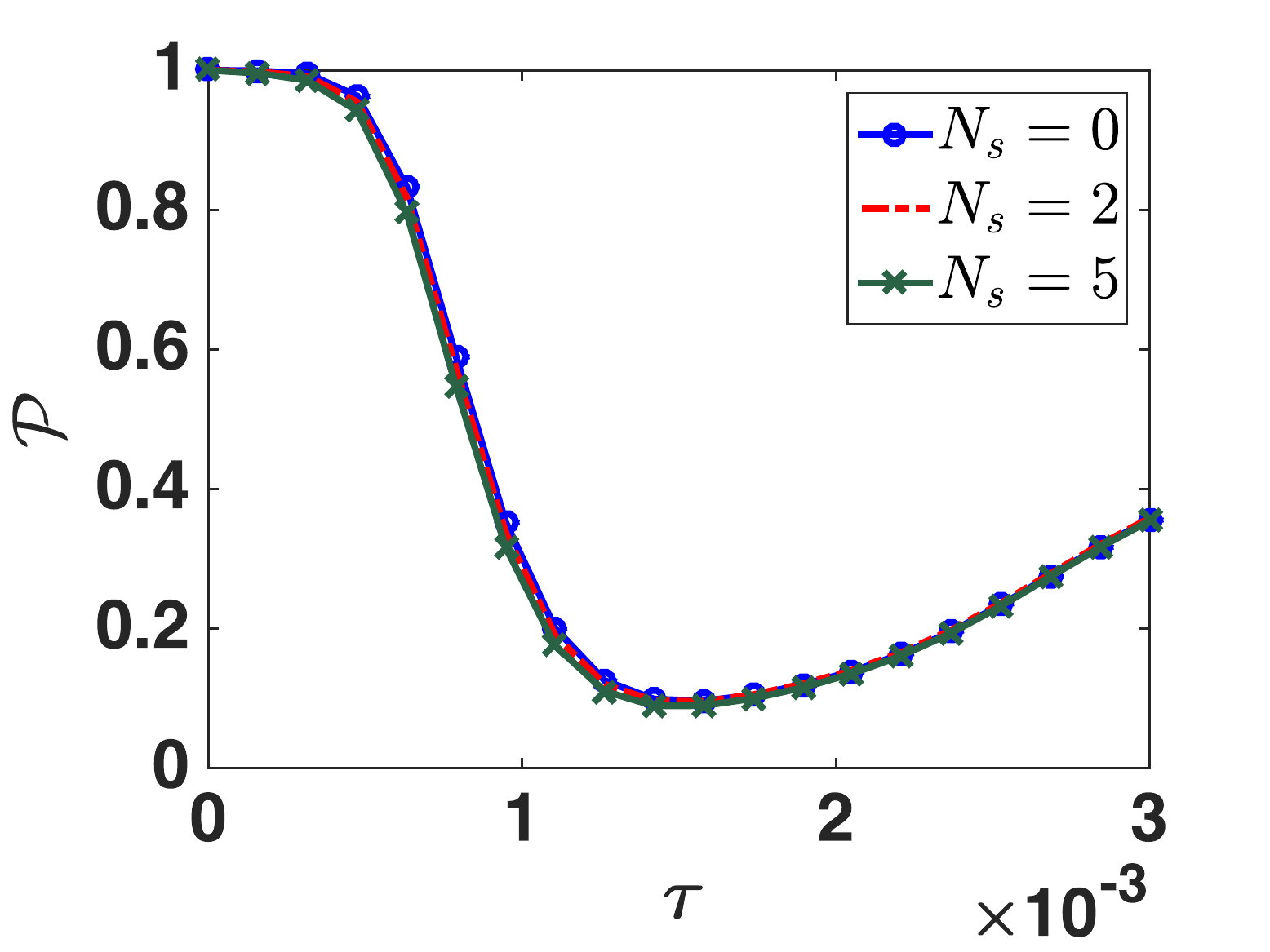}

\caption{The time evolution of the Wigner negativity (left) and purity (right)
at a finite temperature corresponding to $N_{th}=1$, for different
squeezing strengths. Parameters are $g=2.5$, $\lambda/g^{2}=100$,
$\chi=5$, and $M=\sqrt{N_{s}\left(N_{s}+1\right)}$.\label{fig:wigner_neg_purity_nth1}}
\end{figure}

However, the same figure shows a lower Wigner negativity and purity
for a larger squeezing strength $N_{s}=5$, compared to that with
$N_{s}=2$. These results can be understood by realizing that throughout
the simulation, the reservoir squeezing is along the $p'$-quadrature
and anti-squeezed along the $x'$-quadrature, while the steady-state
of the system only reaches two coherent amplitudes with opposite phases
along the $x'$-quadrature after a certain time. In other words, before
the system reaches the steady-state the squeezed reservoir adds noise
with a non-optimal orientation . This is evident from the slightly
counter-intuitive result of \emph{lower} Wigner negativities for \emph{larger}
squeezing amplitudes, as shown in Fig. \ref{fig:wigner_neg_purity_nth0},
before the development of the two probability distribution peaks along
the $x'$-quadrature. In the case of finite temperature, the thermal
bath contributes further extra noise to the system. This is shown
in Fig. \ref{fig:wigner_neg_purity_nth1}, where the application of
the squeezed state with fixed direction has little effect on the negativity.
\begin{figure}[H]
\includegraphics[width=0.51\columnwidth]{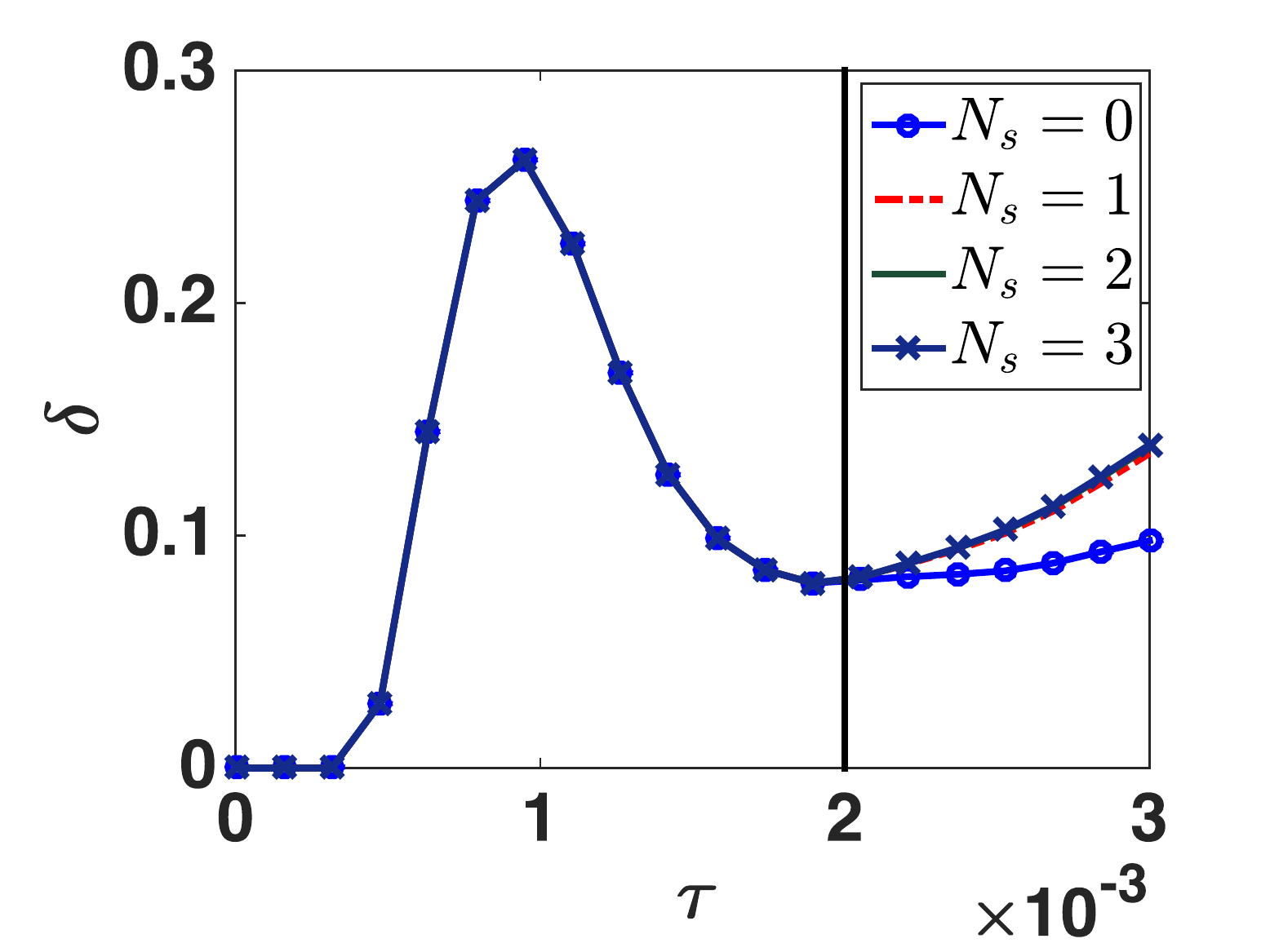}\includegraphics[width=0.51\columnwidth]{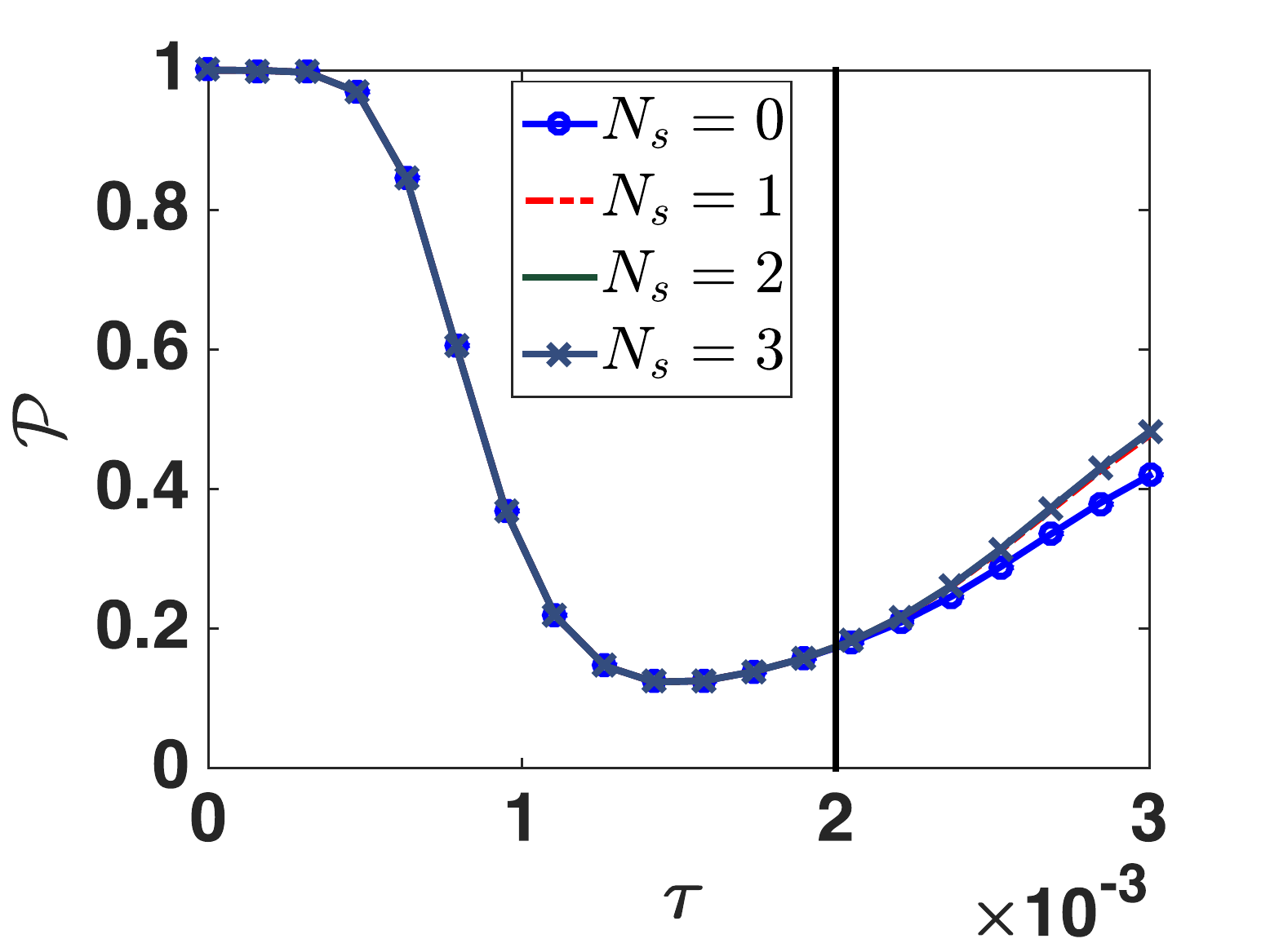}

\caption{The time evolution of the Wigner negativity (left) and purity (right)
at zero temperature, for different squeezing strengths. Here, different
to Fig. \ref{fig:wigner_neg_purity_nth0}, the squeezed reservoir
only interacts with the system at the time of formation of the two
distinct peaks along the $x'$-quadrature. The vertical line at $\tau=0.002$
is the time when the squeezed state is applied.\textcolor{red}{{} \label{wigner_neg_purity_nth1_after.002}}}
\end{figure}

To overcome this effect, we can take two approaches. First we allow
the system to evolve without squeezing of reservoirs until near the
time for the two distinct peaks in $x'$ to form. For the parameters
of Figure \ref{fig:wigner_neg_purity_nth0}, this corresponds to $\tau\sim0.002$.
At that time, we include a squeezed reservoir, the direction of squeezing
being in the $p'$ direction. This models the insertion of a squeezed
state into the input port of a single-ended cavity. The results of
the simulation are shown in Fig. \ref{wigner_neg_purity_nth1_after.002},
where it is seen that the squeezing gives an increased value of the
negativity after that time. The second approach is to apply a squeezed
state that has a varying direction, to match at a given time the direction
$p'$ orthogonal to the rotating amplitudes $\alpha_{0}$.

\begin{figure}
\begin{centering}
\includegraphics[width=0.7\columnwidth]{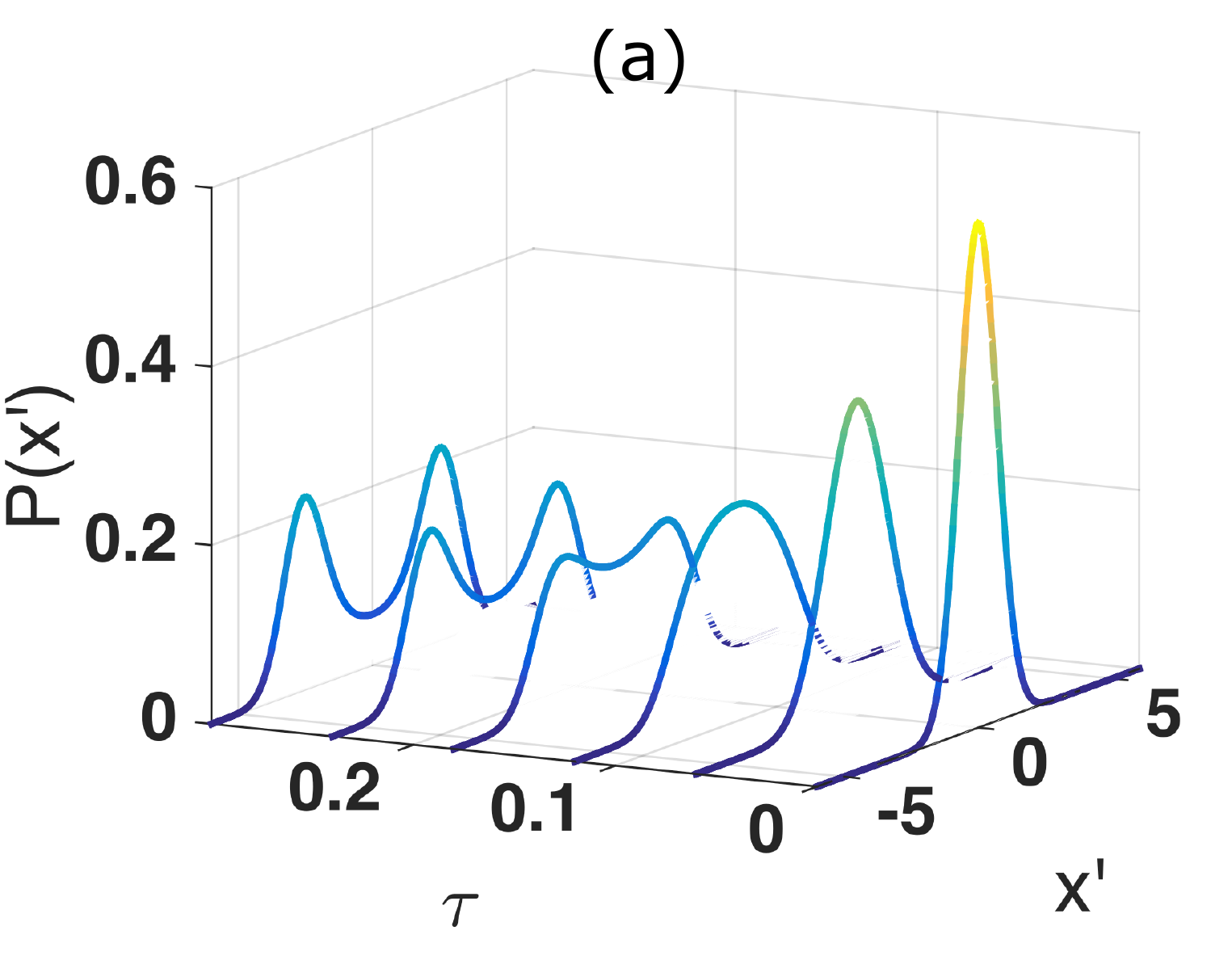}
\par\end{centering}
\includegraphics[width=0.7\columnwidth]{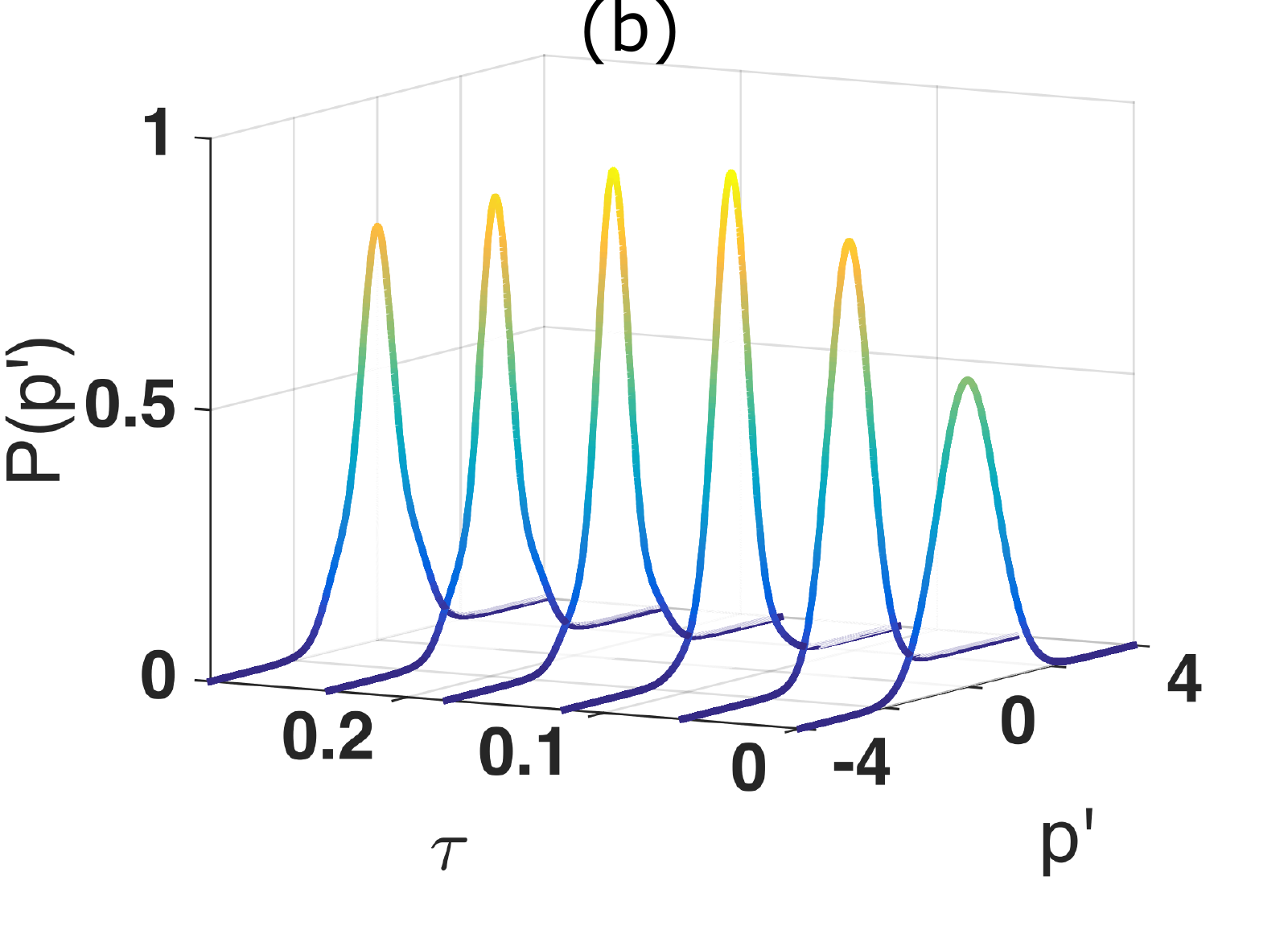}

\includegraphics[width=0.7\columnwidth]{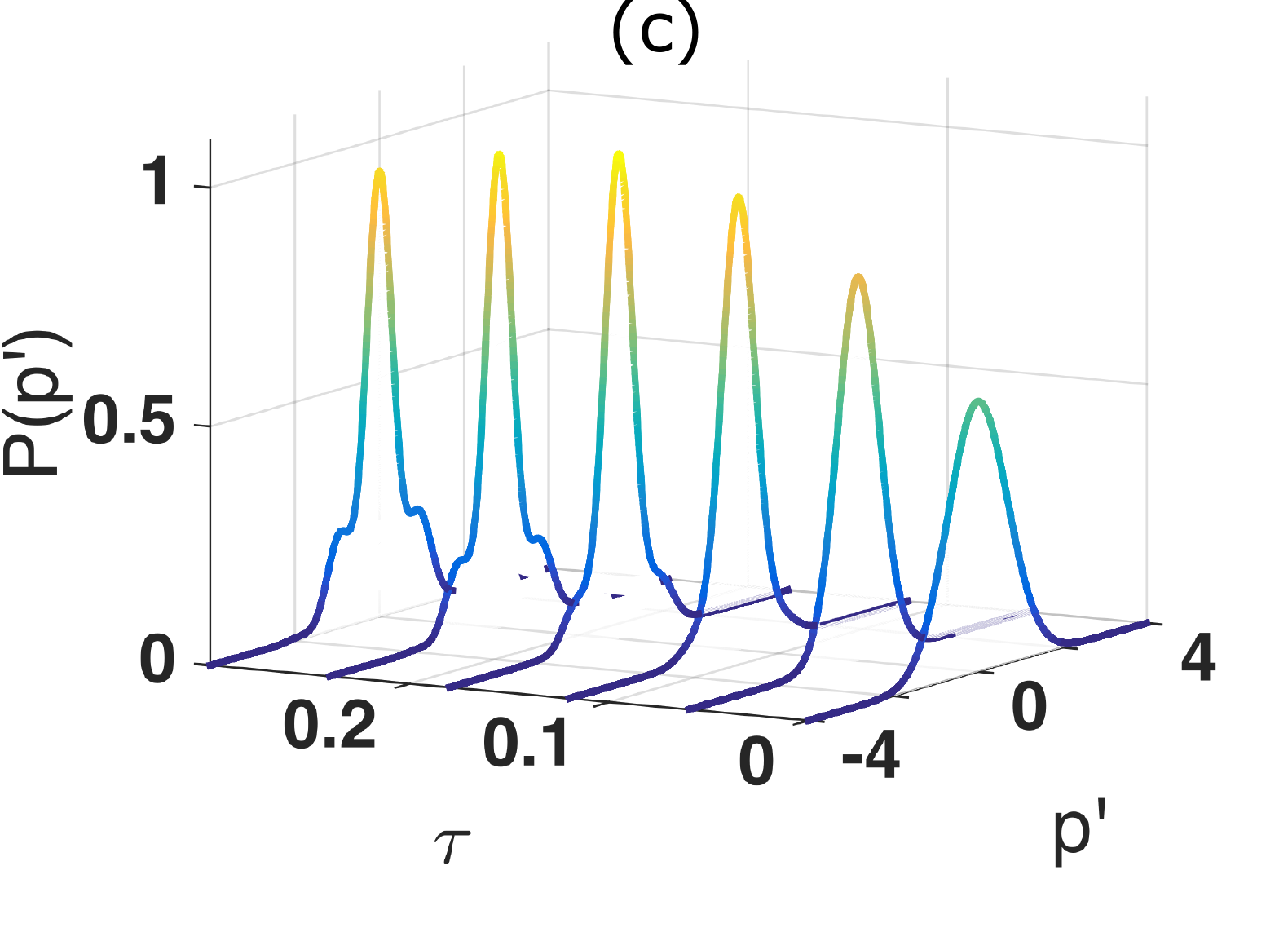}

\caption{The $x'$-quadrature (a) and $p'$-quadrature {[}(b) and (c){]} probability
distributions as a function of dimensionless time $\tau$, at a finite
temperature $N_{th}=0.02$. Here, the coherent amplitude is $|\alpha_{0}|=2$.
The parameters are $g=1.41$ and $\chi'=1.01$, which are taken from
the experiment of Leghtas et al. \citep{Leghtas_Science2015}. Plot
(b) is without squeezing, $N_{s}=0$. Plot (c) is with squeezing $N_{s}=0.5$.
The $x'$-quadrature probability distributions with and without squeezing
are indistinguishable. \label{fig:p_x_experiment}}
\end{figure}

\subsection{Squeezing effects with typical experimental parameters}

Next, we look at the effects of a squeezed reservoir using experimental
parameters for the superconducting circuit experiment of Leghtas et
al.\textcolor{red}{{} }\citep{Leghtas_Science2015}. Here we evaluate
$g=1.41$ and $\chi'=1.01$, giving an estimated $\chi=\chi'/g^{2}=0.5$
and $|\alpha_{0}|=2$. The mean thermal photon number is very low,
at $N_{th}\leq0.05\ll1$, which is set here at $N_{th}=0.02$. We
present the $x'$-quadrature probability distributions in Fig. \ref{fig:p_x_experiment}
and observe that the two peaks along $x'$ do not become that well
separated. The simulation introduces the squeezed reservoir after
$\tau=0.1$, when the two peaks are starting to develop. 

\begin{figure}
\includegraphics[width=0.51\columnwidth]{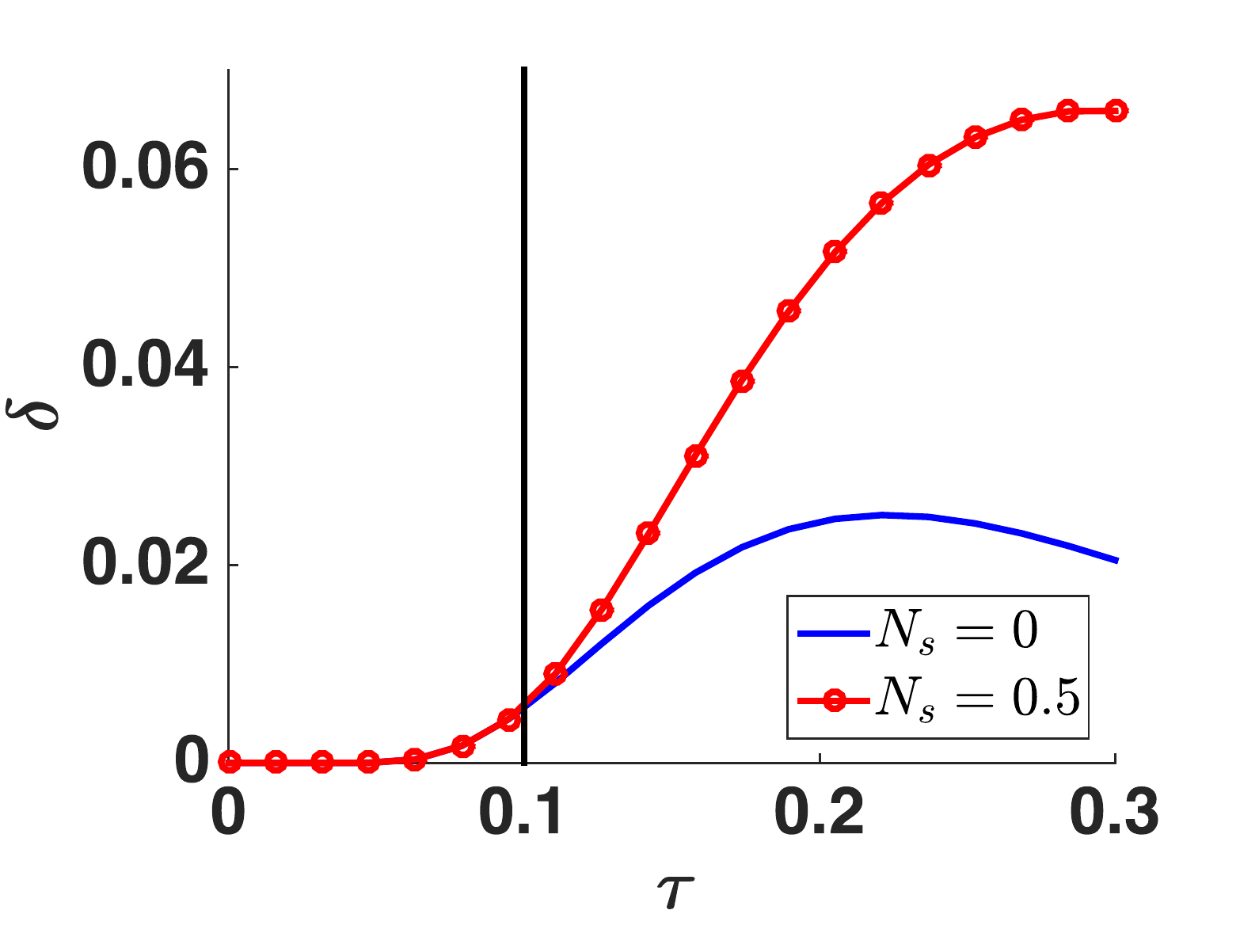}\includegraphics[width=0.51\columnwidth]{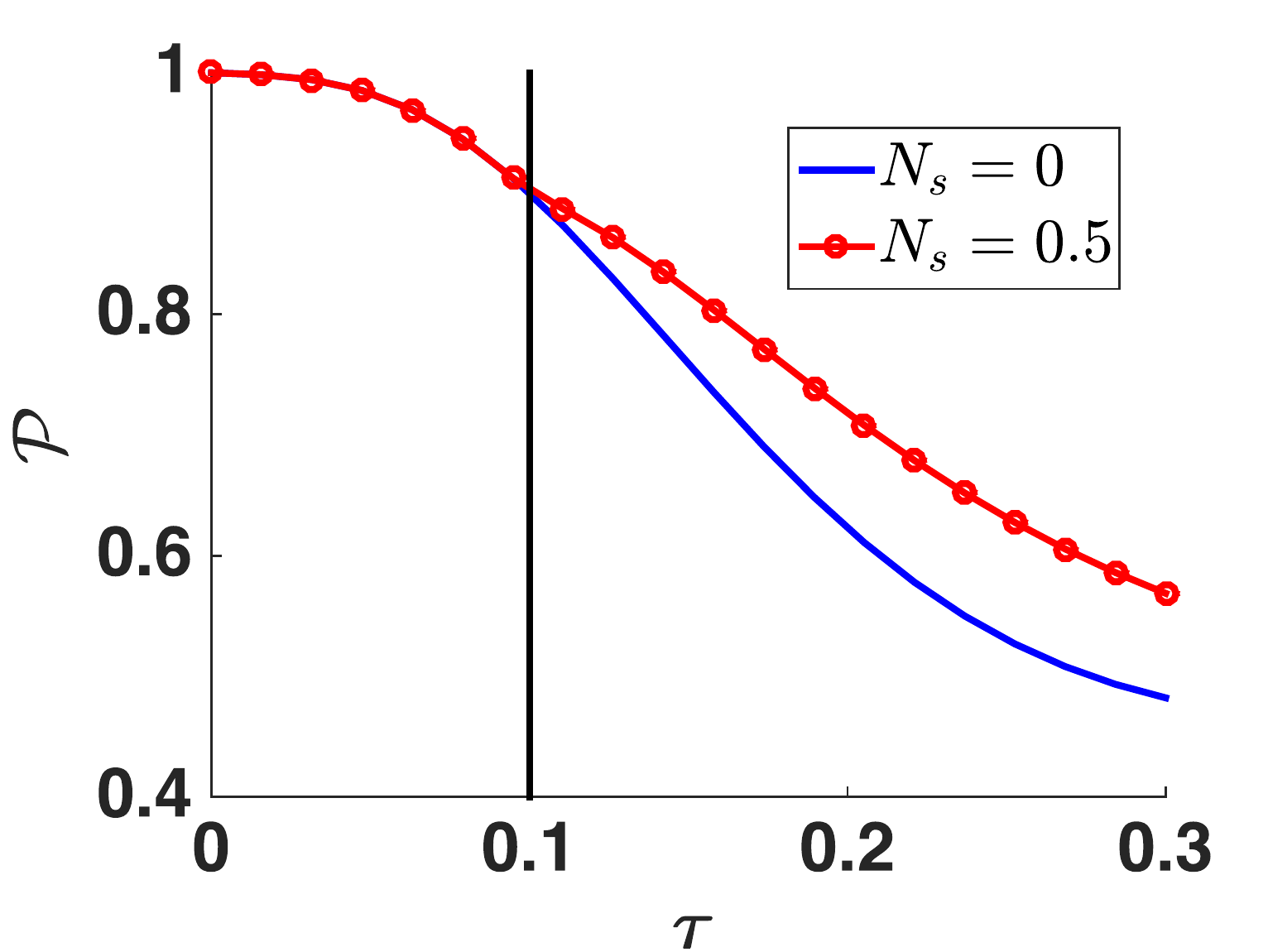}

\caption{The time evolution of (left) the Wigner negativity and (right) purity.
The parameters are identical to those given in Figs. \ref{fig:p_x_experiment}.
\label{fig:wigner_neg_purity_experiment} }
\end{figure}

The effect of adding a squeezed reservoir, which was not present in
the original experiment, can be observed in Fig. \ref{fig:p_x_experiment}
where we show the time evolution of the $p'$-quadrature probability
distribution. Without the squeezed reservoir, the $p'$-quadrature
probability distribution is Gaussian-like throughout the simulation.
The probability distribution with squeezing displays some features
of interference fringes, suggesting a quantum state with larger non-classicality
in the presence of the squeezed reservoir. This is confirmed in Fig.
\ref{fig:wigner_neg_purity_experiment} where we compare the Wigner
negativity and purity for the case without and with squeezing. With
squeezing, the Wigner negativity has a larger value. Similarly, there
is an increase in the purity of the quantum state.

\begin{figure}
\includegraphics[width=0.5\columnwidth]{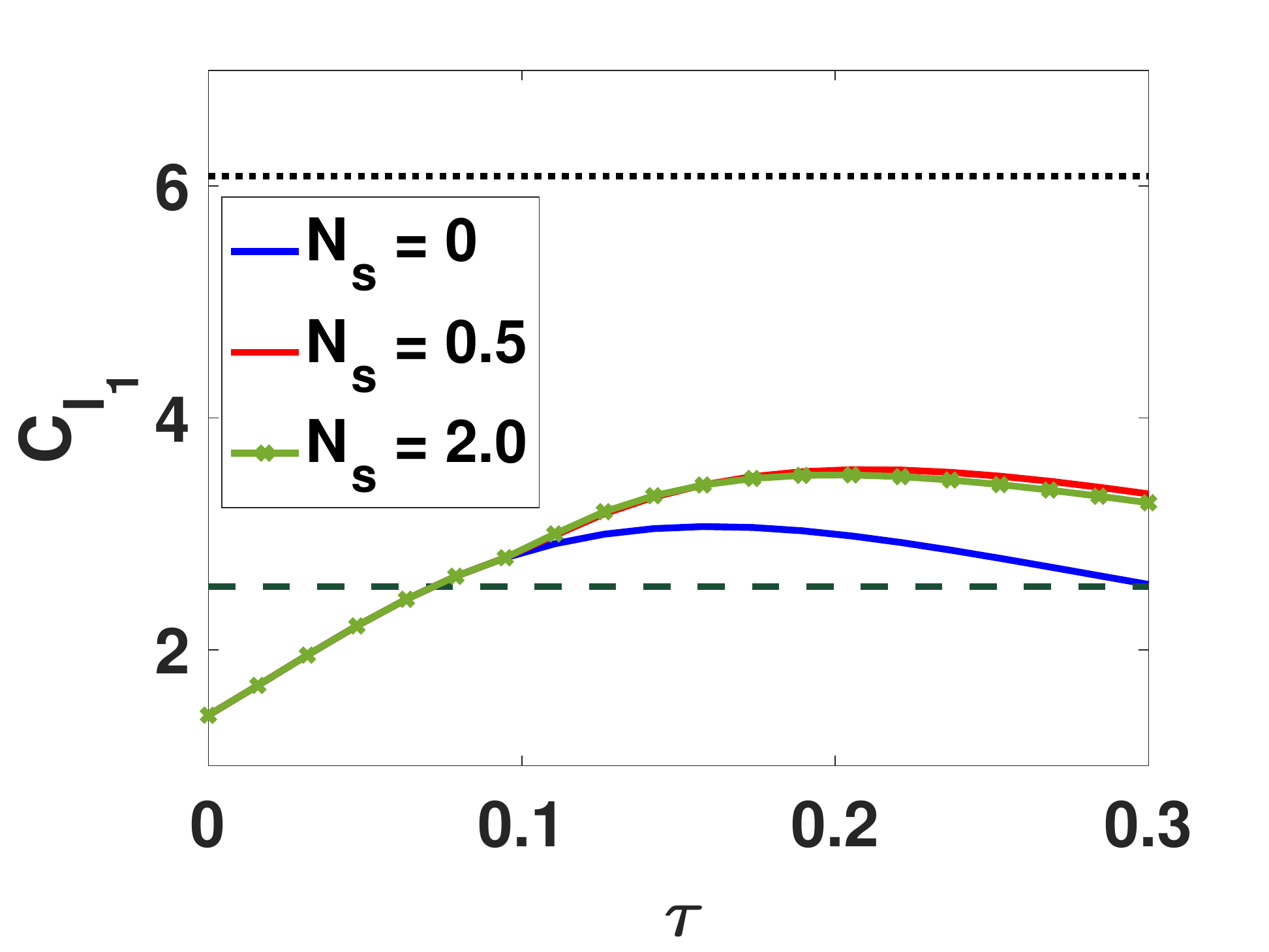}\includegraphics[width=0.5\columnwidth]{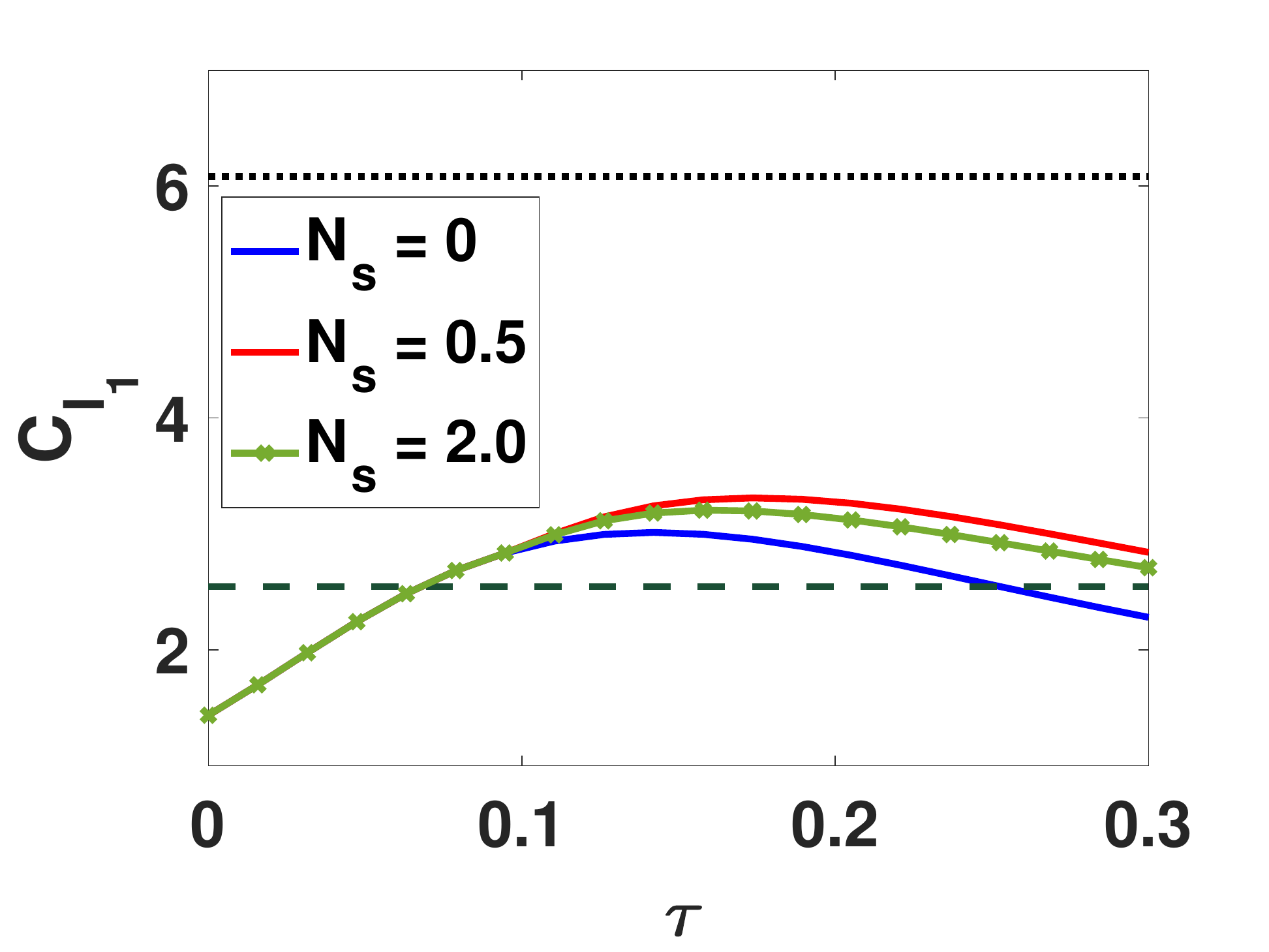}

\caption{The time evolution of quantum coherence $C_{l_{1}}$ for the evolution
of a cat-like state using parameters similar to the experiment of
Ref \citep{Leghtas_Science2015}. For each plot we show results with
squeezing (upper red solid line) and without squeezing (lower blue
solid line). The results for parameters similar to the experiment
($g=1.41$, $N_{th}=0.02$, $\chi'=1.01$) are shown in the figure
on the right. Parameters for the left figure are $g=1.41$, $N_{th}=0$,
$\chi'=0$, and are given for comparison. In both figures, $|\alpha_{0}|=2$
and $M=\sqrt{N_{s}\left(N_{s}+1\right)}$ when $N_{s}\protect\neq0$.
The upper horizontal dashed line gives the $C_{l_{1}}$ value of the
corresponding cat state. The lower dashed horizontal line gives the
$C_{l_{1}}$ value for mixture of coherent states with amplitude $|\alpha_{0}|=2$.\label{fig:coherence_experiment}}
\end{figure}

To probe the properties of the cat-like state that might be generated
in a potential experiment, we next calculate the quantum coherence
$C_{l_{1}}(\rho)$ for the transient cat-like states versus time.
For the purpose of comparison, Figure (\ref{fig:cl1_nth1}) gives
the values of $C_{l_{1}}(\rho)$ for the pure cat state and for the
mixture $\rho_{mix}$ of the two coherent states. As explained in
Section III, the quantum coherence for the mixture arises from the
quantum coherence of the coherent states $|\pm\alpha\rangle$ involved
in the mixture, whereas the quantum coherence for the cat state contains
the extra contribution due to the state being a macroscopic superposition.
Figure \ref{fig:coherence_experiment} gives the $C_{l_{1}}(\rho)$
for the parameters of the Leghtas et al experiment. Here, the quantum
coherence is at 50\% the value for an ideal cat state, significantly
above that of the mixture, and it can be seen that the squeezing enhances
the lifetime for a higher quantum coherence.  When squeezing is present,
the $C_{l_{1}}$ values are higher, implying a higher degree of quantum
coherence. However, consistent with the observation that the threshold
features for the cat state cannot be significantly altered by squeezing,
we note that the effect of squeezing saturates, in this case at $N_{s}\sim0.5$.
\textcolor{red}{ }Using the experimental parameters of Leghtas et
al. \citep{Leghtas_Science2015}, where the signal mode frequency
is $\omega_{s}=2\pi\times4.01\text{GHz}$, and the expression $N_{th}=1/\left[\text{exp}\left(\hbar\omega_{s}/kT\right)-1\right]$,
a mean thermal photon number $N_{th}$ of $0.02$ corresponds to a
temperature $T\approx50\text{mK}$.

\section{Conclusion}

In this paper, we have demonstrated the feasibility of using quantum
squeezed states to enhance the macroscopic quantum coherence of dynamical
cat states prepared in a cavity through a nonlinear parametric interaction.
Here, we have focussed on degenerate parametric oscillation (DPO)
in the limit of an adiabatically eliminated pump mode, where a two-photon
dissipative mechanism dominates. The cat state, which is a superposition
of two coherent states with a $\pi$ phase difference, is predicted
to be created in the cavity signal mode from an initial vacuum state,
in the limit where there are no losses in the cavity mode. A cat state
is also predicted in the microwave regime, where extra nonlinear terms
are present. The loss of photons from the cavity decoheres the cat
state. Thus, whether the cat state is formed or not depends on the
dynamical interplay of the quantum nonlinear effect versus the signal
decoherence.

We have found that it is possible to significantly limit the effects
of decoherence, by squeezing the input vacuum noise that enters the
cavity port at the signal frequency. This amounts to controlling the
fluctuations of the reservoir. The cat state then has a longer lifetime,
which enables the formation of cat states of a higher purity for the
same experimental parameters. Our result is consistent with earlier
work on the decoherence of a pure cat state, and extends previous
treatments of the effect of squeezing on cat states formed in the
DPO system by giving a broader range of parameters and implementing
different signatures, such as negativity and measures of quantum coherence.

We obtain new results for the microwave system, which is relevant
to current experiments, and based on a different Hamiltonian. We find
that the direction of squeezing is important. The enhancement occurs
when the quantum noise of the signal cavity input is squeezed in the
direction orthogonal to the axis connecting the amplitudes of the
two coherent states. This rotates with the dynamics, and hence the
squeezing direction needs to be controlled. Ultimately, the direction
is determined by the type of nonlinearity, and whether an additional
Kerr effect is present.

Thermal noise also induces a decoherence of the cat state and has
a profound effect on the feasibility of creating the cat states in
the dynamical system. In the microwave regime, cooling is necessary
to observe the formation of cat states. We have shown how the presence
of squeezing enhances the quality of the cat state that can be formed
for systems at a temperature corresponding to the microwave experiment
of Leghtas et al \citep{Leghtas_Science2015}. Our results are based
on statistics modeled on a squeezed thermalized reservoir, and are
consistent with earlier treatments \citep{Kennedy_PRA1988,serafini2005quantifying,serafini2004minimum},
showing that while squeezing can inhibit the thermal decoherence,
the thermal decoherence cannot be completely overcome by further increasing
squeezing. This contrasts with the result in the absence of thermal
noise. On the whole, based on analytical calculations where one analyses
the decoherence of an ideal cat state coupled to a reservoir, we anticipate
a more promising situation for a squeezed thermal state input. This
models a squeezed state generated from a source at nonzero temperature,
but requires that the cavity be kept at zero temperature over the
timescales during which the cat state forms in the cavity.

\section*{\textup{\normalsize{}Acknowledgements}}

This work was performed on the OzSTAR national facility at Swinburne
University of Technology. OzSTAR is funded by Swinburne University
of Technology and the National Collaborative Research Infrastructure
Strategy (NCRIS). PDD and MDR thank the hospitality of the Weizmann
Institute of Science. This work was funded through Australian Research
Council Discovery Project Grants DP180102470 and DP190101480, and
through a grant from NTT Phi Laboratories.

\section*{Appendix}

\subsection*{1. Statistical moments for a squeezed thermal state}

A squeezed thermal state has a corresponding density operator \citep{Fearn_JModOpt1988,Kim_squeezed_PRA1989}
\begin{align*}
\rho_{ST} & =S\rho_{T}S^{\dagger}\\
 & =\frac{1}{1+N_{th}}\sum_{n}\left(\frac{N_{th}}{1+N_{th}}\right)^{n}S|n\rangle\langle n|S^{\dagger}\,,
\end{align*}
where $S(\epsilon)=\text{exp}\left[\left(\epsilon^{*}\Gamma^{2}-\epsilon\Gamma^{\dagger2}\right)/2\right]$
is the squeezing operator, $\epsilon=re^{i\phi}$ and $N_{th}$ is
the mean photon number of a thermal state. The mean photon number
for the squeezed thermal state is 
\begin{align*}
\langle\Gamma^{\dagger}\Gamma\rangle & =N_{th}\cosh2r+\sinh^{2}r\,,
\end{align*}
while other statistical moments are given by
\begin{align*}
\langle\Gamma^{2}\rangle & =-\frac{1}{2}\sinh2re^{i\phi}\left(2N_{th}+1\right)
\end{align*}
and 
\begin{align*}
\langle\Gamma^{\dagger2}\rangle & =-\frac{1}{2}\sinh2re^{-i\phi}\left(2N_{th}+1\right)\,.
\end{align*}
Compare these equations with the statistical moments for the environment
mode operators in Eq. (\ref{eq:noise_statistics}), we see that, for
a squeezed thermal state,
\begin{align*}
N & =N_{th}\cosh2r+\sinh^{2}r\\
M & =-\frac{1}{2}\sinh2re^{i\phi}\left(2N_{th}+1\right)\\
 & =\frac{1}{2}\left(2N_{th}+1\right)\sinh2re^{i\left(\phi-\pi\right)}\,.
\end{align*}
In particular, for $\theta=0$, the variances of $X$ and $P$ quadratures
as defined in Eq. (\ref{eq:noise_X_P-1}) have the expressions:\textcolor{blue}{}
\begin{align}
\Delta^{2}X & =\left(N_{th}\cosh2r+\sinh^{2}r+\frac{1}{2}\right)\nonumber \\
 & +\frac{1}{2}\sinh2r\left(2N_{th}+1\right)\cos\left(\phi-\pi\right)\nonumber \\
\Delta^{2}P & =\left(N_{th}\cosh2r+\sinh^{2}r+\frac{1}{2}\right)\nonumber \\
 & -\frac{1}{2}\sinh2r\left(2N_{th}+1\right)\cos\left(\phi-\pi\right)\,.\label{eq:XP_var}
\end{align}
In order to get a squeezing in the $P$ quadrature, we set $\phi=\pi$
($M$ is real with amplitude $\left(N_{th}+1/2\right)\sinh2r$) and
Eq. (\ref{eq:XP_var}) is then
\begin{align}
\Delta^{2}X & =\left(N_{th}+\frac{1}{2}\right)e^{2r}\nonumber \\
\Delta^{2}P & =\left(N_{th}+\frac{1}{2}\right)e^{-2r}\,.\label{eq:XP_var_pi}
\end{align}

We note that the thermal contribution is squeezed (anti-squeezed)
by the factor of $e^{-2r}$ ($e^{2r}$). This means that such thermal
squeezing (if in the direction so as to squeeze fluctuations in $P$)
would give significant enhancement of the observed macroscopic quantum
coherence. We do not give explicit calculation of such enhancement
in this paper, but the calculations could be done in principle by
substituting for $N$ and $M$.

\subsection*{2. Statistical moments for a thermalized squeezed state}

Fearn and Collett \citep{Fearn_JModOpt1988} defined a thermalized
squeezed state with a density operator in the coherent basis (Glauber-P
representation) as follows:
\begin{align*}
\rho_{TS} & =\frac{1}{\pi N_{th}}\intop d^{2}\beta\,\text{exp}\left(-\frac{\left|\beta\right|^{2}}{N_{th}}\right)|\beta,\epsilon\rangle\langle\beta,\epsilon|\,,
\end{align*}
where $|\beta,\epsilon\rangle=D(\beta)S(\epsilon)|0\rangle=S(\epsilon)D(\gamma)|0\rangle$
is a displaced squeezed state, $D(\beta)$ is the displacement operator,
$S(\epsilon)$ is the squeezing operator as defined in the previous
section and $\gamma=\beta\cosh r+\beta^{*}e^{i\phi}\sinh r$.

The mean photon number for the thermalized squeezed state is
\begin{align*}
\langle\Gamma^{\dagger}\Gamma\rangle & =\sinh^{2}r+N_{th}
\end{align*}
and the statistical moments $\langle\Gamma^{2}\rangle$ and $\langle\Gamma^{\dagger2}\rangle$
are
\begin{align*}
\langle\Gamma^{2}\rangle & =\frac{1}{2}\sinh2re^{i\left(\phi-\pi\right)}\\
\langle\Gamma^{\dagger2}\rangle & =\frac{1}{2}\sinh2re^{-i\left(\phi-\pi\right)}\,.
\end{align*}
Comparing these equations with the statistical moments for the environment
mode operators in Eq. (\ref{eq:noise_statistics}), we see that, for
a thermalized squeezed state
\begin{align}
N & =N_{th}+\sinh^{2}r\nonumber \\
M & =\frac{1}{2}\sinh2re^{i\left(\phi-\pi\right)}\,.\label{eq:NM_thermal_squeezed_state}
\end{align}
 For $\theta=0$ and $\phi=\pi$, the variances of $X$ and $P$ quadratures
as defined in Eq. (\ref{eq:noise_X_P-1}) have the expressions:\textcolor{blue}{}
\begin{align}
\Delta^{2}X & =2N_{th}+e^{2r}\nonumber \\
\Delta^{2}P & =2N_{th}+e^{-2r}\,.\label{eq:XP_TS}
\end{align}
As opposed to the squeezed thermal state, the thermal number $N_{th}$
in Eq. (\ref{eq:XP_TS}) is not affected by the squeezing operation,
implying that the squeezing is weaker for a thermalized squeezed state
with the same squeezing strength $r$ as in the squeezed thermal state.\textcolor{blue}{}

\subsection*{3. Effect of thermal noise on decoherence of a system prepared initially
in a cat state}

Kennedy and Walls \citep{Kennedy_PRA1988} investigated the effect
of squeezing on the probability distribution interference fringes
visibility. They considered an initial cat state with a density operator
$\rho(0)=\mathcal{N}\left(e^{-i\pi/4}|\alpha\rangle+e^{i\pi/4}|-\alpha\rangle\right)\left(e^{i\pi/4}\langle\alpha|+e^{-i\pi/4}\langle-\alpha|\right)$.
The term in the probability distribution for $x_{\theta}$ that contributes
to the interference fringes is found to be proportional to $\left|\langle\alpha|-\alpha\rangle\right|^{\eta}$:
\begin{align}
\langle x_{\theta}|\rho(t)|x_{\theta}\rangle & \propto\left|\langle\alpha|-\alpha\rangle\right|^{\eta}\nonumber \\
 & =\exp\left(-2|\alpha|^{\eta}\right)\,
\end{align}
where 
\begin{align*}
\eta & =1-\frac{e^{-2\gamma t}}{\left\{ 1+2\left[N+|M|\cos\left(2\theta+\Phi\right)\right]\left(1-e^{-2\gamma t}\right)\right\} }\,,\\
\end{align*}
$\theta$ determines the phase quadrature $x_{\theta}$, and $\Phi$
is the phase of the complex parameter $M$, as previously defined.
Here, $N=N_{th}+N_{s}$. For the cat state considered, it is optimal
to choose $\cos\left(2\theta+\phi\right)=-1$.\textcolor{blue}{}

\begin{figure}
\includegraphics[width=0.8\columnwidth]{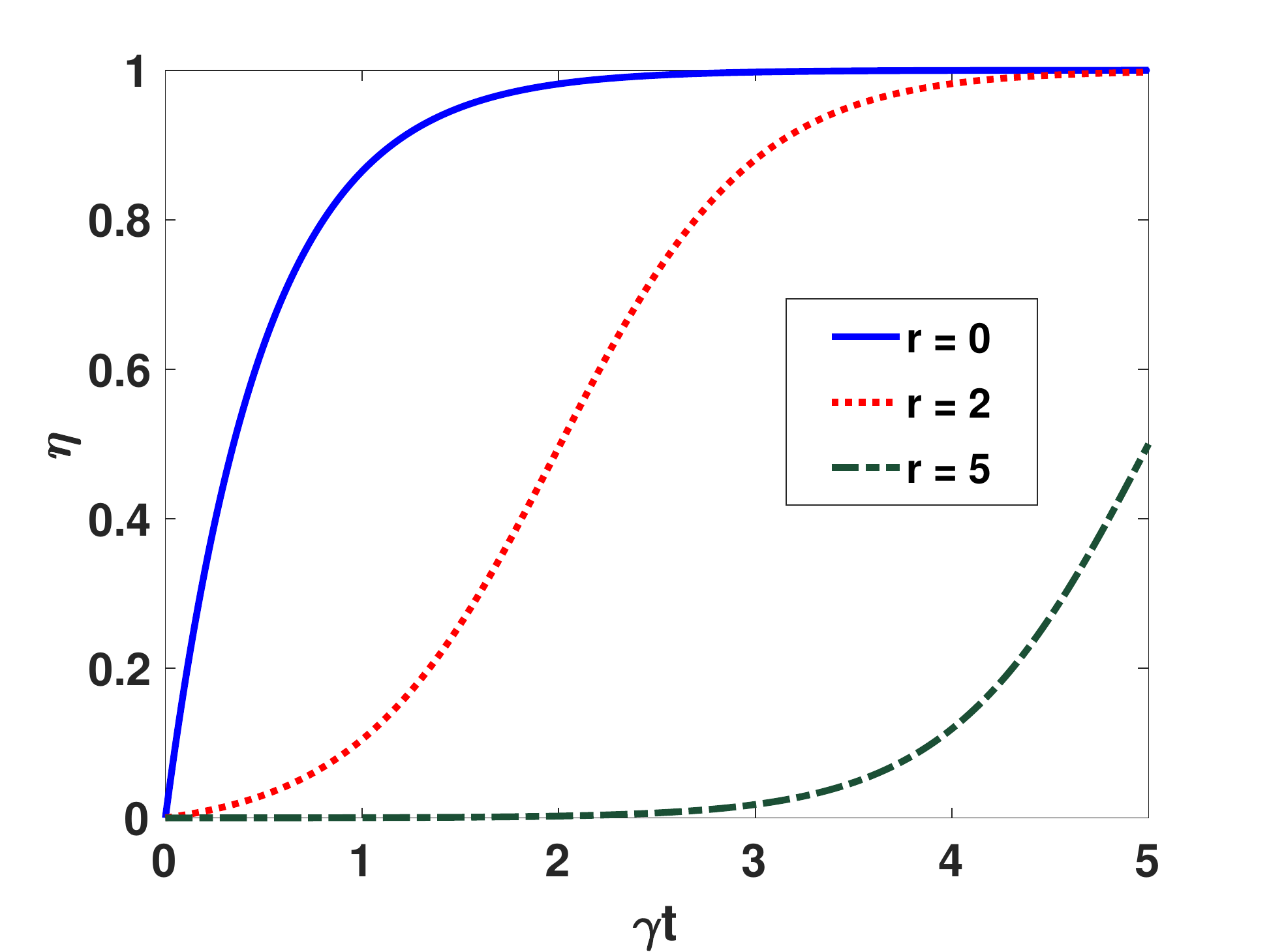}

\caption{The term that contributes to the interference pattern $\eta$ as a
function of $\gamma t$. $\eta=0$ demonstrates full visibility of
the interference fringes. The visibility is zero when $\eta=1$. This
figure corresponds to a system with zero temperature. \label{fig:thermal1}Squeezing
increases with $r$, the squeeze parameter.}
\end{figure}

\begin{figure}
\includegraphics[width=0.8\columnwidth]{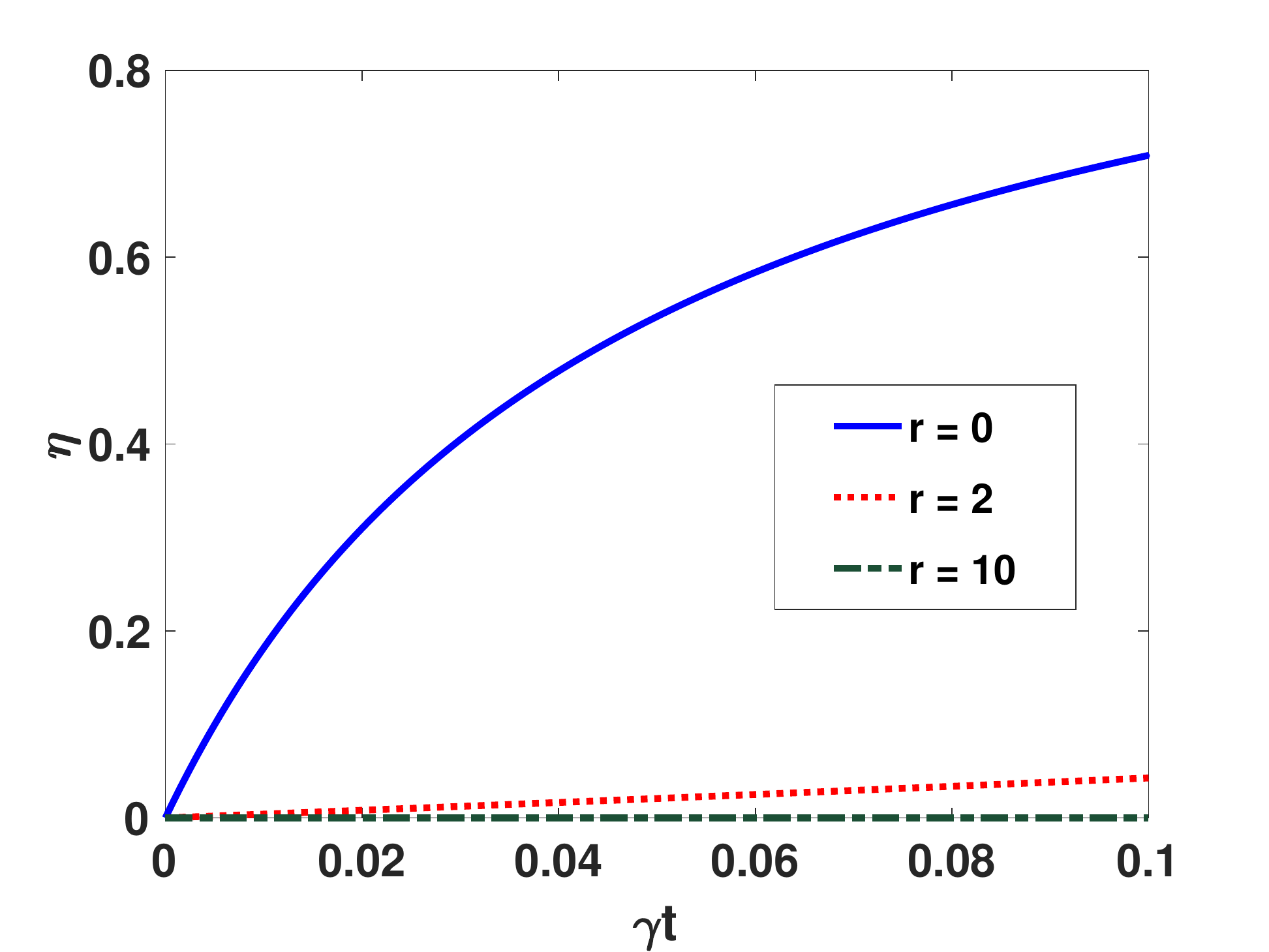}

\caption{Here, $\eta=0$ demonstrates full visibility of interference fringes.
The visibility is zero when $\eta=1$. This figure corresponds to
a squeezed thermal state at a finite temperature with $N_{th}=5$.
\label{fig:thermal2}}
\end{figure}

\begin{figure}
\includegraphics[width=0.8\columnwidth]{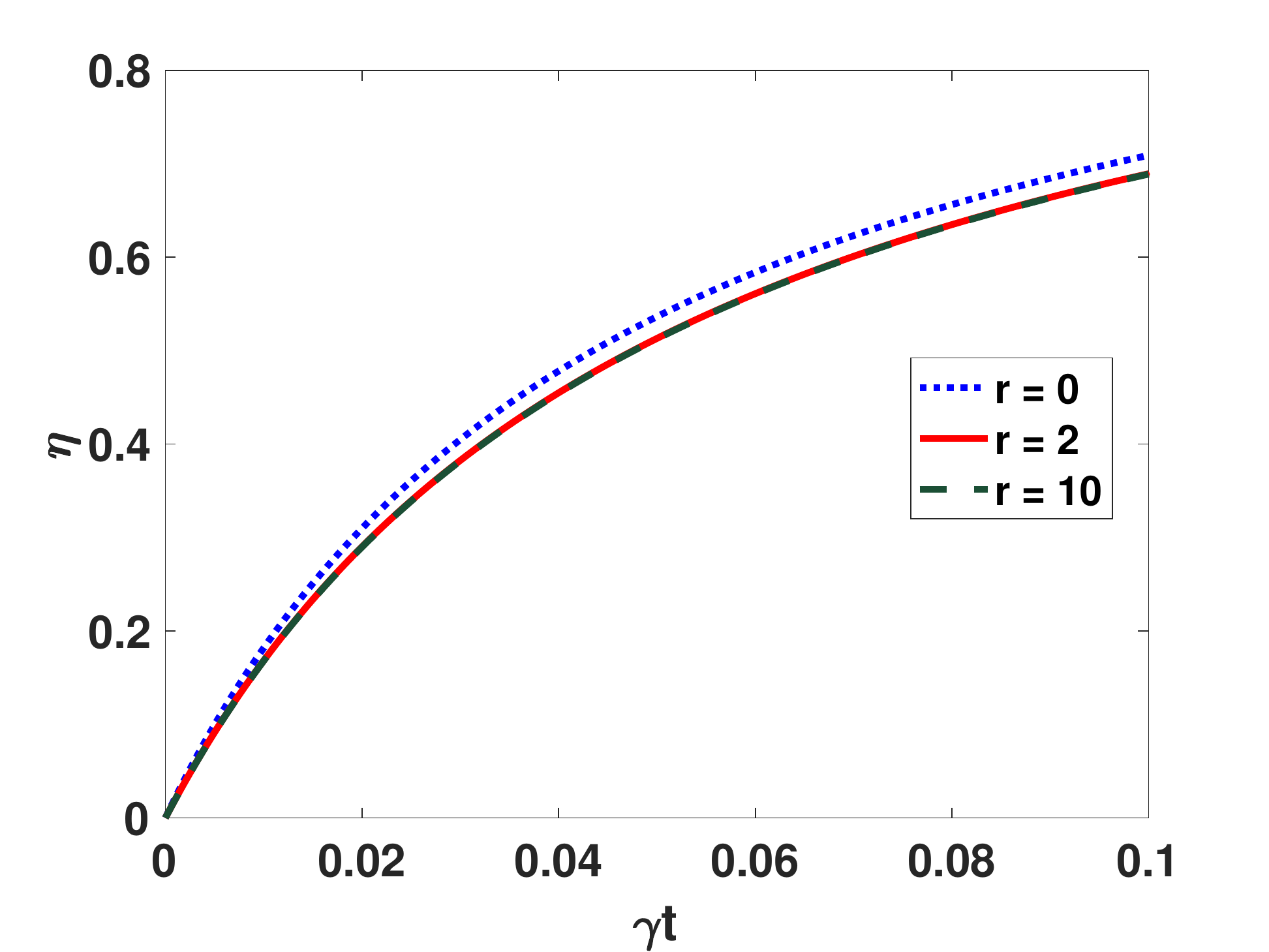}

\caption{Here, $\eta=0$ demonstrates full visibility of interference fringes.
The visibility is zero when $\eta=1$. This figure corresponds to
a thermalized squeezed state at a finite temperature with $N_{th}=5$.\label{fig:thermal3}\textcolor{red}{}}
\end{figure}
We now summarize the results for the two types of thermal states considered
above. We examine results for the type of squeezing that comes from
a DPO, denoting the squeezing parameter by $r$, and where $M=\sqrt{N_{s}\left(N_{s}+1\right)}$.
For a thermalized squeezed bath, $N=N_{th}+\sinh^{2}r=N_{th}+N_{s}$,
$r$ is the squeezing strength and $\left|M\right|=\frac{1}{2}\sinh2r$
as calculated in the above. For a squeezed thermal bath, $N=N_{th}\cosh2r+\sinh^{2}r$
and $\left|M\right|=\frac{1}{2}\left(2N_{th}+1\right)\sinh2r$. Figure
\ref{fig:thermal1} shows the decoherence where there is damping but
no thermal noise, so that $N_{th}=0$. In this limit, the decoherence
can be totally suppressed for longer times for sufficiently strong
squeezing, where $N_{s}\rightarrow\infty$.

The results for a squeezed thermal state are given in Figure \ref{fig:thermal3}.
Here, we see on comparing (\ref{eq:XP_var_pi}) with (\ref{eq:XP_TS})
that the effect on the squeezing variances is such that the effect
of the thermal noise term $N_{th}$ for the squeezed thermal state
is itself reduced by increasing squeezing parameter. This arises because
the technique of preparation is to reduce, or squeeze, the thermal
noise at the squeezing source. From the figure, we see that with a
sufficient amount of squeezing it is possible to completely suppress
the decoherence due to thermal noise for this type of squeezed thermal
input.

Looking at the thermalized squeezed state, we find for finite $N_{th}$
and infinite squeezing
\begin{align}
\eta & \rightarrow1-\frac{e^{-2\gamma t}}{\left\{ 1+2N_{th}\left(1-e^{-2\gamma t}\right)\right\} }
\end{align}
For long times, $\eta\rightarrow1-\frac{e^{-2\gamma t}}{\left\{ 1+2N_{th}\right\} }$.
This shows that thermal noise (where $N_{th}$ is fixed) remains
significant in increasing decoherence. For short times, we obtain
a linear response of $\eta$ with $\gamma t$, $\eta\sim4N_{th}\gamma t$.
These features are observed in Figure \ref{fig:thermal2}. Consistent
with results reported by Serafini et al \citep{serafini2004minimum,serafini2005quantifying,serafini2004decoherence},
we also note that increasing squeezing does not always lead to a decrease
of decoherence over a certain time.

\bibliography{DPO_squeeze_cat}

\end{document}